\documentclass[11pt,a4paper]{article}
\pdfoutput=1
\usepackage{jheppub}

\usepackage{multirow, graphicx,amssymb,url,mathrsfs,amsmath}
\usepackage{wrapfig,boxedminipage,setspace,subfigure,epsfig}
\usepackage{amsxtra,amstext,latexsym,dsfont,amsfonts}
\usepackage{color,eucal}
\usepackage[dvipsnames]{xcolor}
\usepackage{float}
\usepackage{slashed,comment}
\usepackage{kotex}

\usepackage{tikz}
\usetikzlibrary{calc,patterns,angles,quotes}
\usetikzlibrary{decorations.pathreplacing,decorations.markings,snakes}

\usepackage{tabularx, array}
\newcolumntype{L}[1]{>{\raggedright\arraybackslash}p{#1}}
\newcolumntype{C}[1]{>{\centering\arraybackslash}p{#1}}
\newcolumntype{R}[1]{>{\raggedleft\arraybackslash}p{#1}}

\DeclareMathOperator{\csch}{csch}
\DeclareMathOperator{\sech}{sech}

\newcommand{\dd}{\mathrm{d}}

\newcommand{\mO}{{\mathcal{O}}}

\title{Spectral and Krylov Complexity in Billiard Systems}

\author[a]{Hugo A. Camargo,}
\author[a]{Viktor Jahnke,}
\author[b,c]{Hyun-Sik Jeong,}
\author[a,d]{Keun-Young Kim,} 
\author[e]{and Mitsuhiro Nishida}

\emailAdd{hugo.camargo@gist.ac.kr}
\emailAdd{viktorjahnke@gist.ac.kr}
\emailAdd{hyunsik.jeong@uam.es}
\emailAdd{fortoe@gist.ac.kr}
\emailAdd{nishida124@postech.ac.kr}

\preprint{IFT-UAM/CSIC-23-61}

\affiliation[a]{Department of Physics and Photon Science, Gwangju Institute of Science and Technology,\\123 Cheomdan-gwagiro, Gwangju 61005, Korea}
\affiliation[b]{Instituto de F\'isica Te\'orica UAM/CSIC, Calle Nicol\'as Cabrera 13-15, 28049 Madrid, Spain}
\affiliation[c]{Departamento de F\'isica Te\'orica, Universidad Aut{\'o}noma de Madrid, Campus de Cantoblanco, 28049 Madrid, Spain}
\affiliation[d]{Research Center for Photon Science Technology, Gwangju Institute of Science and Technology, 123 Cheomdan-gwagiro, Gwangju 61005, Korea}
\affiliation[e]{Department of Physics, Pohang University of Science and Technology, Pohang 37673, Korea}

\abstract{
In this work, we investigate spectral complexity and Krylov complexity in quantum billiard systems at finite temperature. We study both circle and stadium billiards as paradigmatic examples of integrable and non-integrable quantum-mechanical systems, respectively. We show that the saturation value and time scale of spectral complexity may be used to probe the non-integrability  of the system since we find that when computed for the circle billiard, it saturates at a later time scale compared to the stadium billiards. This observation is verified for different temperatures. Furthermore, we study the Krylov complexity of the position operator and its associated Lanczos coefficients at finite temperature using the Wightman inner product. We find that the growth rate of the Lanczos coefficients saturates the conjectured universal bound at low temperatures. Additionally, we also find that even a subset of the Lanczos coefficients can potentially serve as an indicator of integrability, as they demonstrate erratic behavior specifically in the circle billiard case, in contrast to the stadium billiard. Finally, we also study Krylov entropy and verify its early-time logarithmic relation with Krylov complexity in both types of billiard systems.
}

\begin{document}
\maketitle

%%%%%%%%%%%%%%%%%%%%%%%%%%%%% 
%%%%%%%%%%%%%%%%%%%%%%%%%%%%
\section{Introduction}

The precise characterization of chaos in the quantum realm is an outstanding problem. The challenges involved in combining quantum mechanics and chaos were pointed out by Einstein in 1917, who noted
that the  Bohr--Sommerfeld quantization rule of the old quantum theory did not work in the case of classically chaotic systems \cite{einstein1917a}. In the case of integrable systems, the orbits in phase space lie on a torus and the quantization condition can be imposed by requiring the cross-sectional area of the torus to be an integral multiple value of Planck's constant. For chaotic systems, however, the orbits are irregular and unpredictable, exploring a constant-energy surface in phase space uniformly. In particular, the orbits do not lie in a torus and hence  there is no natural area to enclose an integral multiple of Planck's constant. It was not until the early 1970s that the significance of Einstein's observation was fully appreciated, and the fundamental difficulty in semiclassically quantizing chaotic systems was addressed with the development of periodic-orbit theory \cite{gutzwiller1971}.

Perhaps one of the most useful ways to characterize quantum chaotic behavior is through the statistics of the spacing between consecutive energy eigenvalues, the so-called level spacing statistics, which (normally) differs significantly depending on whether the system is chaotic or integrable. In chaotic systems, the level spacing statistics obeys a Wigner-Dyson distribution, mimicking the behavior observed in random matrix theories, while in integrable systems the level spacing statistics is expected to follow a Poisson distribution. In systems that have a well-defined classical limit, it is possible to obtain information about the spectrum from a complete enumeration of the periodic orbits using Gutzwiller trace formula \cite{gutzwiller1971,gutzwiller1990}. One particular successful application of the Gutzwiller trace formula was its use in the derivation of the level spacing statistics of integrable and chaotic systems \cite{berry1977, berry1985}. Level spacing statistics described by random matrix theory (RMT) also occurs in non-integrable systems without a strict classical limit. Because of that RMT universality is usually considered as a defining feature of chaotic systems.\footnote{There are, however, a few systems for which this criterion does not work. This is the case of geodesic flows on hyperbolic surfaces, which are classically chaotic, but their level spacing statistics exhibit properties that are more similar to integrable systems. See, for instance, \cite{PhysRevLett.69.1477, PhysRevLett.69.2188}.} However, a general mechanism explaining the emergence of RMT universality in chaotic systems without reference to semiclassical concepts is still missing.

Another useful way to diagnose chaos is through the spectral form factor  (SFF)
\begin{equation}
   \textrm{SFF}(t)\approx |Z(\beta+it)|^2 = \sum_{n,m} e^{-\beta(E_n+E_m)+it(E_n-E_m)} \,,
\end{equation}
which, like the level spacing statistics, only depends on the spectrum of the system. Here $\beta=1/T$ is the inverse temperature, and $E_n$ are the eigenvalues of the system. In random matrix theories, SFF as a function of time displays a typical slope-dip-ramp-plateau behavior. The same behavior is expected to occur in chaotic systems, while in integrable systems one expects a different behavior.\footnote{For many-body localized systems, the ramp is absent \cite{PhysRevResearch.3.L012019}, while for non-interacting disordered models, one finds an exponential ramp \cite{Winer:2020mdc}.}

Out-of-time-order correlators (OTOCs) \cite{Larkin1969QuasiclassicalMI} have received a lot of attention recently, due to their connection with black hole physics, and also for providing a useful way to diagnose scrambling, especially in the context of many-body quantum systems. See, for instance, the recent review \cite{Xu:2022vko}. In chaotic systems with a well-defined form of semiclassical limit, OTOCs are expected to decay exponentially with time \cite{Richter:2022sik}, in a time scale that is much smaller than the time scale at which the slope-ramp-plateau appears in the SFF. Because of that, OTOCs are usually considered to be an early-time notion of quantum chaos, while the level spacing statistics and the SFF are considered to be intermediate or late-time notions of quantum chaos. The exponential behavior of OTOCs resembles the divergence of the distance between initially nearby trajectories in phase space of classically chaotic systems, but the parameter characterizing such divergence in the quantum case, known as quantum Lyapunov exponent, is in general different from its classical counterpart \cite{Rozenbaum:2016mmv}. Moreover, the exponential behavior of OTOCs also occurs in integrable systems near a saddle point \cite{Xu:2019lhc, Hashimoto:2020xfr}. This suggests that one should distinguish the notions of chaos and scrambling. In fact, it has been suggested that scrambling is a necessary condition for chaos, but itself is not a sufficient condition \cite{Dowling:2023hqc}. In many-body systems that do not have a well-defined classical limit, like spin chains and quantum circuits, scrambling is defined in terms of the late-time vanishing of OTOCs, which happens in a non-exponential fashion. In those cases, one can distinguish chaotic from integrable systems based on the scrambling properties of the system (integrable systems do not scramble). Such criterion, however, encounters limitations as one decreases the system size. In those cases, finite-size effects prevent OTOCs to decay to zero, but their residual value can still provide useful insights into the chaotic dynamics \cite{Huang:2017fng}. 

In recent years, the notion of Krylov complexity has emerged as a promising tool in the study of quantum chaos, showing connections both with OTOCs and with SFF, and hence providing a bridge between early-time and late-time manifestations of quantum chaotic behavior. In its original formulation, the Krylov operator complexity measures the rate of growth of operators under Heisenberg time evolution. In this case, the problem can be mapped to a one-dimensional tight-binding model. The position $n$ of the particle in the chain, known as Krylov chain, measures the complexity of the time-evolved operator. The model is characterized by hopping terms $b_n$ which control the probability of the particle moving in the chain. The hopping terms are known as Lanczos coefficients, and it has been proposed that these coefficients grow as fast as possible in chaotic models as one moves along the Krylov chain \cite{Parker:2018yvk}. In the context of lattice models, it is possible to show that the maximal possible growth of the Lanczos coefficients is linear, i.e., $b_n = \alpha n + \gamma$. Under a few assumptions (thermalizing system), one can show that the maximal growth of Lanczos coefficients implies an exponential growth of Krylov complexity, i.e., $C_K (t) \sim e^{2 \alpha t}$. The connection of Krylov operator complexity with OTOCs comes from the fact that at infinite temperature one can prove that the rate of growth of the Lanczos coefficients bounds the quantum version of the Lyapunov exponent \cite{Parker:2018yvk}
\begin{equation}\label{INEQLAT}
    \lambda_L \leq 2 \, \alpha \,.
\end{equation}
The authors of \cite{Parker:2018yvk} conjecture that the above inequality is also valid at finite temperature. Moreover, it has been suggested that the following relation should be valid \cite{Avdoshkin:2022xuw}
\begin{equation}\label{INEQLAT2}
    \lambda_L \leq 2 \, \alpha \leq 2 \pi T\,,
\end{equation}
which provides a tighter bound than the Maldacena--Shenker--Stanford (MSS) bound on chaos \cite{Maldacena:2015waa}.  In QFTs on non-compact space, the maximal growth of Lanczos coefficients occurs universally due to the unbounded continuous spectra \cite{Dymarsky:2021bjq, Avdoshkin:2022xuw, Camargo:2022rnt}, and the exponential growth of Krylov operator complexity is not a suitable measure of chaos in this case. In 2d holographic CFTs on a compact space $S^1$, the Krylov operator complexity is sensitive to scaling dimensions of primary states \cite{Kundu:2023hbk}.

The notion of Krylov state complexity, also known as spread complexity, measures the spread of a wavefunction minimized over all possibles basis of the Hilbert space \cite{Balasubramanian:2022tpr}. The minimum is uniquely attained by the Krylov basis. For a maximally entangled state, the Krylov state complexity only depends on the spectrum of the Hamiltonian and it was shown to be a suitable probe of late-time chaos, having non-trivial connections with the SFF \cite{Balasubramanian:2022tpr, Erdmenger:2023shk}. Moreover, at early times, Krylov state complexity matches the so-called {\it spectral complexity}, whose definition is inspired by the result obtained for the regularized volume of the black hole interior in 2-dimensional models of quantum gravity~\cite{Iliesiu:2021ari} and which shares also intimate connections with other spectral quantities such as the SFF.

This non-trivial connection with OTOCs and SFF and potential connections with holography led to a wealth of studies of Krylov complexity for states and/or operators in different settings, ranging from quantum many-body systems~\cite{Barbon:2019wsy,Avdoshkin:2019trj,Dymarsky:2019elm,Rabinovici:2020ryf,Cao:2020zls,Kim:2021okd,Rabinovici:2021qqt,Trigueros:2021rwj,Balasubramanian:2022tpr,Fan:2022xaa,Heveling:2022hth,Bhattacharjee:2022vlt,Caputa:2022eye,Muck:2022xfc,Rabinovici:2022beu,He:2022ryk,Hornedal:2022pkc,Alishahiha:2022nhe,Alishahiha:2022anw, Bhattacharyya:2023dhp}, gauge theories~\cite{Magan:2020iac}, holographic models~\cite{Jian:2020qpp, Rabinovici:2023yex}, conformal field theories (CFT)~\cite{Dymarsky:2021bjq,Caputa:2021ori}, Lie groups~\cite{Caputa:2021sib,Patramanis:2021lkx, Patramanis:2023cwz, Chattopadhyay:2023fob}, matrix models~\cite{Iizuka:2023pov}, models of quantum quenches~\cite{Pal:2023yik} and open quantum systems~\cite{Bhattacharya:2022gbz,Liu:2022god,Bhattacharjee:2022lzy, Bhattacharya:2023zqt}. 

Krylov complexity has been extensively studied in the context of many-body quantum systems. In particular, the original paper \cite{Parker:2018yvk} has in mind the thermodynamic limit with many degrees of freedom. However, it remains largely unexplored in the context of simple quantum mechanical models with a few fields which are classically chaotic. This includes, for example, the case of dynamical billiards, which are paradigms for the study of classical and quantum chaos. At the same time, the surprising connection between  Krylov state complexity and spectral complexity may be seen as an indication that the latter could also be used to understand the chaotic properties of quantum many-body systems. In this work, we study Krylov operator complexity as well as spectral complexity for the stadium and circle billiards at finite temperatures. 

Furthermore, in addition to investigating Krylov complexity, we also examine Krylov entropy~\cite{Barbon:2019wsy}.\footnote{The physical meaning of the Krylov entropy may not be clear. For instance, see \cite{Barbon:2019wsy}, where authors examined Krylov entropy within the scrambling and post-scrambling regime for chaotic systems.} While these two quantities have been explored in the literature in various scenarios, we aim to further contribute to the understanding of their dynamics within simple quantum mechanical models that have a well-established history in the study of quantum chaos: billiard systems.

%While these two quantities have been explored in the literature in various scenarios, for instance~\cite{Caputa:2021sib,Jian:2020qpp,Rabinovici:2020ryf,Dymarsky:2021bjq,Kim:2021okd,Patramanis:2021lkx}, we aim to further contribute to the understanding of their dynamics within simple quantum mechanical models that have a well-established history in the study of quantum chaos: billiard systems. \HC{I think that most of these references already appear in the previous paragraph. So, perhaps we can just say: "While these two quantities have been explored in the literature in various scenarios, we aim to further contribute to the understanding of their dynamics within simple quantum mechanical models that have a well-established history in the study of quantum chaos: billiard systems.". What do you think?}

\paragraph{Summary of our results.}
Note that the billiard systems are bosonic systems whose Hilbert spaces are infinite dimensional spaces. To extract the features of chaos with finite degrees of freedom, we truncate the high-energy eigenstates of the system. This truncation introduces a saturation of the Lanczos coefficients, but the first Lanczos coefficients display a linear behavior ($b_n = \alpha n + \gamma$) from which we can extract the slope $\alpha$. Notably, we observe that the universal bound (\ref{INEQLAT2}) is saturated at finite and sufficiently low temperatures both for the stadium and circle billiard. However, as we decrease the temperature we note a striking difference between the two billiards: while the linear behavior of the Lanczos coefficients persists in the chaotic case (stadium billiard), the behavior of the Lanczos coefficients becomes completely erratic and non-linear with $n$ in the integrable case (circle billiard). We also study the corresponding early-time behavior of the Krylov complexity and Krylov entropy for different choices of operators. We checked that our numerical results satisfy basic consistency checks, including the normalization condition of wave functions, the Ehrenfest theorem, and universal relations with Krylov entropy. We observe that the operator growth tends to increase slowly at low temperatures and the qualitative features of our results do not depend on the choice of operator.

Studying the time behavior of the spectral complexity for both circle and stadium billiards, we find that the spectral complexity initially grows as $C_S \propto  c_1 \log \cosh (c_2 t/\beta)$, where $c_1$ and $c_2$ are constants, and then saturates (oscillates wildly around a constant value) in both cases. However, the saturation occurs much earlier in the chaotic case (stadium billiard), as compared to the integrable case (circle billiard). The phenomenon can be explained in terms of the tendency of level clustering (repulsion) in the spectrum of integrable (chaotic) systems. We also discuss possible generalizations of the concept of spectral complexity, which do not involve the full spectrum of the system, but only some symmetry sectors.

This paper is organized as follows. 
In section \ref{SEC2LAB}, we review the method to calculate the spectral complexity, the Krylov complexity and  Krylov entropy.
In section \ref{SEC3LAB}, using the formulas derived in section \ref{SEC2LAB}, we study the time evolution of the spectral and Krylov complexity/entropy for the billiard systems. 
Section \ref{sec:conclusion} is devoted to conclusions.\\

\paragraph{Note added:} While this paper was in preparation, ~\cite{Hashimoto:2023swv} appeared, which has some overlap with the present work. In both works, some part is devoted to the study of Krylov operator complexity in billiard systems. Our work focuses on finite temperature effects, which allow us to verify non-trivial bounds on the rate of growth of Lanczos coefficients, while \cite{Hashimoto:2023swv} focuses on the infinite temperature case. Moreover, we also introduce the analysis of spectral complexity as a novel measure of probing quantum chaos.

%Nevertheless, our conclusions seem to be consistent with each other.

%%%%%%%%%%%%%%%%%%%%%%%%%%%%%
%    
%%%%%%%%%%%%%%%%%%%%%%%%%%%%%
\section{Review of the formalism}\label{SEC2LAB}

%%%%%%%%%%%%%%%%%%%%%%%%%%%%%
%    
%%%%%%%%%%%%%%%%%%%%%%%%%%%%%
\subsection{Spectral complexity}

One of the earlier attempts to resolve the puzzle of the growth of the black hole interior for times beyond usual thermalization scales resulted in the well-known holographic ``complexity=volume'' proposal~\cite{Susskind:2014moa,Susskind:2014rva}. In it, the computational complexity of the thermofield double (TFD) state was proposed to be dual to the volume of a maximal codimension-$1$ slice of the black hole interior, i.e. the Einstein--Rosen Bridge (ERB). While both quantities exhibit an initial early-time growth, computational complexity is expected to saturate at times which are exponentially large in the entropy of the system~\cite{Balasubramanian:2019wgd,Balasubramanian:2021mxo,Haferkamp:2021uxo}. However, until recently it had been an open problem to establish whether the volume of the black hole interior would also exhibit a saturation at late times.

This was shown in~\cite{Iliesiu:2021ari}, where authors computed the non-perturbative volume of the black hole interior via the gravitational path-integral in $2$-dimensional models of gravity including higher-topology contributions. The regularized volume, which exhibits the aforementioned late-time saturation, led authors to define a new spectral quantity in the boundary theory dual to the volume of the ERB. To be precise, for quantum systems with discrete non-degenerate energy spectrum, authors defined the following quantity

%Motivated by the connection between the volume of the Einstein--Rosen Bridge (ERB) and complexity in holographic models of two-dimensional gravity, authors in~\cite{Iliesiu:2021ari} proposed a quantity which depends only on the spectral data of quantum systems with discrete energy levels and which they found to be the complexity of the thermofield double state (TFD), dual to the volume of the ERB. %Authors argued that this quantity, known as \emph{spectral complexity}, could be defined for any quantum system with a discrete spectrum according to

%
\begin{align}\label{eq:SpectralComplexityDef}
\begin{split}
  C_{S}(t):= \frac{1}{D Z(2\beta)}\sum_{p\neq q}\left[\frac{\sin\left(t(E_{p}-E_{q})/2\right)}{(E_{p}-E_{q})/2}\right]^2e^{-\beta(E_{p}+E_{q})}~,
\end{split}
\end{align}
where $Z(\beta)$ is the thermal partition function, $\lbrace E_{p} \rbrace$ are the discrete energy eigenvalues and $D$ is the dimension of the Hilbert space $\mathcal{H}$. This quantity, dubbed the \emph{spectral complexity} of the TFD state, is expected to have a similar behaviour to the volume of the ERB for chaotic systems with random-matrix (spectral) statistics, namely an initial early-time linear growth and a late-time saturation. This was shown in the case of the Sachdev--Ye--Kitaev (SYK) model.

It should be noted that the microcanonical definition of the spectral complexity (ERB volume)
%
%\begin{align}\label{eq:MicroSpectralComplexityDef}
%\begin{split}
 % C^{\textrm{micro}}_{S}(t)\approx -\int_{0}^{\infty}\textrm{d}E\int_{-2E}^{2E}\textrm{d}\omega\frac{\langle\rho(E+\omega/2)\rho(E-\omega/2)\rangle}{2\omega^{2}}\sin^{2}(t\omega/2)~,
%\end{split}
%\end{align}
%
is related to the microcanonical SFF
%
%\begin{align}\label{eq:MicroSFFDef}
%\begin{split}
 % \textrm{SFF}^{\textrm{micro}}(t)\approx \int_{0}^{\infty}\textrm{d}E\int_{-2E}^{2E}\textrm{d}\omega\frac{\langle\rho(E+\omega/2)\rho(E-\omega/2)\rangle}{2}\sin^{2}(t\omega/2)~,
%\end{split}
%\end{align}
%
in the following way
\begin{align}\label{eq:SFFSpectralComp}
\begin{split}
  \textrm{SFF}(t)=\frac{\textrm{d}^{2}C_{S}(t)}{\textrm{d}t^{2}}+\textrm{constant} \,.
\end{split}
\end{align}
Note, however, that this quantity depends only on the energy spectrum of the Hamiltonian, and more specifically, on the difference between distinct energy levels. 

While spectral complexity was introduced in the context of two-dimensional-models of holography, one may take the approach of studying this spectral quantity, beyond holography, in quantum systems at finite number of degrees of freedom. Furthermore, {following the original motivation, one may view this quantity directly as a measure of the computational complexity of the canonically-defined TFD state of the system.} As we will discuss in Sec.~\ref{subsec:CSresults} of this work, the saturation time-scale and value of this quantity may in fact be sensitive to the breaking of integrability in quantum mechanical systems such as billiards.

%Additionally, this quantity was shown to be related to the \emph{Krylov state complexity} of the TFD state via an Ehrenfest theorem~\cite{Erdmenger:2023shk}.
%In systems with a spectral statistics determined by random matrices, this quantity is expected to have a similar behavior to the volume of the ERB: i.e. an initial linear growth followed a late-time plateau.  

%%%%%%%%%%%%%%%%%%%%%%%%%%%%%%

%%%%%%%%%%%%%%%%%%%%%%%%%%%%%

%%%%%%%%%%%%%%%%%%%%%%%%%%%%%
\subsection{Krylov complexity and entropy}

In this section, we review two operator growth quantities, Krylov complexity~\cite{Parker:2018yvk} and Krylov entropy~\cite{Barbon:2019wsy}.

%%%%%%%%%%%%%%%%%%%%%%%%%%%%%
%
%%%%%%%%%%%%%%%%%%%%%%%%%%%%%
\subsubsection{Lanczos algorithm and operator growth}
\label{sec.LanczosAlgorithm}

%In this section, we provide a brief overview of operator growth, as well as of the concepts of Krylov complexity and entropy. 
\paragraph{Operator growth in the Heisenberg picture.}
In a quantum system with Hamiltonian $H$, the time evolution of an operator $\mO(t)$ is governed by the Heisenberg equation
\begin{align}\label{SP}
\begin{split}
     \partial_t \mO(t) = i [H, \mO(t)]   \,, \qquad \mO(t)  \,=\, e^{i H t} \,\mO_0\, e^{-i H t} \,,
\end{split}
\end{align}
where $\mO_0 := \mO(0)$.
Formally, we can rewrite the operator $\mO(t)$ as a power-series
\begin{align}\label{OTEXP}
\begin{split}
    \mO(t)  \,=\,& \mO_0 + i t  [H, \mO_0] + \frac{(it)^2}{2!} [H, [H,\mO_0]]  + \dots \,, \\  
    \,:=\,& \tilde{\mO}_0 + i t  \tilde{\mO}_1 + \frac{(it)^2}{2!} \tilde{\mO}_2  + \dots \,,
\end{split}
\end{align}
where the initial operator $\tilde{\mO}_0\equiv \mO_0$ can be seen as evolving in time within the operator space spanned by the operators $\lbrace \tilde{\mO}_0,\tilde{\mO}_1,\tilde{\mO}_2,\ldots\rbrace$. This expression shows how an initial ``simple" operator may grow increasingly more ``complex" through its commutation with the Hamiltonian. Additionally, a more ``chaotic" $H$ may lead to a more complex $\mO(t)$ as compared to a less ``chaotic" one.

\paragraph{Krylov basis and Lanczos algorithm.}

%Given an operator $\tilde{\mO}_n$ in \eqref{OTEXP},  one may construct the basis $|\tilde{\mO}_n)$ in the operator space, which is in general not orthonormal. One can define a set of orthonormal bases, $|{\mO}_n)$, in the operator space which is called \textit{Krylov basis}. Krylov basis is constructed by imposing an iterative Gram-Schmidt orthogonalization for some operator $\mO_n$. This process is known as the \textit{Lanczos algorithm}.

Given the initial operator $\tilde{\mO}_0$ in \eqref{OTEXP}, one typically studies its evolution by using the Gelfand--Naimark--Segal (GNS) construction of the operator algebra, whereby one constructs a Hilbert space $\mathcal{H}_{\mathcal{O}}$ spanned by the states $\lbrace |\tilde{\mO}_n) \rbrace$, obtained by the successive application of the \emph{Liouvillian} super-operator $\mathcal{L}:=[H,\cdot]$ on the initial state $\vert \tilde{\mO}_0)$. This Hilbert space, known as the \emph{Krylov space}, admits an orthonormal basis $|{\mO}_n)$, known as the \textit{Krylov basis}, which can be obtained by performing the Gram--Schmidt orthogonalization procedure of the original basis. This procedure is known as the \emph{Lanczos algorithm} and in order to perform it we require an inner product in Krylov space. Since we are interested in understanding the physics at finite temperature and we would like to connect our results to the study of quantum chaos, we will work with the \emph{Wightman inner product}.

%Before introducing the Lanczos algorithm, we need to determine {the inner product} used in the operator space. We take the \textit{Wightman inner product}:
%
\begin{align}\label{}
\begin{split}
(A|B) = \frac{1}{Z} \text{Tr}\left( e^{-H\beta/2} A^{\dagger} e^{-H\beta/2} B \right) \,, \,\,\, Z = \text{Tr} \left( e^{-\beta H} \right) \,,
\end{split}
\end{align}
where $\beta$ is the inverse temperature.
After inserting the completeness relation $\sum_{\ell} \left| \ell \right>\left<\ell\right|=\mathbb{I}$, we obtain the expression
\begin{align}\label{WMIP}
\begin{split}
  (A|B) = \frac{1}{Z} \sum_{m, \ell} e^{-\frac{\beta}{2} (E_m + E_\ell)}  A_{m\ell}^{\dagger} B_{\ell m} \,,
\end{split}
\end{align}
where we introduced the short-hand notation $A_{m\ell} = \left<m|A|\ell\right>$ for the matrix elements.

With the definition of the inner product, we can construct the Krylov basis following the Lanczos algorithm:
\begin{enumerate}\label{}
\item{Define $\vert A_0):=\vert\tilde{\mO}_0)$ and normalize it with $b_0:=\sqrt{(A_0|A_0)}$ to find the orthonormal basis element $\vert \mO_0 ) := b_{0}^{-1} \vert A_0 )$. }
\item{Define $\vert A_1 ):=\mathcal{L}\vert \tilde{\mO}_0 )$, and normalize it with $b_1 = \sqrt{(A_1|A_1)}$. Define the normalized basis element $\vert \mO_1 ) := b_{1}^{-1} \vert A_1 )$.}
\item{For $n\geq2$, given $\vert \mO_{n-1} )$ and $\vert \mO_{n-2} )$, construct the successive basis element
\begin{align}\label{}
\begin{split}
  \vert A_n ) = \mathcal{L}\vert \mO_{n-1}) - b_{n-1} \vert \mO_{n-2} ) \,.
\end{split}
\end{align}
It can be normalized by setting $b_n = \sqrt{(A_n|A_n)}$ and by defining the $n$-th basis element as $\vert \mO_n ) := b_{n}^{-1} \vert A_n )$.
}
\item{Stop the algorithm when $b_n$ becomes zero.}
\end{enumerate}
The Lanczos algorithm thus yields the Krylov basis $\lbrace |\mO_n) \rbrace$ and the normalization coefficients $\lbrace b_n \rbrace$ known as the \textit{Lanczos coefficients}. \\

\paragraph{Krylov complexity and entropy.}
Once we have obtained the Krylov basis $\lbrace |\mO_n) \rbrace$ with the Lanczos algorithm, we can use it to write the GNS state $\vert \mO(t))$ as
\begin{align}\label{OKB1}
\begin{split}
  |\mO(t)) := \sum_{n} \, i^n \, \varphi_n(t) \, |\mO_n) \,,
\end{split}
\end{align}
where the functions $\lbrace \varphi_n(t)\rbrace $ are known as the transition amplitudes. These describe the distribution of the operator across the Krylov basis, illustrating how the operator is allocated within the Krylov space. By multiplying both sides of \eqref{OKB1} with $(\mO_n|$, we obtain an explicit expression for $\varphi_n(t)$
\begin{align}\label{EQVA}
\begin{split}
  \varphi_n(t) = i^{-n} (\mO_n|\mO(t)) \,, \qquad \sum_n |\varphi_n(t)|^2 =1 \,,
\end{split}
\end{align}
where the second relation shows the conservation of probability in the Krylov basis.

Note that using \eqref{WMIP}, the inner product in \eqref{EQVA} can be further expressed as
\begin{align} \label{OnOt}
(\mO_n|\mO(t))=\frac{1}{Z} \sum_{m, \ell} e^{-\frac{\beta}{2} (E_m + E_\ell)}  e^{it(E_\ell-E_m)}  \left<m|\mO^\dagger_n|\ell\right> \left<\ell|\mO_0|m\right> \,,
\end{align}
where we used $ \mO(t)  \,=\, e^{i H t} \,\mO_0\, e^{-i H t}$ in \eqref{SP}.
It is important to note that the computation of the Wightman inner product \eqref{OnOt} only requires the numerical integration of $\left<m|\mO_n|\ell\right>$, where $\lbrace \mO_n \rbrace$ are obtained via the Lanczos algorithm. As a result, we can efficiently compute the transition amplitudes $\varphi_n(t)$ in \eqref{EQVA}.
%Note that in order to calculate the value of the Wightman inner product \eqref{OnOt}, we only need to numerically integrate $\left<m|\mO_n|\ell\right>$ where $\mO_n$ is constructed by Lanczos algorithm. 
%
%Consequently, then, we can compute the amplitude $\varphi_n(t)$ in \eqref{EQVA}.

The dynamics of the operator growth may be conceptualized as a particle moving along a chain. As the particle progresses further along the chain, it engages with increasingly complex states within the Krylov basis. This can be seen by noting that by substituting~\eqref{OKB1} into the Heisenberg equation~\eqref{SP} leads to the recursion relation
%Together with \eqref{OKB1}, one can also find that the Heisenberg time evolution equation \eqref{SP} gives rise to the following recursion relation
%
\begin{align}\label{DCEQ}
\begin{split}
  \partial_t \varphi_n(t) = b_n \varphi_{n-1}(t) - b_{n+1} \varphi_{n+1}(t) \,,
\end{split}
\end{align}
subject to the boundary condition $\varphi_{n}(0) = \delta_{n0}$ and $b_0 = \varphi_{-1}=0$.
The Lanczos coefficients can be interpreted as hopping amplitudes that allow the initial operator to traverse the ``Krylov chain", while the functions $\varphi_n(t)$ can be visualized as wave packets moving along this chain~\cite{Parker:2018yvk}.
%The physics of the growth can be understood as a motion of a particle on a chain. The further in the chain in the particle is, the more complex state in the Krylov basis is employed. 
This motivates a natural definition of complexity as average position on the chain as well as the entropy
\begin{align}\label{KCF}
\begin{split}
  &\text{Krylov complexity:} \quad  C_{K} := \sum_n n |\varphi_n(t)|^2 \,, \qquad \\
  & \text{Krylov entropy:} \quad S_{K} := -\sum_n |\varphi_n(t)|^2 \log |\varphi_n(t)|^2 \,,
\end{split}
\end{align}
where the amplitudes $\varphi_n(t)$ are given by \eqref{EQVA}.
Krylov complexity computes the average position of the distribution on the Krylov basis, whereas Krylov entropy quantifies the level of randomness within the distribution~\cite{Rabinovici:2020ryf}.\\
%K-complexity computes the average position of the distribution on the ordered Krylov basis, while K-entropy computes how randomized the distribution is.\\

\paragraph{Numerical procedure for Krylov complexity and entropy.}
In order to compute the transition amplitudes~\eqref{EQVA}, and consequently the Krylov complexity and Krylov entropy~\eqref{KCF}, we need to find the matrix elements $\langle m\vert \mO^{\dagger}_{n}\vert \ell\rangle$ and $\langle \ell\vert \mO_{0}\vert m\rangle$. This requires that we find the energy eigenvalues and eigenfunctions for a given Hamiltonian $H$, i.e. solve the Schr\"{o}dinger equation. Thus, the numerical approach that we will follow can be summarized as follows:
\begin{enumerate}\label{}
\item{Numerically solve the Schr\"{o}dinger equation and obtain the corresponding energy eigenvalues $\{E_n\}$ and eigenfunctions $\{\psi_n\}$.}
\item{For different $m$ and $\ell$, compute $\left<m|\mO_{0}|\ell\right> = \int \,\dd \mathbf{x} \, \psi_{m}^{\dagger}\, \mO_{0} \,\psi_\ell$ numerically.}
\item{Implement the Lanczos algorithm by writing the Wightman inner product involved in each step as a function of $\left<m|\mO_{0}|\ell\right>$, then substitute in $\left<m|\mO_{0}|\ell\right>$ obtained by step 2.}
\item{Through the Lanczos algorithm, obtain the Krylov basis $|\mO_n)$ and Lanczos coefficients $b_n$.}
\item{Calculate the transition amplitudes $\varphi_n$ using \eqref{EQVA}. The corresponding Krylov complexity and entropy can then be obtained by substituting $\varphi_n$ into \eqref{KCF}.}
\end{enumerate}
In this work, we will choose the initial operator to be the position operator $\mO_0\equiv x$. We checked that we obtain qualitatively similar results for other choices of operators, namely, $x^2$, $p$, and $p^2$. 

\paragraph{Analytic expression for the operator growth at early times.}

By solving the equation \eqref{DCEQ} at early times, the authors in \cite{Fan:2022xaa} found analytic expressions for Krylov complexity and entropy in the following way. First, by imposing the boundary condition at $t=0$ in \eqref{DCEQ}, one finds that
\begin{align}\label{}
\begin{split}
  \varphi_n(0) = \delta_{n0} \,,\quad \dot{\varphi}_n(0)  = b_1 \delta_{n1} \,, \quad \ddot{\varphi}_n(0) = -b_1^2 \delta_{n0} + b_1 b_2 \delta_{n2} \,,
\end{split}
\end{align}
which yields
\begin{align}\label{}
\begin{split}
  C_{K}(0) &= 0 \,, \quad \dot{C}_{K}(0) = 0\,, \quad \ddot{C}_{K}(0) = 2 b_1^2 \,, \\
  S_{K}(0) &= 0 \,, \quad \dot{S}_{K}(0) = 0\,, \quad \ddot{S}_{K}(0) = -2 b_1^2 \log \left(b_1^2 t^2 \right) -4 b_1^2 \,.
\end{split}
\end{align}
%
%Then, together with these condition, \eqref{KCF} gives rise to 
These initial conditions together with~\eqref{KCF} give rise to the equations
\begin{align}\label{KCF2}
\begin{split}
  C_{K} = b_1^2 t^2 + \dots \,, \quad 
  S_{K} = -b_1^2 t^2 \log \left(b_1^2 t^2\right) + b_1^2 t^2 + \dots \,,
\end{split}
\end{align}
where the ellipses denote higher-order corrections in $t$.
Note that using \eqref{KCF2}, one can also find the universal logarithmic relation between $C_K$ and $S_K$ at early times as
\begin{align}\label{KCF3}
\begin{split}
  S_{K} = -C_K(t) \log C_K(t) + C_K(t) + \dots \,.
\end{split}
\end{align}
In section~\ref{sec:OperatorGrowthResults}, we will analyze our numerical findings alongside the corresponding analytic formulas given by equations \eqref{KCF2}-\eqref{KCF3}.

%%%%%%%%%%%%%%%%%%%%%%%%%%%%%
%%%%%%%%%%%%%%%%%%%%%%%%%%%%
\section{The billiard systems}\label{SEC3LAB}

\subsection{Quantum mechanics of stadium billiards}
\begin{figure}[]
\centering
     \includegraphics[width=8.0cm]{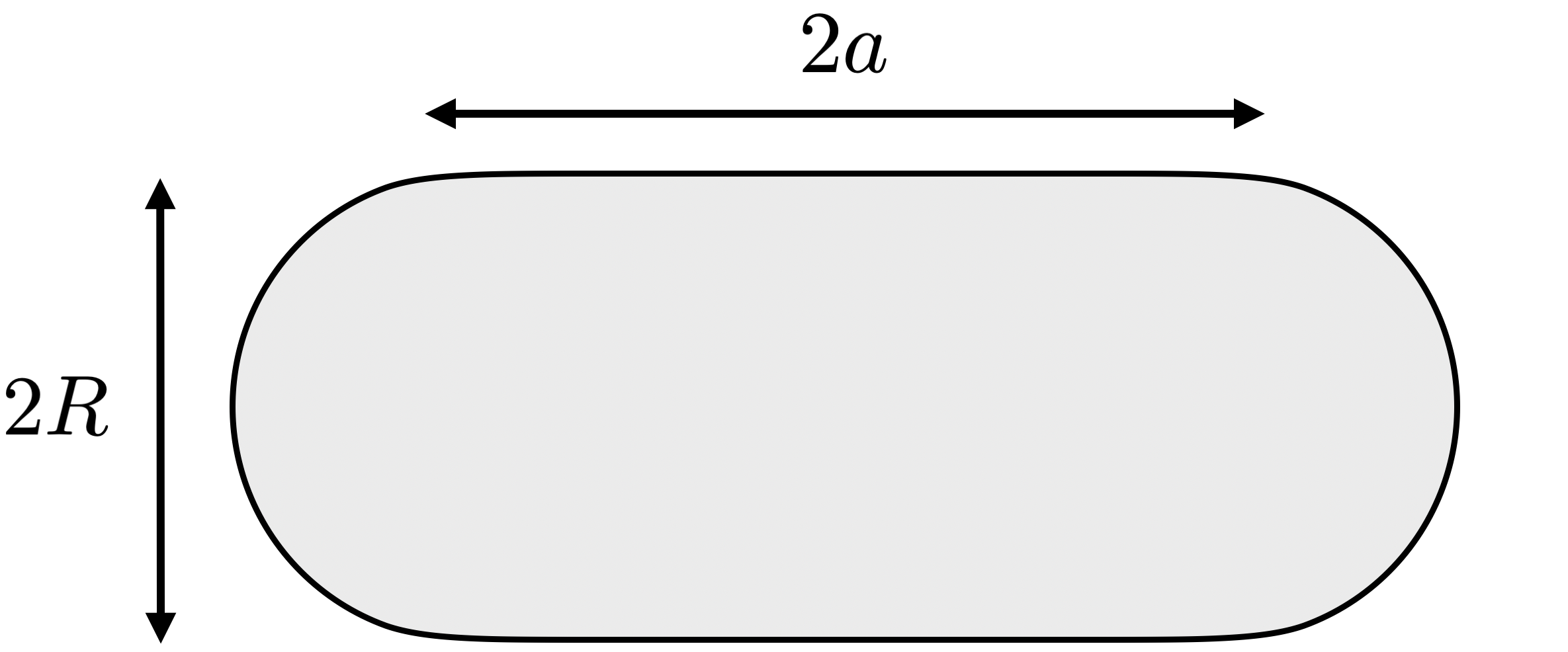}
 \caption{Geometry of the stadium billiard.}\label{skaa}
\end{figure}

%In this section, we study the complexity and entropy in 2-dimensional quantum mechanical systems. In particular, we consider a stadium billiard~\cite{Sinai_1970,Bunimovich_1975,Bunimovich_1979,Bunimovich_1991,Benettin:1978aa} as a  representative example of the non-integrable (i.e., chaotic) system:
We consider a stadium billiard~\cite{Sinai_1970,Bunimovich_1975,Bunimovich_1979,Bunimovich_1991,Benettin:1978aa} as a representative example of a 2-dimensional non-integrable (i.e., chaotic) system. The Hamiltonian of such system is given by
\begin{align}\label{BILLMO}
\begin{split}
  H = p_x^2 + p_y^2 + V_{\text{stad}}(x,y) \,,\qquad V_{\text{stad}}(x,y) =  
   \begin{cases}
       0 & (x,y)\in \Omega \\
       \infty & \textrm{else}
   \end{cases}
\,,
\end{split}
\end{align}
where the domain $\Omega$ is shown in Fig. \ref{skaa}.
The stadium consists of semicircles with radii $R$ together with straight lines of length $2a$. 
The area $A$ of the stadium is given by
\begin{align}\label{}
\begin{split}
 A = \pi R^2 + 4 a R \,.
\end{split}
\end{align}
%
%In this paper, following the previous literature~\cite{Hashimoto:2017oit,Hashimoto:2023swv} studying the Krylov complexity (or thermal OTOC) of the stadium billiard, we also set $A=1$ in the computations.
We fix the area of the billiard to $1$ following existing literature which deals with the study of Krylov complexity and thermal OTOCs in the same system~\cite{Hashimoto:2017oit,Hashimoto:2023swv}.

It is worth noting that the parameter $a/R$, where $a/R=0$ corresponds to the circle billiard, serves as a meaningful and relevant deformation parameter in the context of the stadium billiards~\cite{BenettinStochastic,McDonaldSpectrum,CasatiSpectra}. It can be observed that the circle billiard displays characteristics of an integrable system, such as a vanishing Lyapunov exponent~\cite{Hashimoto:2017oit} while the stadium billiard ($a/R\neq0$) is considered non-integrable with a finite Lyapunov exponent.

\begin{figure}[]
\centering
     \subfigure[Stadium billiard ($a/R=1$)]
     {\includegraphics[width=7.0cm]{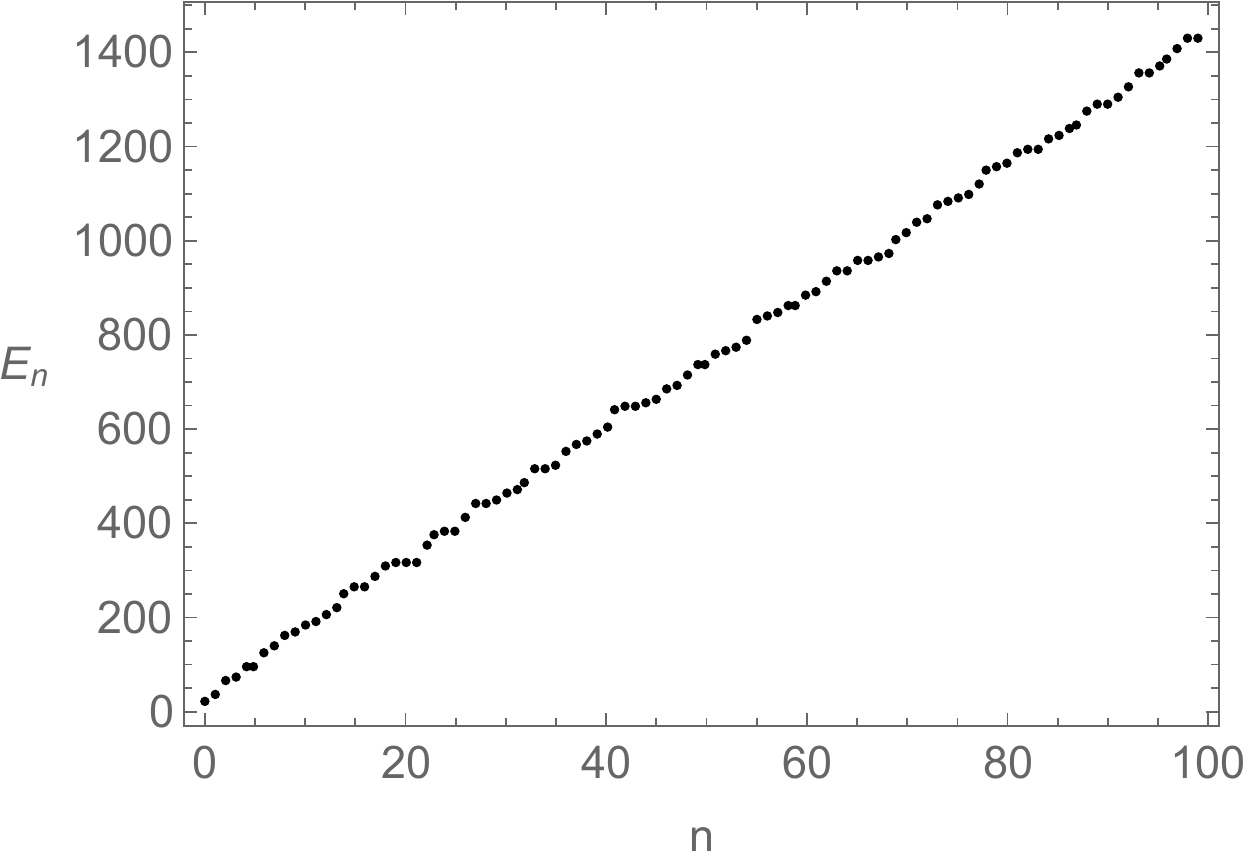} \label{}}
     \subfigure[Circle billiard ($a/R=0$)]
     {\includegraphics[width=7.0cm]{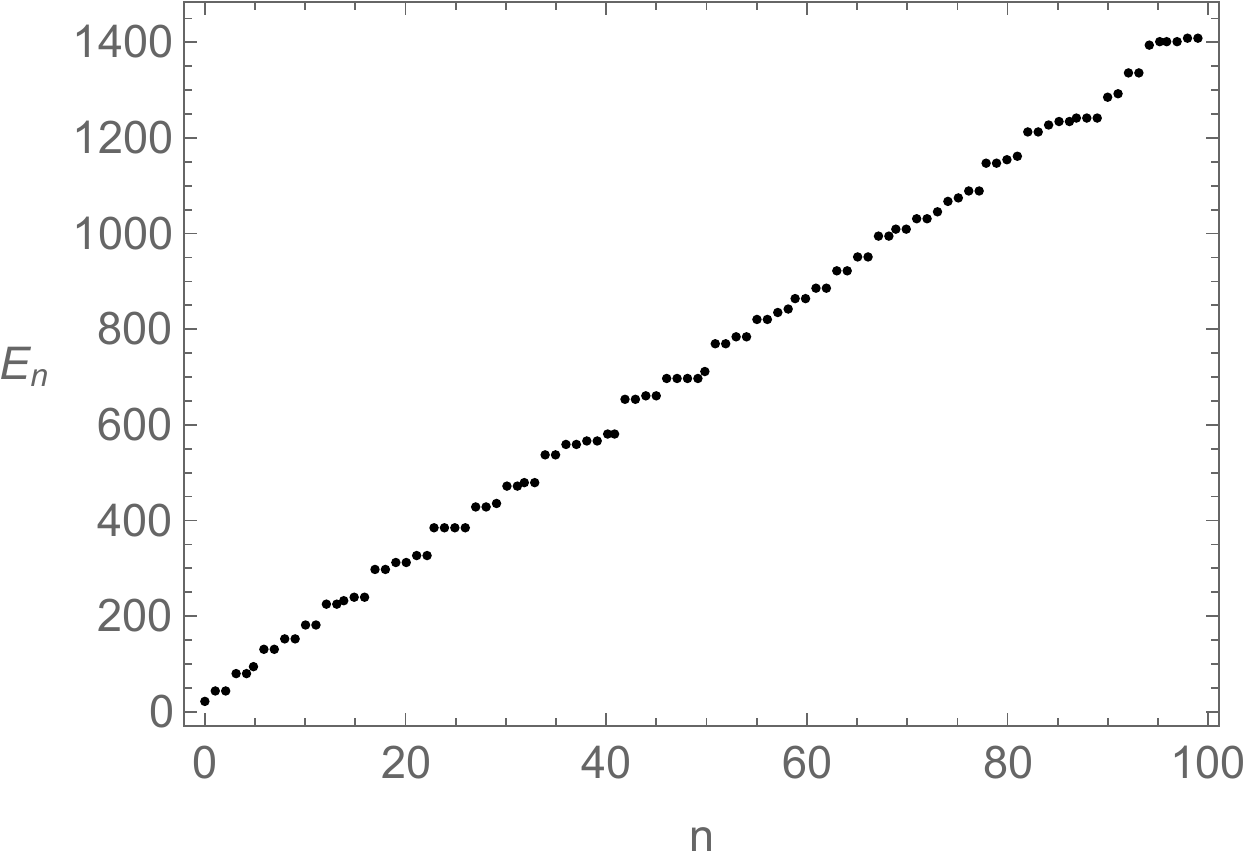} \label{}}
 \caption{Eigenvalues of the quantum billiard systems. \textbf{Left:} the stadium billiard ($a/R=1$). \textbf{Right:} the circle billiard ($a/R=0$).}\label{EIGENSTA}
\end{figure}

\begin{figure}[]
\centering
     \subfigure[Stadium billiard: $E_1=22.7$]
     {\includegraphics[width=4.8cm]{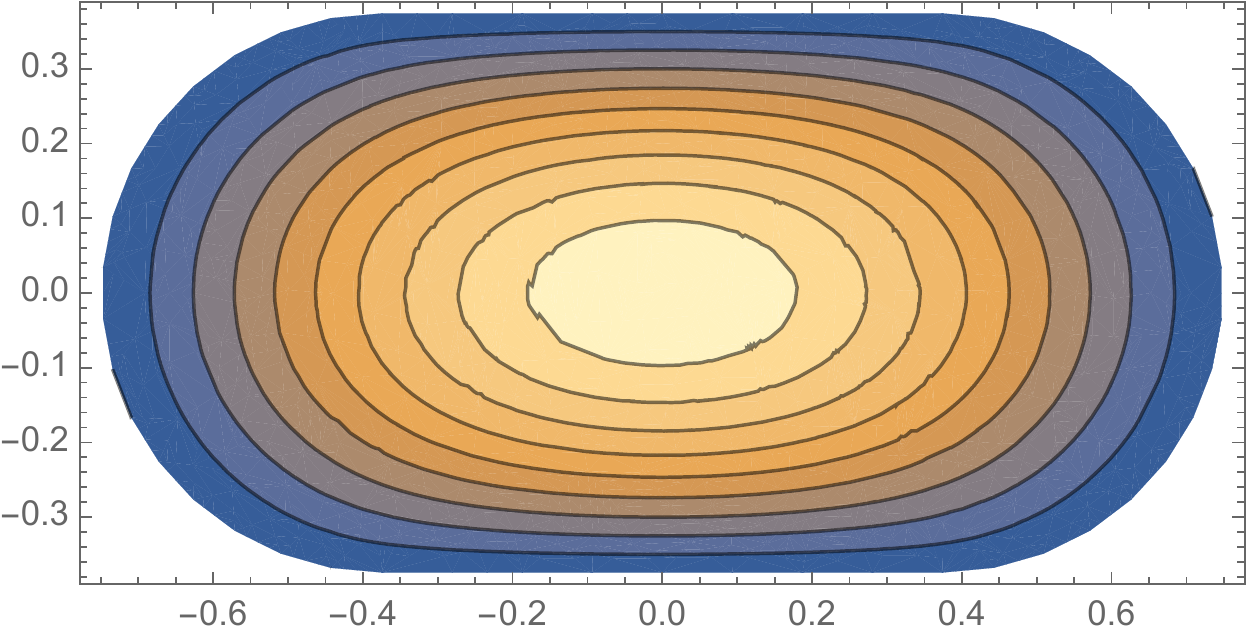} \label{}}
     \subfigure[Stadium billiard: $E_7=123.5$]
     {\includegraphics[width=4.8cm]{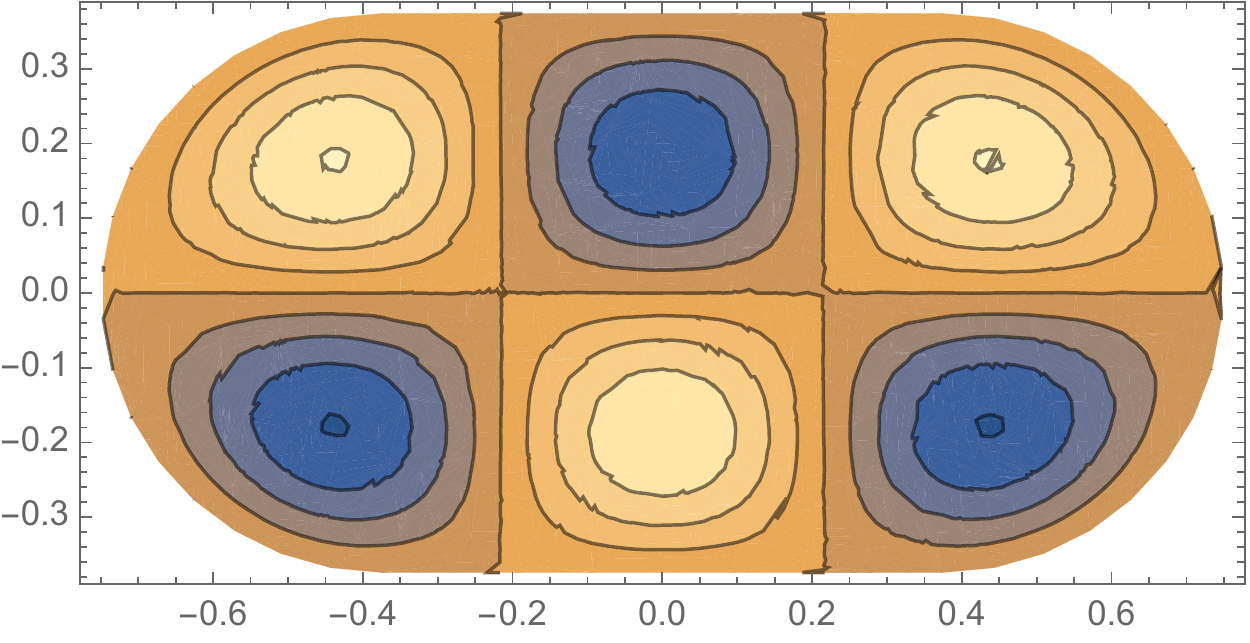} \label{}}
     \subfigure[Stadium billiard: $E_{50}=734.1$]
     {\includegraphics[width=4.8cm]{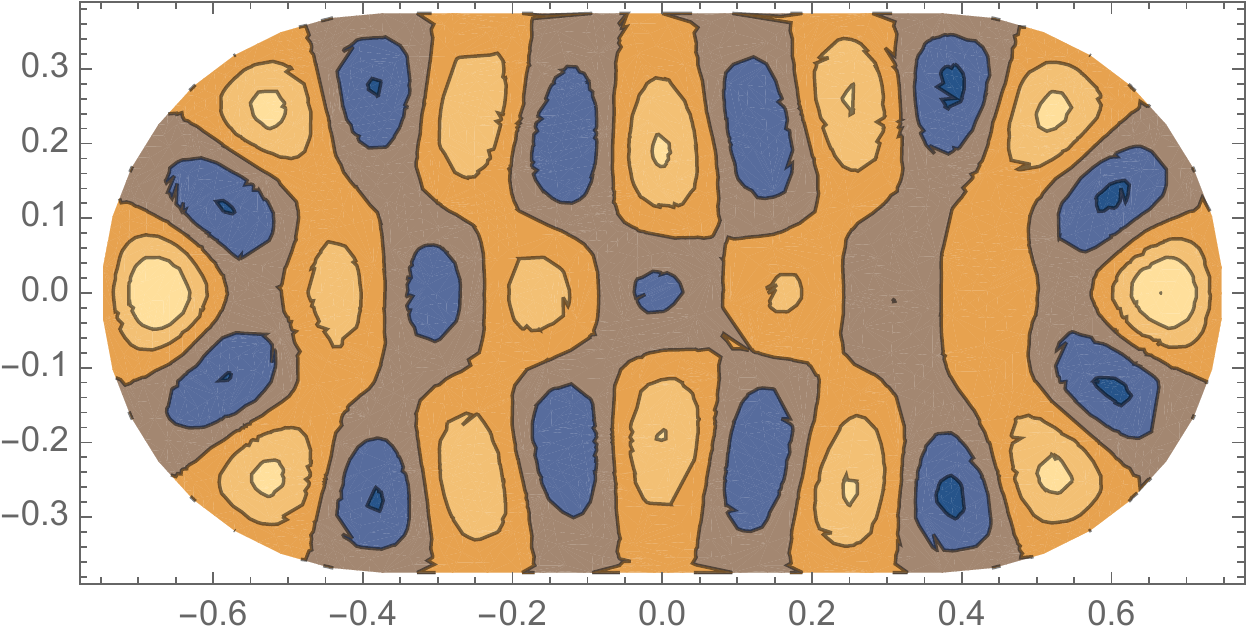}\label{}}

     \vspace{0.30cm}
     
     \subfigure[Circle billiard: $E_1=18.1$]
     {\includegraphics[width=4.8cm]{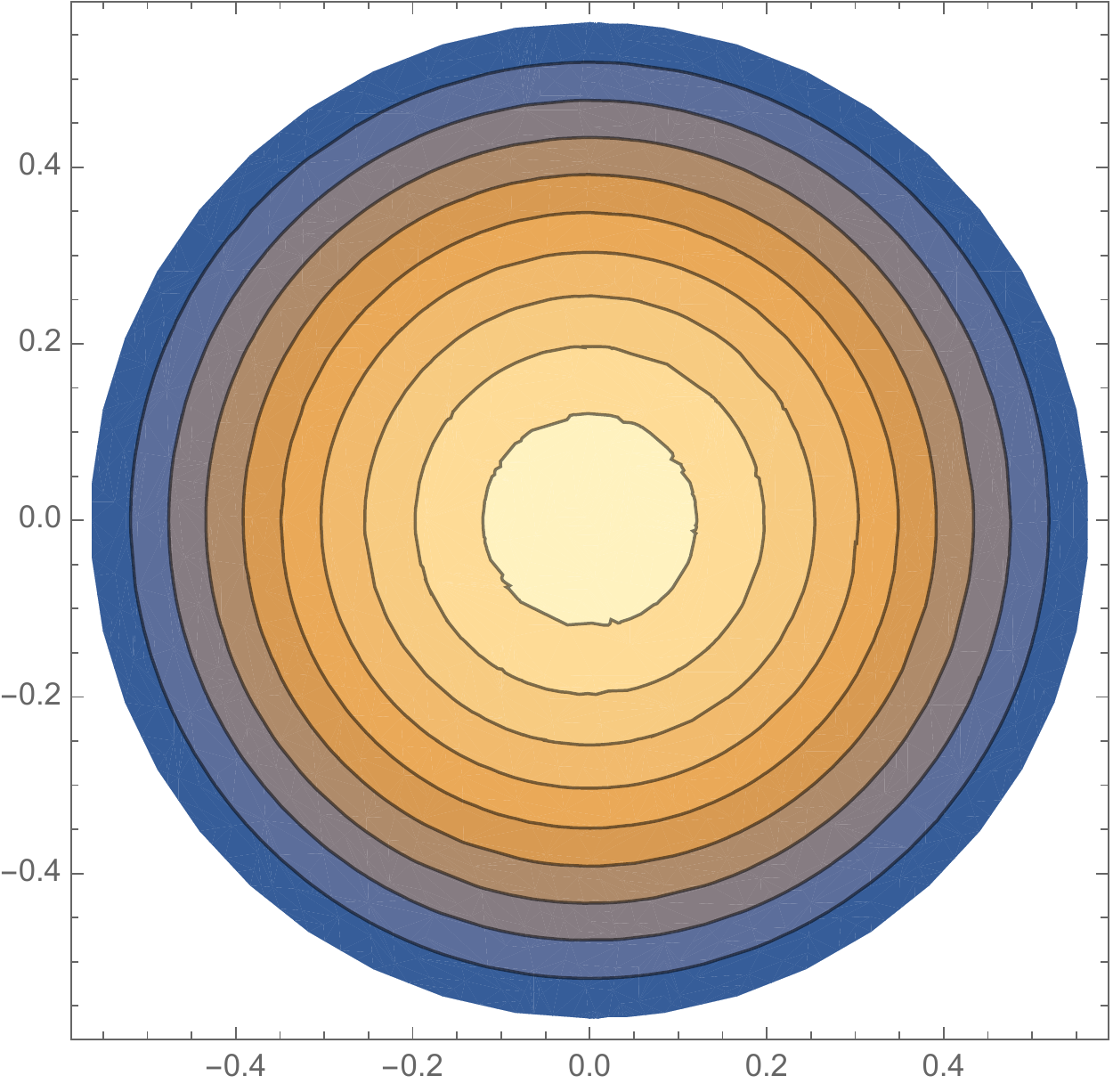} \label{}}
     \subfigure[Circle billiard: $E_7=127.9$]
     {\includegraphics[width=4.8cm]{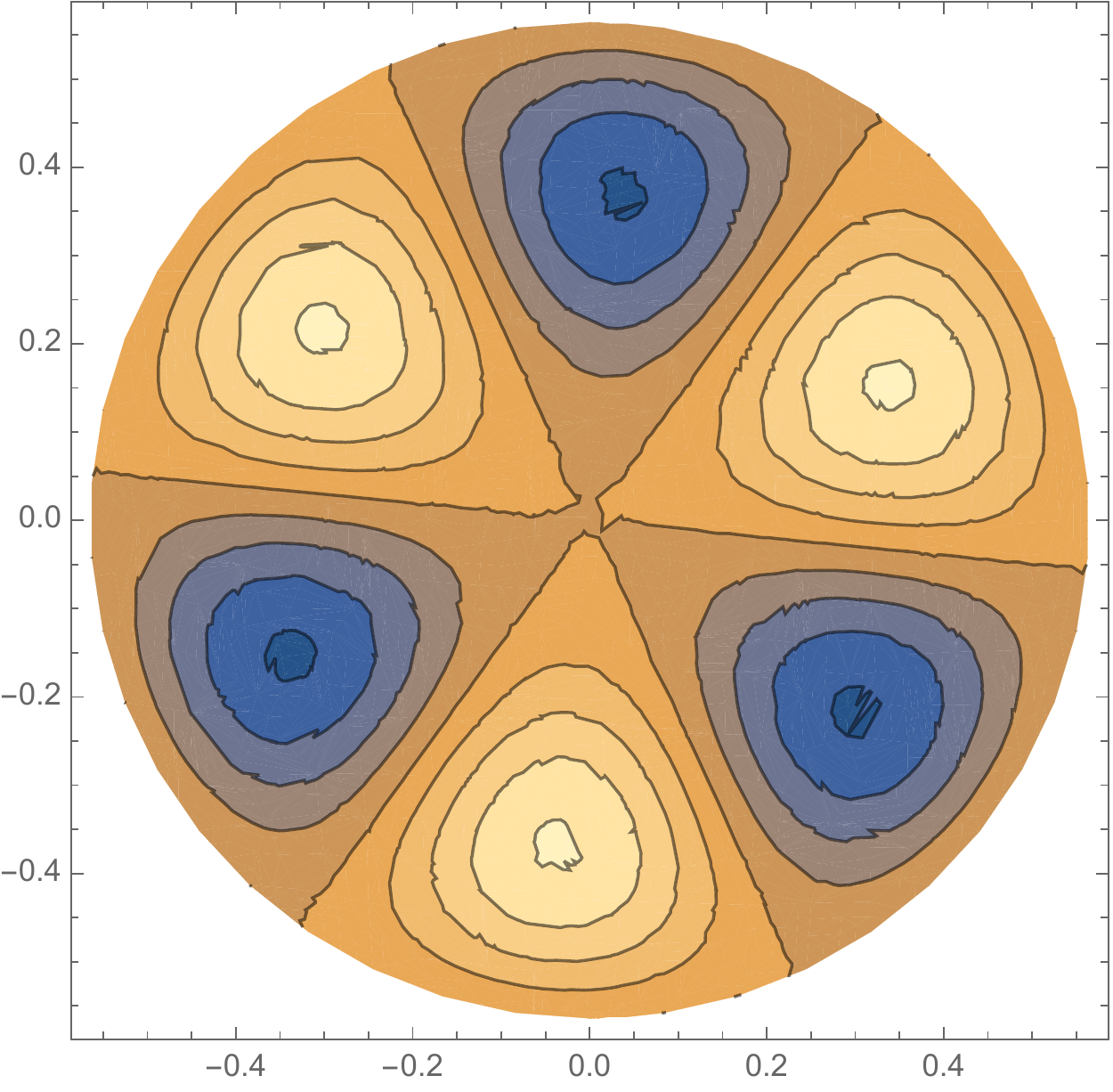} \label{}}
     \subfigure[Circle billiard: $E_{50}=696.9$]
     {\includegraphics[width=4.8cm]{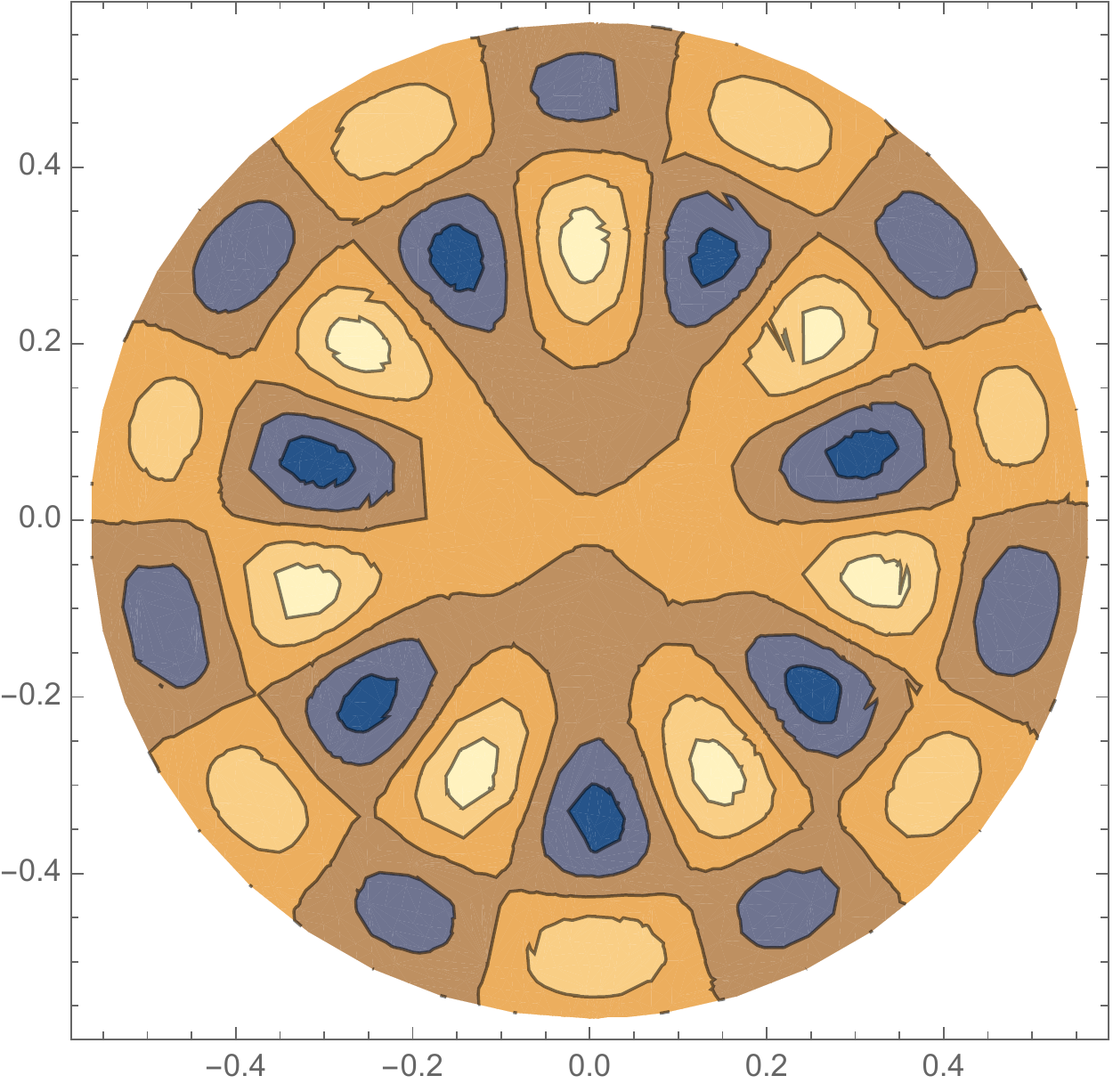} \label{}}
 \caption{Eigenfunctions of the quantum billiard systems for $n=1, 7, 50$. \textbf{Upper panels (a)-(c):} the stadium billiard ($a/R=1$). \textbf{Lower panels (d)-(f):} the circle billiard ($a/R=0$).}\label{EIGENstaSTA}
\end{figure}

\paragraph{The quantum mechanics in the billiard.}
In order to study the quantum mechanics in the billiard, we consider the Schr\"{o}dinger equation
\begin{align}\label{SCHREQ}
\begin{split}
  - \frac{\dd^2}{\dd x^2} \psi_n (x, y)  - \frac{\dd^2}{\dd y^2} \psi_n (x, y)   + V_{\text{stad}}(x,y)  \psi_n (x, y) = E_n \psi_n (x, y) \,,
\end{split}
\end{align}
where the potential is given in \eqref{BILLMO}.
We numerically solve this equation and obtain the energy eigenvalues $E_n$ and eigenstates $\psi_n (x, y)$.
In Fig. \ref{EIGENSTA}, we display the energy eigenvalues for the stadium billiard with $a/R=1$ and circle billiard with $a/R=0$.
We also plot the first few eigenstates of standard/circle billiards in Fig. \ref{EIGENstaSTA}.

Note that in order to calculate the Krylov complexity and entropy of the quantum billiards, we need to truncate the sum, for instance, in \eqref{KCF}. In this paper, we set $n\leq N_{\text{trunc}}=100$ as in \cite{Hashimoto:2017oit,Hashimoto:2023swv}. Furthermore, we also choose $a/R=1$ for the representative result of the stadium billiard, i.e.,
\begin{align}
\begin{split}
  \text{Stadium billiard:} \,\,\, \frac{a}{R} = 1 \,, \qquad \text{Circle billiard:} \,\,\, \frac{a}{R} = 0 \,.
\end{split}
\end{align}
%

%%%%%%%%%%%%
\subsection{Spectral complexity}
\label{subsec:CSresults}

In this section we discuss how the spectral complexity~\eqref{eq:SpectralComplexityDef} behaves in the billiard systems. The essential ingredient is the energy spectrum, which we obtain numerically following Eq.~\eqref{SCHREQ} for $N_{\textrm{trunc}}=100$.

\begin{figure}[]
\centering
     \subfigure[Spectral complexity]
     {\includegraphics[width=7.3cm]{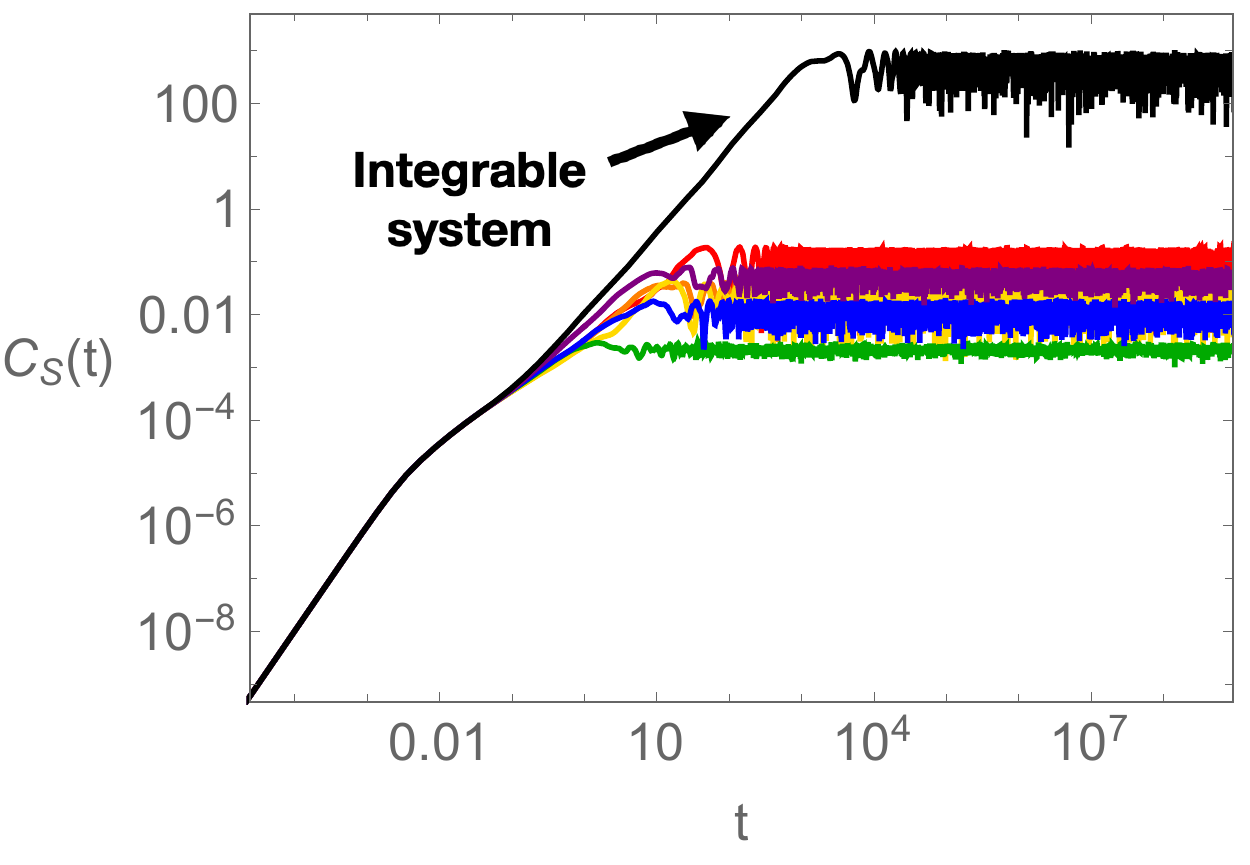} \label{}}
     \subfigure[Spectral complexity at $t = 10^{10}$]
     {\includegraphics[width=7.3cm]{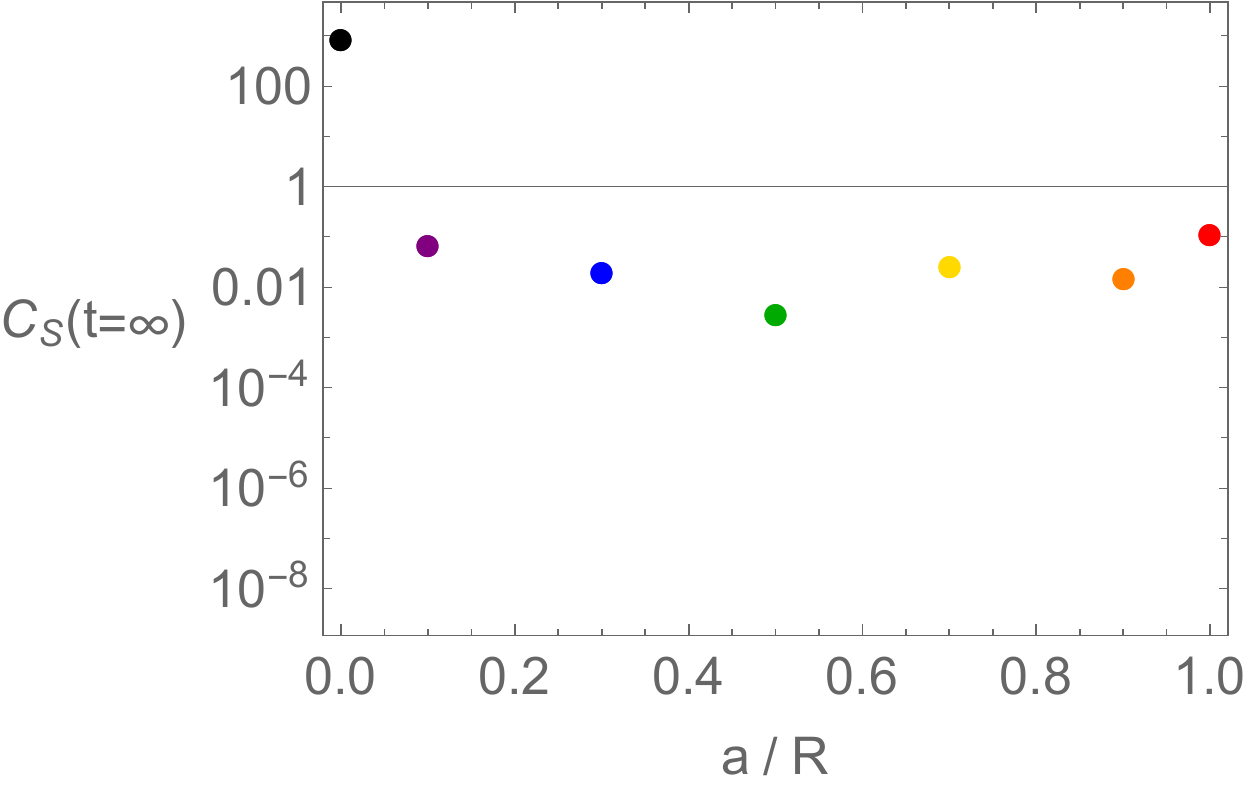} \label{}}
 \caption{Spectral complexity at infinite temperature with $a/R = 0, 0.1, 0.3, 0.5, 0.7, 0.9, 1$ (black, purple, blue, green, yellow, orange, red).}\label{SCFIG11}
\end{figure}
In Fig.~\ref{SCFIG11}, we show how its behavior at infinite temperature changes as we tune the value of $a/R$. Here we observe a distinct feature in the late-time regime: the spectral complexity saturates at a later time scale for the circle (integrable) billiards as compared to the stadium (non-integrable) billiards. In particular, this shows that even a small breaking of integrability, by virtue of a change of the ratio $a/R$, leads to a change in the saturation time scale and value of the spectral complexity. As a consequence, this finding may reveal the importance of the saturation time-scales and value of spectral complexity as a distinguishing factor between integrable and non-integrable systems.

The late-time behavior of the spectral complexity can be understood as follows. At infinite temperature ($\beta=0$), the spectral complexity is given by
\begin{align}
\begin{split}
  C_{S}(t)\vert_{\beta=0}= \frac{1}{D^2}\sum_{p\neq q}\left[\frac{\sin\left(t(E_{p}-E_{q})/2\right)}{(E_{p}-E_{q})/2}\right]^2\leq \frac{1}{D^2}\sum_{p\neq q}\left[\frac{1}{(E_{p}-E_{q})/2}\right]^2.
\end{split}
\end{align}
This expression shows that the complexity is bounded by the energy difference $\Delta E := E_p-E_q$. In other words, the saturation value of the complexity can be larger when $\Delta E$ is small. However, in chaotic systems, the presence of level repulsion prevents $\Delta E$ from becoming zero. As a result, the saturation value of the complexity for the stadium billiard can be smaller compared to the circle billiard.

We note, however, a stark contrast with results for Krylov complexity in spin systems with integrability-breaking terms~\cite{Rabinovici:2021qqt,Rabinovici:2022beu} as well as for Krylov state complexity for maximally entangled states in billiards at infinite temperature~\cite{Hashimoto:2023swv}. In the first case, the Krylov complexity achieves a maximum saturation value for chaotic Hamiltonians with random matrix spectral statistics given by $K/2$ where $K$ is the dimension of the Krylov space~\cite{Rabinovici:2020ryf}. In the absence of integrability-breaking terms, the saturation value of Krylov complexity is smaller than this maximum value. In the latter case, the Krylov state complexity for the circle billiard has the lowest saturation value compared to the stadium billiards, which do not appear to have a monotonically decreasing behavior as the ratio $a/R$ is increased. We comment more on these differences in Sec.~\ref{sec:conclusion}.

Regarding the saturation value of complexity, it should be noted that even in the case of operator growth in spin chains, the saturation value does not strictly follow a monotonous behavior as the parameter which introduces the integrability-breaking term is increased. This can be understood from the perspective of the energy level statistics, which does not have a smooth transition from Poissonian (integrable) to GOE (non-integrable) statistics as measured, for example, by the so-called $r$-parameter (see e.g. Fig. 2 of~\cite{Rabinovici:2022beu}). This also occurs in billiards (see e.g. Fig. 2 of~\cite{Hashimoto:2023swv}).

Another difference is the time-scale at which the saturation occurs: the saturation of Krylov complexity for states and operators occurs at the same time-scale, as computed in the aforementioned works, whereas for spectral complexity the saturation time-scale for the non-integrable billiards occurs several orders of magnitude earlier than the one for the integrable one.

%This implies that the spectral complexity may be able to capture distinctive features related to chaos which are not captured by the state complexity alone in the late-time regime (also recall the table. \ref{TABKO}).
%Here we observe a distinct feature in the late-time regime: the spectral complexity exhibits a maximum for the integrable systems. This finding may reveal the significance of the spectral complexity as a distinguishing factor between integrable and non-integrable systems. This implies that the spectral complexity may be able to capture distinctive features related to chaos which are not captured by the state complexity alone in the late-time regime (also recall the table. \ref{TABKO}).

\begin{figure}[]
\centering
     \subfigure[Stadium billiard ($a/R=1$)]
     {\includegraphics[width=7.3cm]{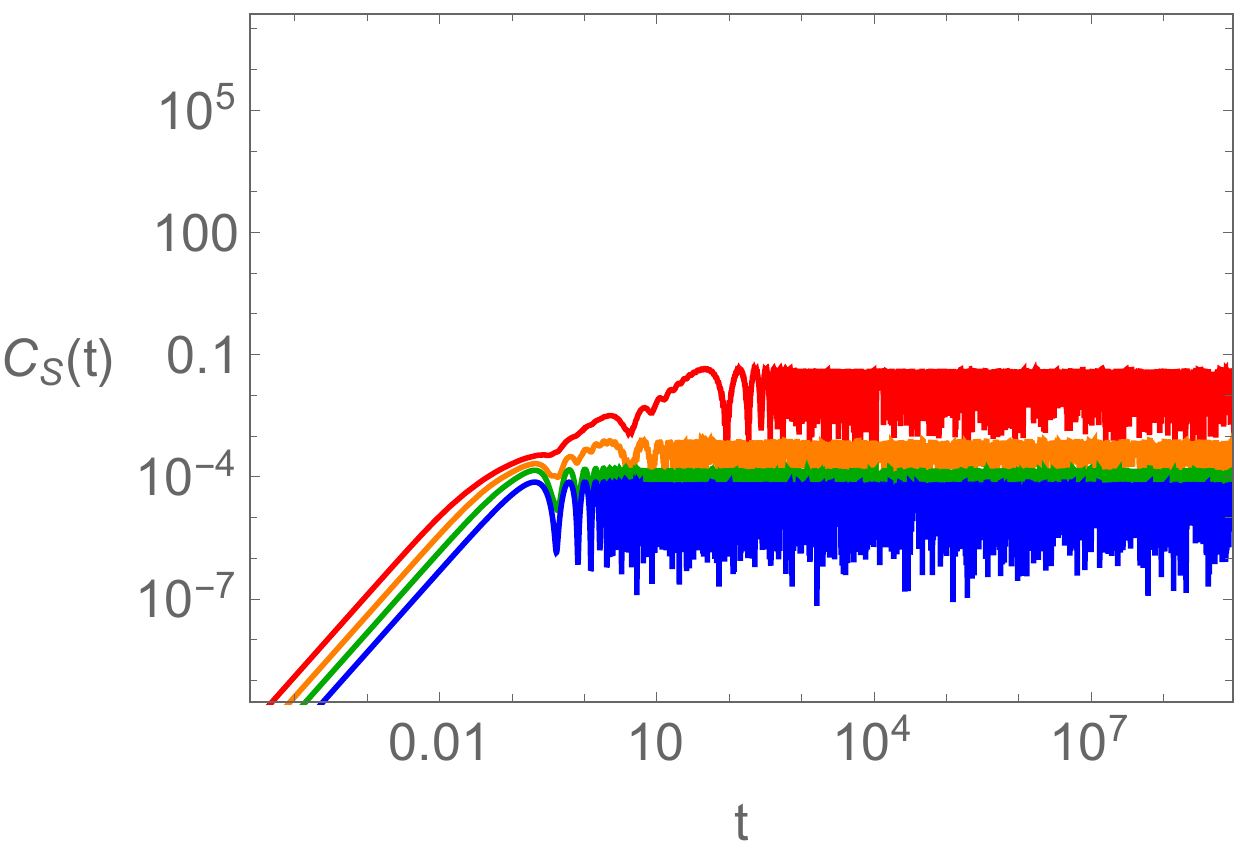} \label{}}
     \subfigure[Circle billiard ($a/R=0$)]
     {\includegraphics[width=7.3cm]{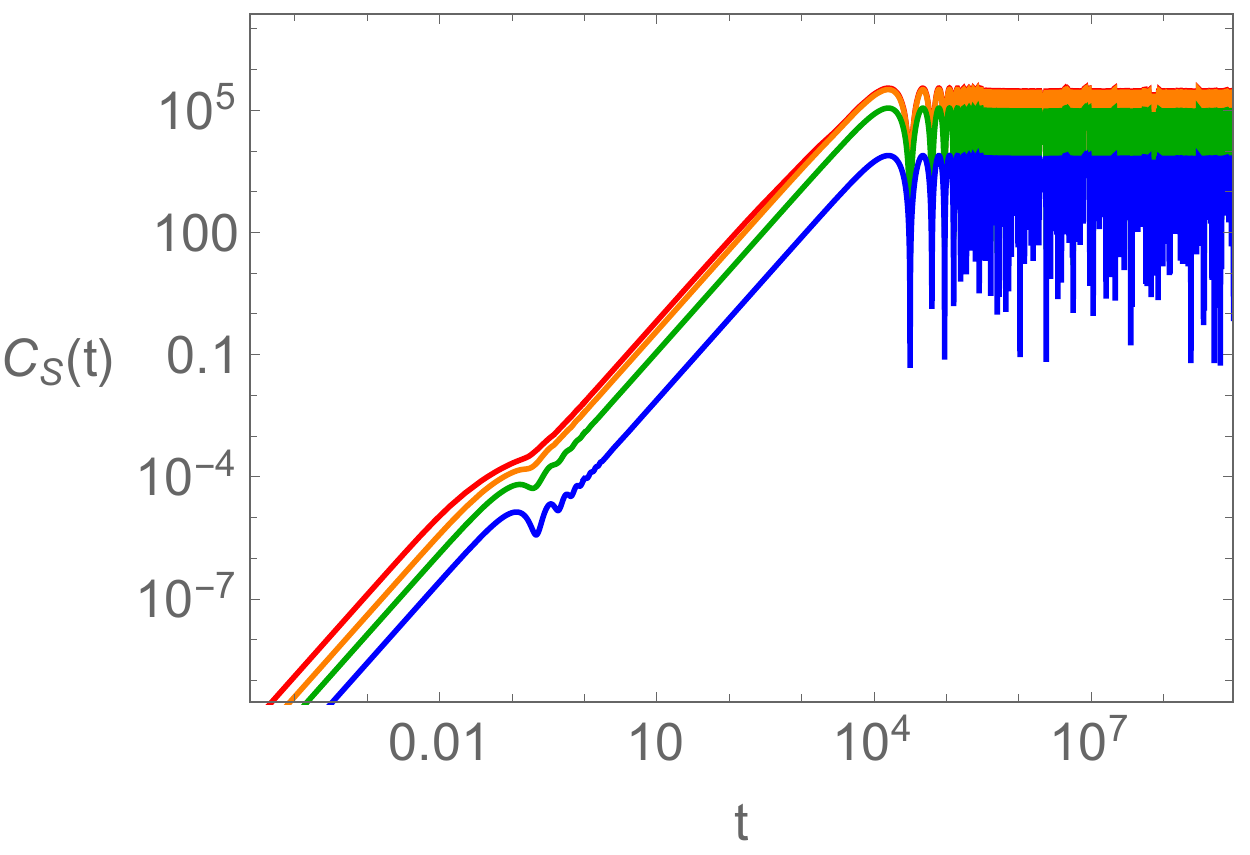} \label{}}
 \caption{Temperature dependence of spectral complexity at $T = 10, 20, 40, 100$ (blue, green, orange, red).}\label{SCFIG22}
\end{figure}

In Fig. \ref{SCFIG22}, we compare the temperature dependence of spectral complexity for the stadium ($a/R=1$) and circle ($a/R=0$) billiards.
Finally, we find that the spectral complexity decreases as the temperature is lowered for both the stadium and circle billiards.
This decreasing behavior is a common feature shared with the operator complexity explored in the following section and is consistent with the fact that lower temperatures have a higher exponential suppression according to~\eqref{eq:SpectralComplexityDef}.

Furthermore, in both cases we also find that the early-time behavior of the spectral complexity is given by $C_{S}\propto c_{1}\log(\cosh(c_{2}t/\beta))$ at finite temperature.

%%%%%%%%%%%%
\subsection{Operator growth}
\label{sec:OperatorGrowthResults}
%While we primarily focus on the case where the initial operator is ${x}$, we also examined cases involving other operators such as ${x}^2$ or ${p}$. However, we found that the qualitative findings remain consistent across these different choices.

\subsubsection{Lanczos coefficients and universal bound on its growth}

\begin{figure}[]
\centering
     \subfigure[Stadium billiard ($a/R=1$)]
     {\includegraphics[width=7.0cm]{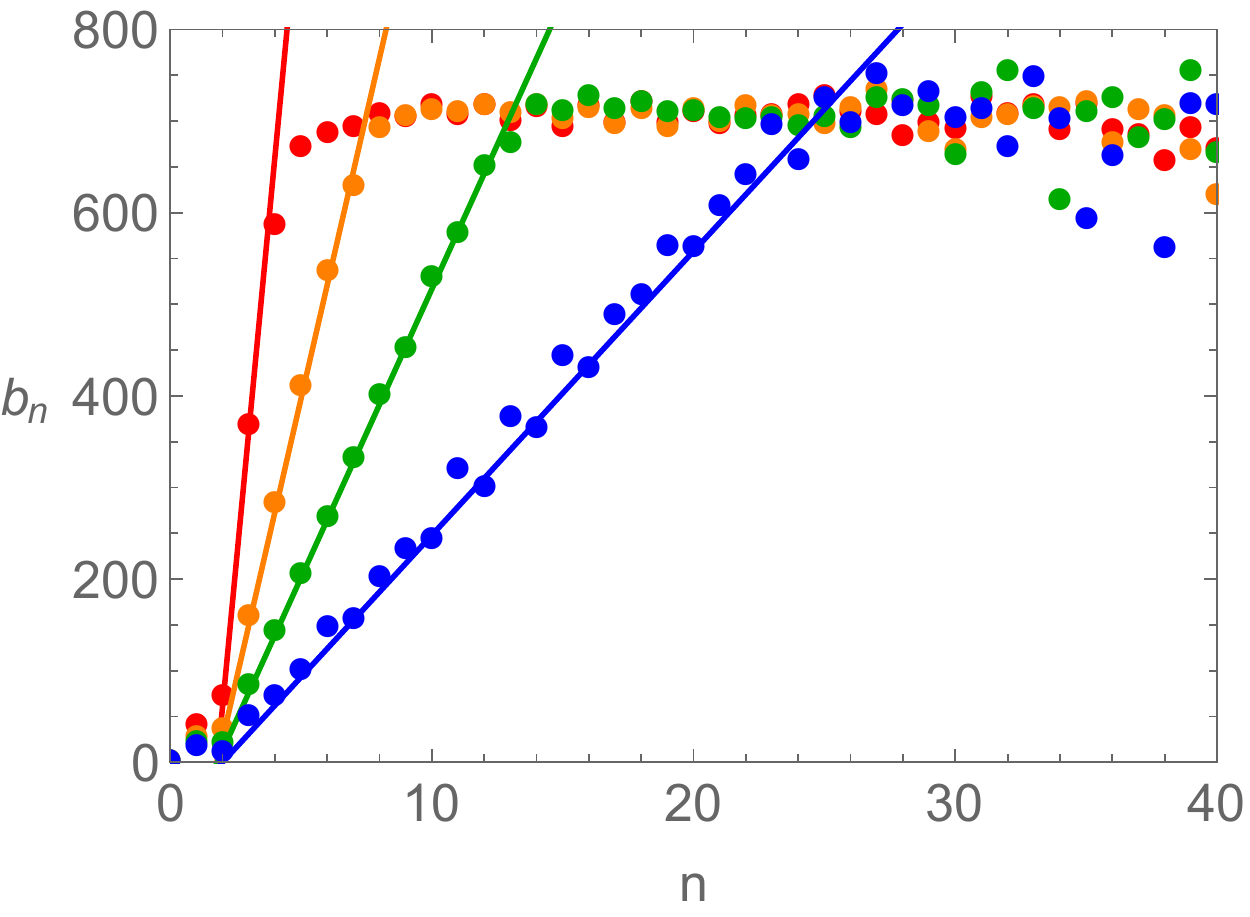} \label{LCF1a}}
     \subfigure[Circle billiard ($a/R=0$)]
     {\includegraphics[width=7.0cm]{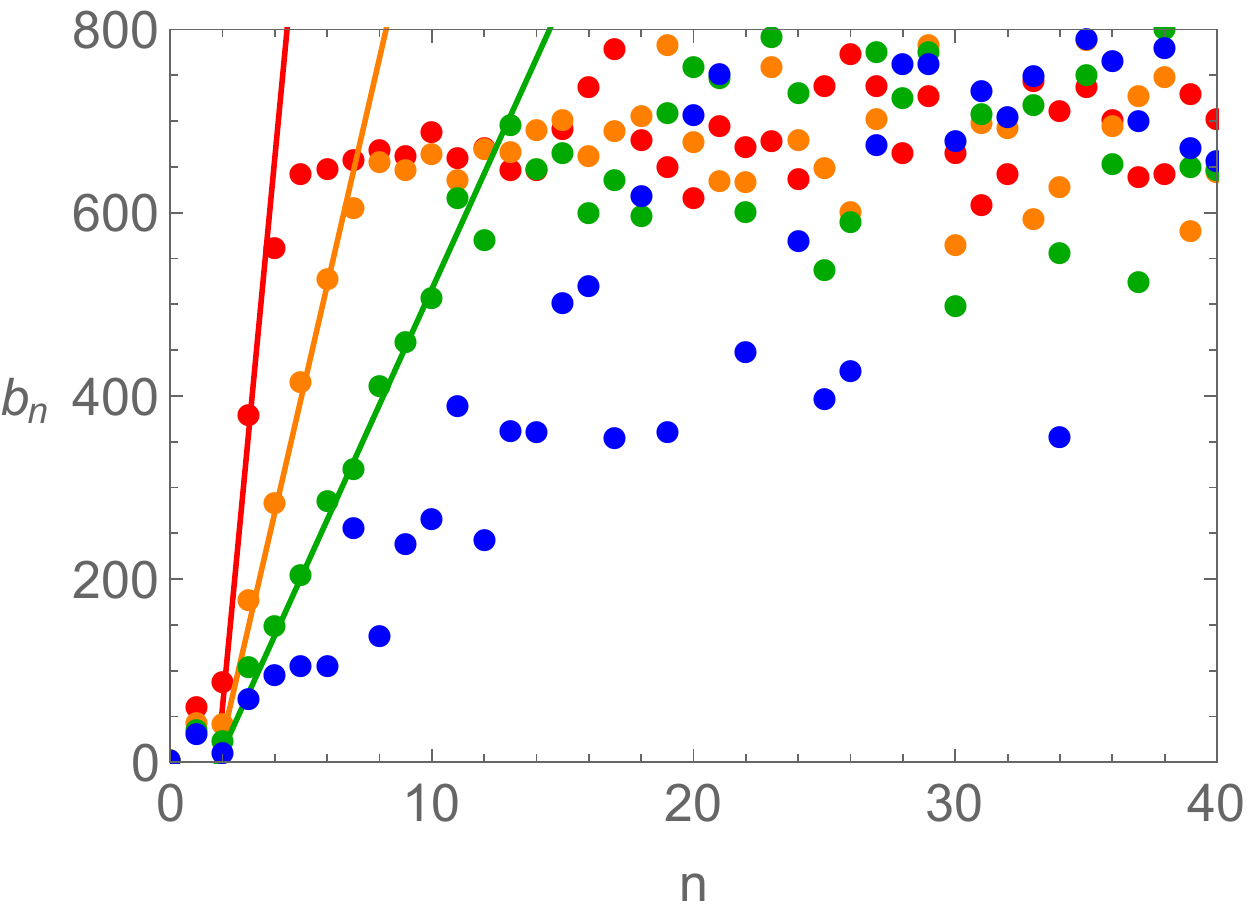} \label{LCF1b}}
 \caption{Lanczos coefficients $b_n$ at $T = 10, 20, 40, 100$ (blue, green, orange, red). The solid lines are linear fitting curves \eqref{FITAL}.}\label{LCF1}
\end{figure}

By numerically implementing the Lanczos algorithm (presented in Section~\ref{sec.LanczosAlgorithm}) we are able to obtain the Lanczos coefficients ($b_n$) for the stadium and circle billiards at different temperatures. The results are summarized in Fig.~\ref{LCF1}.
We find that the $b_n$ grow for small $n$ and reach a saturation value (approximately of $E_{\text{max}}/2 \approx 700$) for large $n$, in both types of billiards.\footnote{In the case of the circle billiard, the $b_n$ fluctuate heavily at low temperatures. We also find such a strong fluctuation by another method, namely the moment method. This is discussed in Appendix~\ref{appmoma}.}

To validate our numerical approach for computing the Lanczos coefficients, we compare our numerical approach with analytical methods in two scenarios where the latter approach exists. First, in Appendix \ref{shore}, we consider the simple harmonic oscillator to demonstrate the consistency between our numerical results and the corresponding analytical outcomes. Furthermore, in Appendix \ref{appmoma}, we illustrate the agreement between the Lanczos coefficients obtained using the Lanczos algorithm in Section~\ref{sec.LanczosAlgorithm} and those acquired through the moment method.

Fig. \ref{LCF1} reveals an additional feature: the fluctuations in $b_n$ are less pronounced in the stadium billiard compared to the circle billiard. This becomes more apparent at lower temperatures (see for instance the blue data corresponding to $T=10$).
In other words, our results support the observation made in \cite{Hashimoto:2023swv} regarding the ``detectability" of different billiard types (e.g., stadium vs. circle) through an analysis of the variance of the Lanczos coefficients. Notably, our findings extend this observation to finite temperature and small $n$, given that the original observation was made at infinite temperature.

\begin{figure}[]
\centering
     {\includegraphics[width=7.0cm]{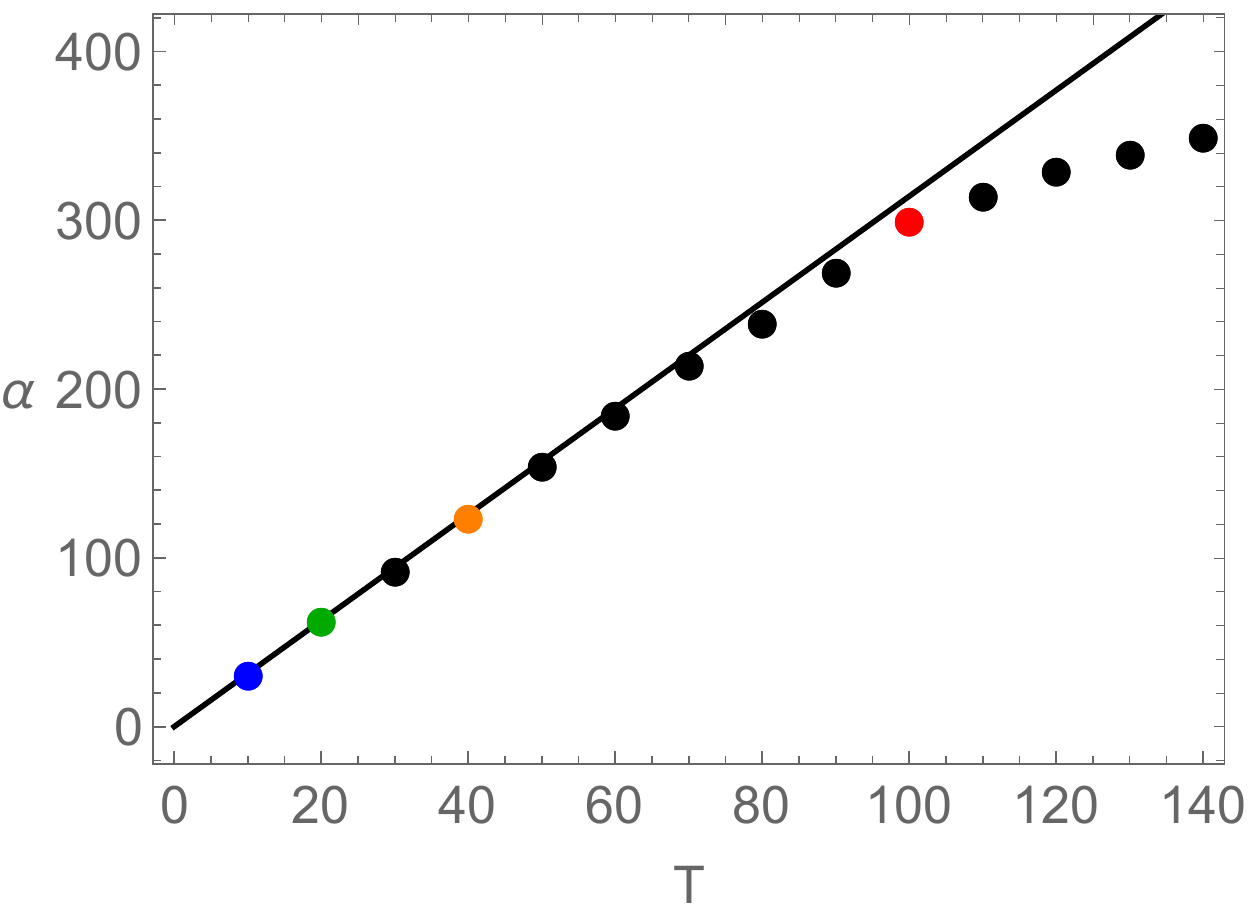} \label{}}
 \caption{The temperature dependence of the growth rate of $b_n$. The solid line is $\alpha=\pi T$ and dots are obtained from the linear fitting of $b_n$. The bound is saturated at low temperature.}\label{LCF22}
\end{figure}

Furthermore, we also find a regime in $n$ for which the $b_n$ exhibit a linear growth, whose growth rate satisfies the generalized chaos bound~\cite{Parker:2018yvk,Avdoshkin:2019trj}, i,e., 
\begin{align}\label{FITAL}
\begin{split}
  b_n \approx \alpha n \,, \qquad \alpha \leq \pi T \,.
\end{split}
\end{align}
In both the stadium and circle billiards, we find that the growth rate $\alpha$ saturates the bound (\ref{INEQLAT2}) at low temperatures $T$, i.e., $\alpha = \pi T$, see Fig. \ref{LCF22}. 
Of course, due to the heavy fluctuations, as mentioned above, one may not simply read off the linearity for the case of circle billiard at low temperature. 
%In this sense, \eqref{FITAL} may potentially serve as a good diagnostic indicator for characterizing billiard systems, even at low temperatures. \HC{What do we mean by this?}
%Note that our result also implies that the quantum billiard model \eqref{BILLMO} shows the saturation of the quantum Lyapunov exponent, $\lambda_{L}=2\pi T$,  as in \eqref{INEQLAT}.
%
%It is also worth noting that the growth rate analysis can be complementary to \cite{Hashimoto:2023swv} in that they specifically focused on the case of infinite temperature.
Our analysis of the growth rate of the Lanczos coefficients at finite temperature uncovers a novel aspect, \eqref{FITAL}, that was not investigated in the analysis conducted at infinite temperature in~\cite{Hashimoto:2023swv}.

\subsubsection{Krylov complexity and entropy}
%Next, we study the Krylov complexity and entropy via \eqref{KCF}. We first discuss the normalization condition  \eqref{EQVA}, $\sum |\varphi_n(t)|^2 =1$, in order to specify the relevant time scale for Krylov complexity and entropy in numerical calculation.

We now turn to the study of Krylov complexity and entropy. In order to determine the relevant time window in which we can trust our numerical results for these two quantities, we first discuss the normalization condition~\eqref{EQVA}, namely $\sum |\varphi_n(t)|^2 =1$.

\begin{figure}[]
\centering
     \subfigure[Stadium billiard ($a/R=1$)]
     {\includegraphics[width=7.0cm]{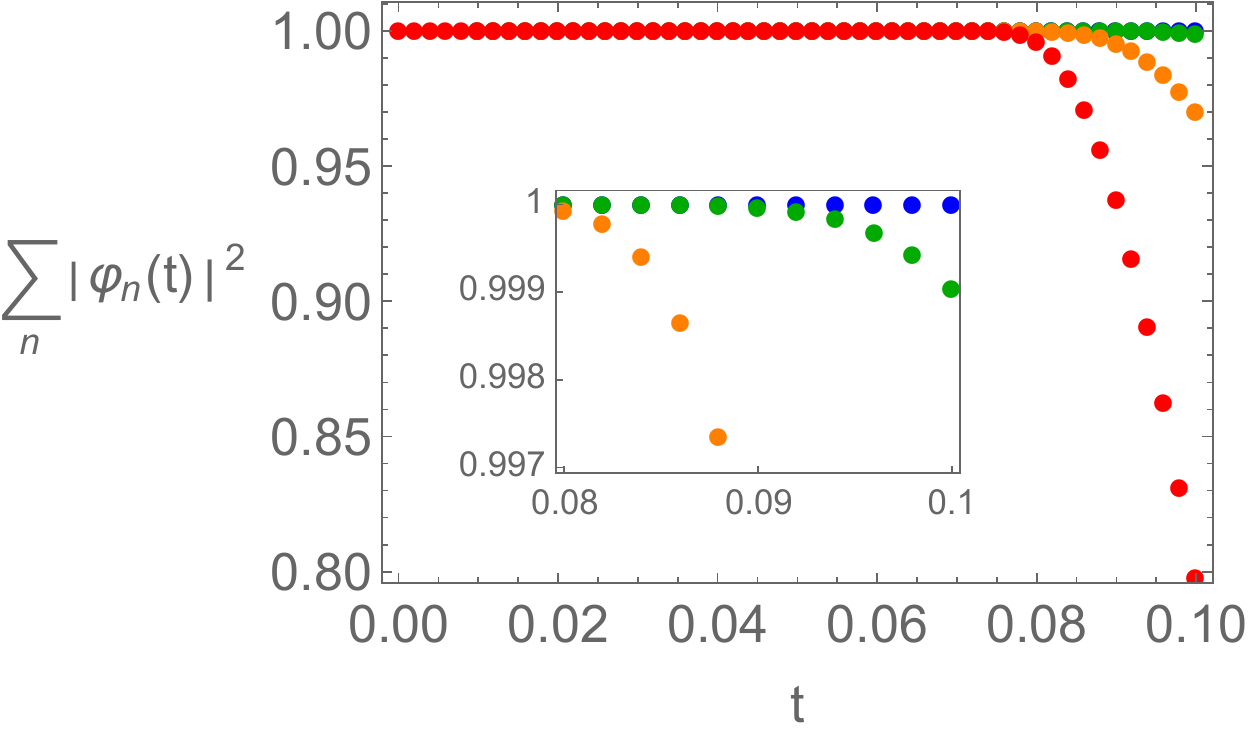} \label{}}
     \subfigure[Circle billiard ($a/R=0$)]
     {\includegraphics[width=7.0cm]{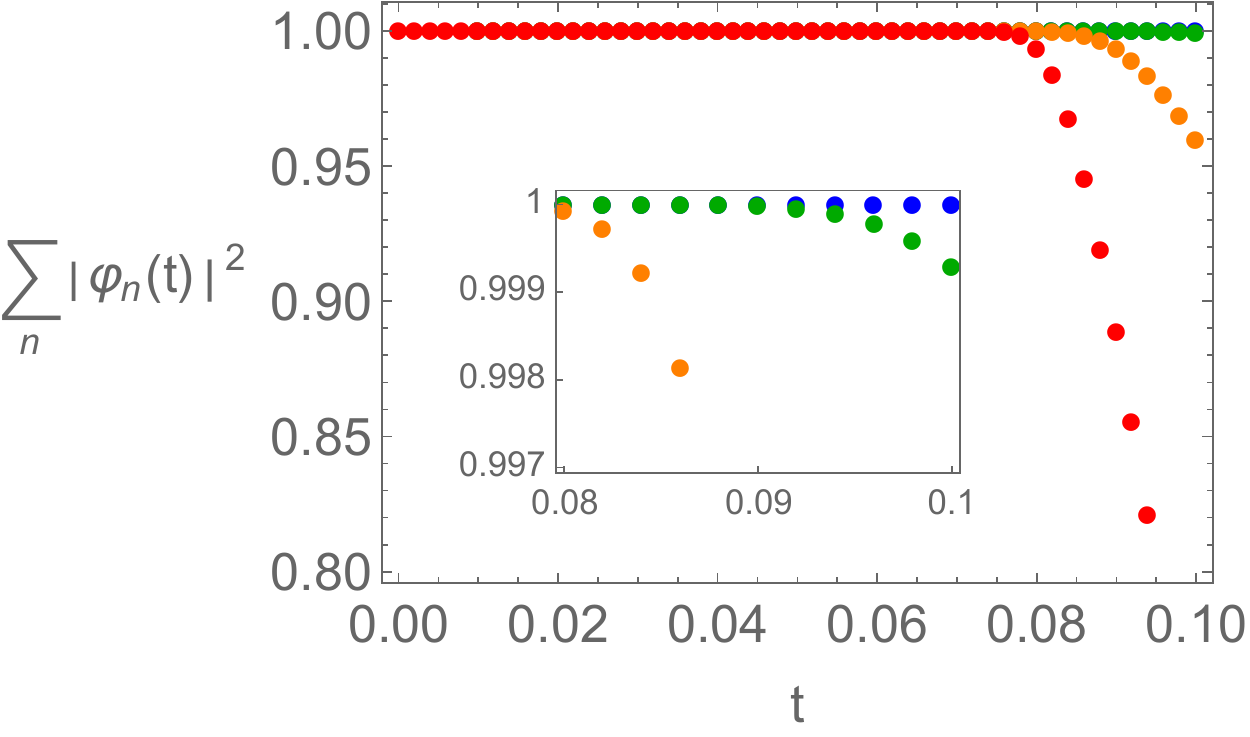} \label{}}
 \caption{The normalization condition at $T = 10, 20, 40, 100$ (blue, green, orange, red). The insets show near $t\approx0.1$.}\label{TSSTA}
\end{figure}
In Fig. \ref{TSSTA}, we show the normalization condition of both billiards and find that  the normalization condition is satisfied up to $t\leq0.08$. It is important to mention that in our numerical calculations, we have set the maximum value of $n$ for $b_n$ as $n_{\text{max}}=100$. However, it is noteworthy that by considering larger values of $n_{\text{max}}$, one can observe an expanded time window as well as the emergence of vanishing $b_n$ for higher $n$. Additional insights on this matter can be found in Appendix \ref{appmomb} and \cite{Hashimoto:2023swv}. This implies that our chosen cutoff value of $n_{\text{max}}=100$ is sufficient for the purpose of our early-time analysis.% which focuses on the early-time behavior, within the scope of this paper.

\begin{figure}[]
\centering
     \subfigure[Stadium billiard ($a/R=1$)]
     {\includegraphics[width=7.0cm]{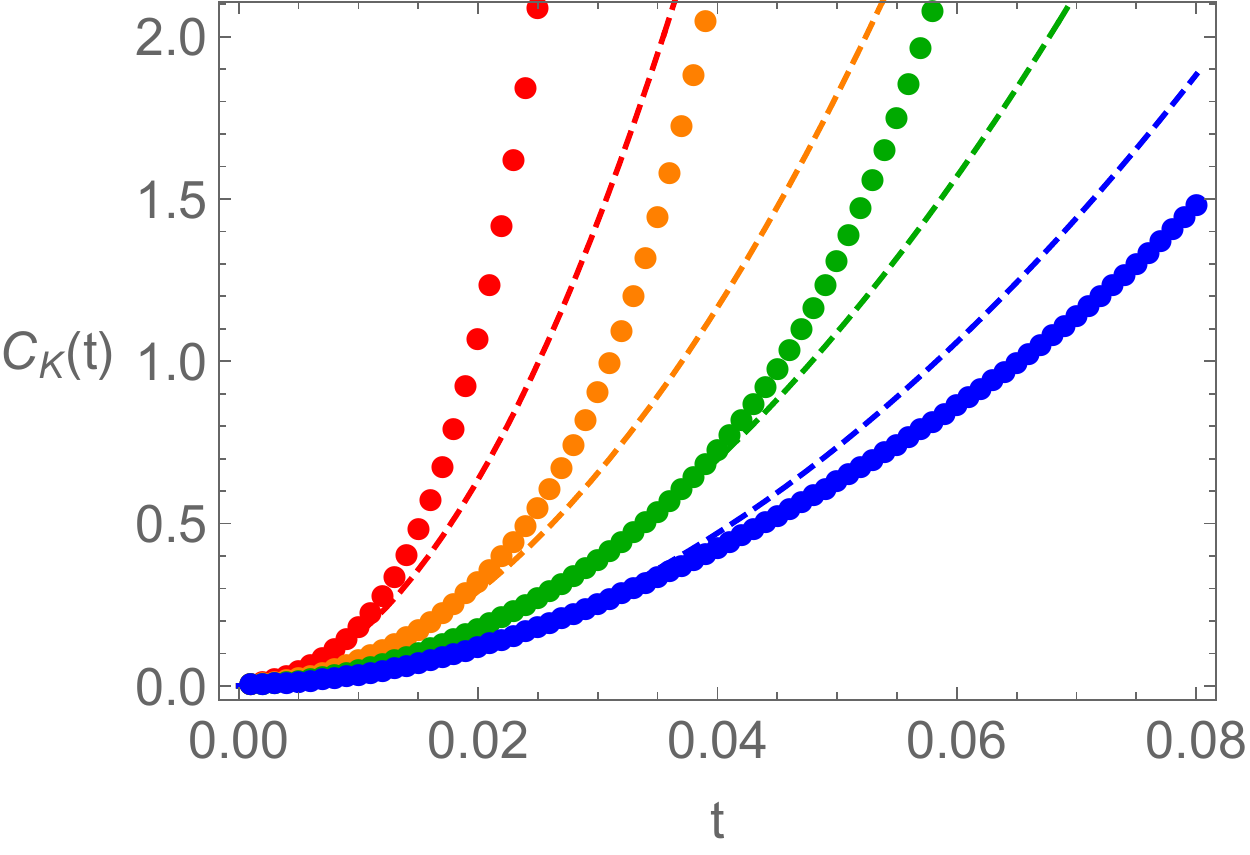} \label{}}
     \subfigure[Circle billiard ($a/R=0$)]
     {\includegraphics[width=7.0cm]{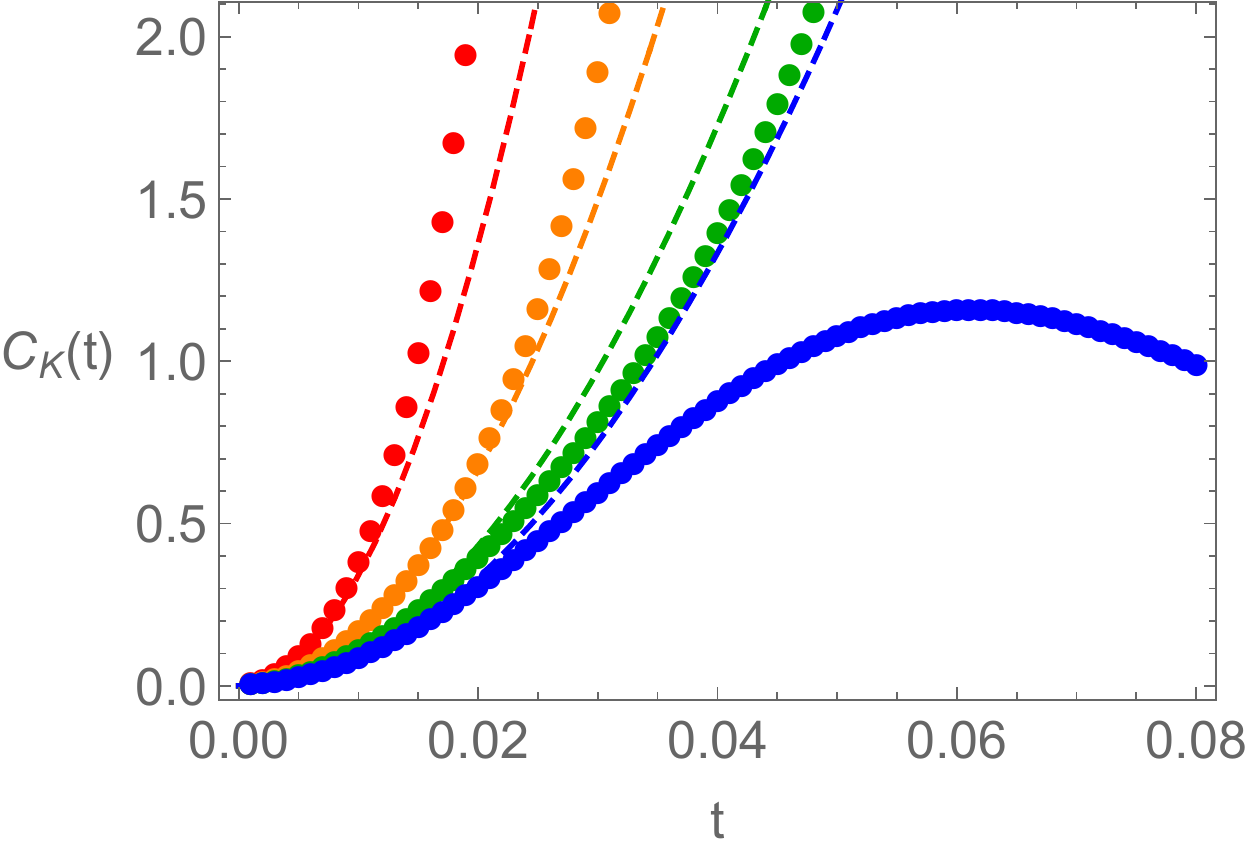} \label{}}
 \caption{Krylov complexity at $T = 10, 20, 40, 100$ (blue, green, orange, red). Dots are numerical data and the dashed lines are the analytic early-time results \eqref{KCF2}.}\label{KRYST1}
\end{figure}

\begin{figure}[]
\centering
     \subfigure[Stadium billiard ($a/R=1$)]
     {\includegraphics[width=7.0cm]{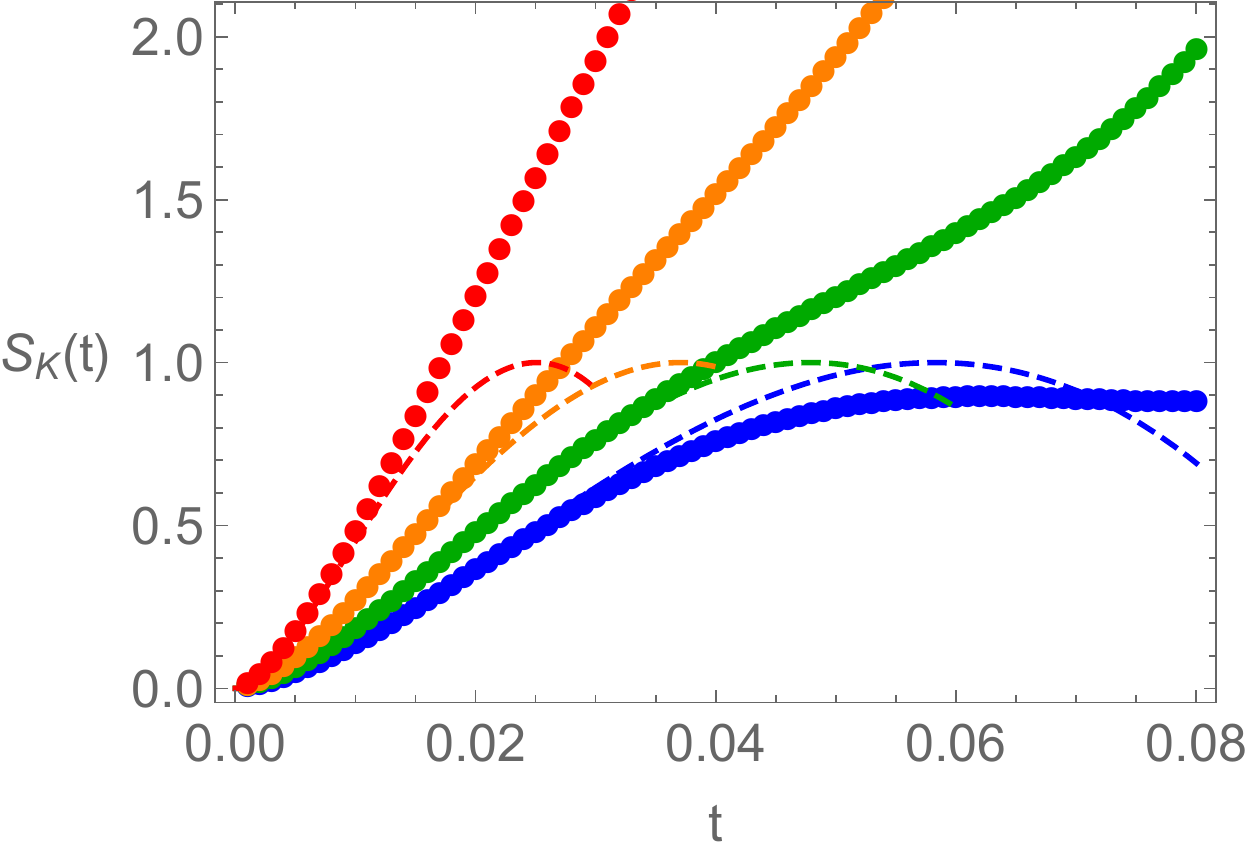} \label{}}
     \subfigure[Circle billiard ($a/R=0$)]
     {\includegraphics[width=7.0cm]{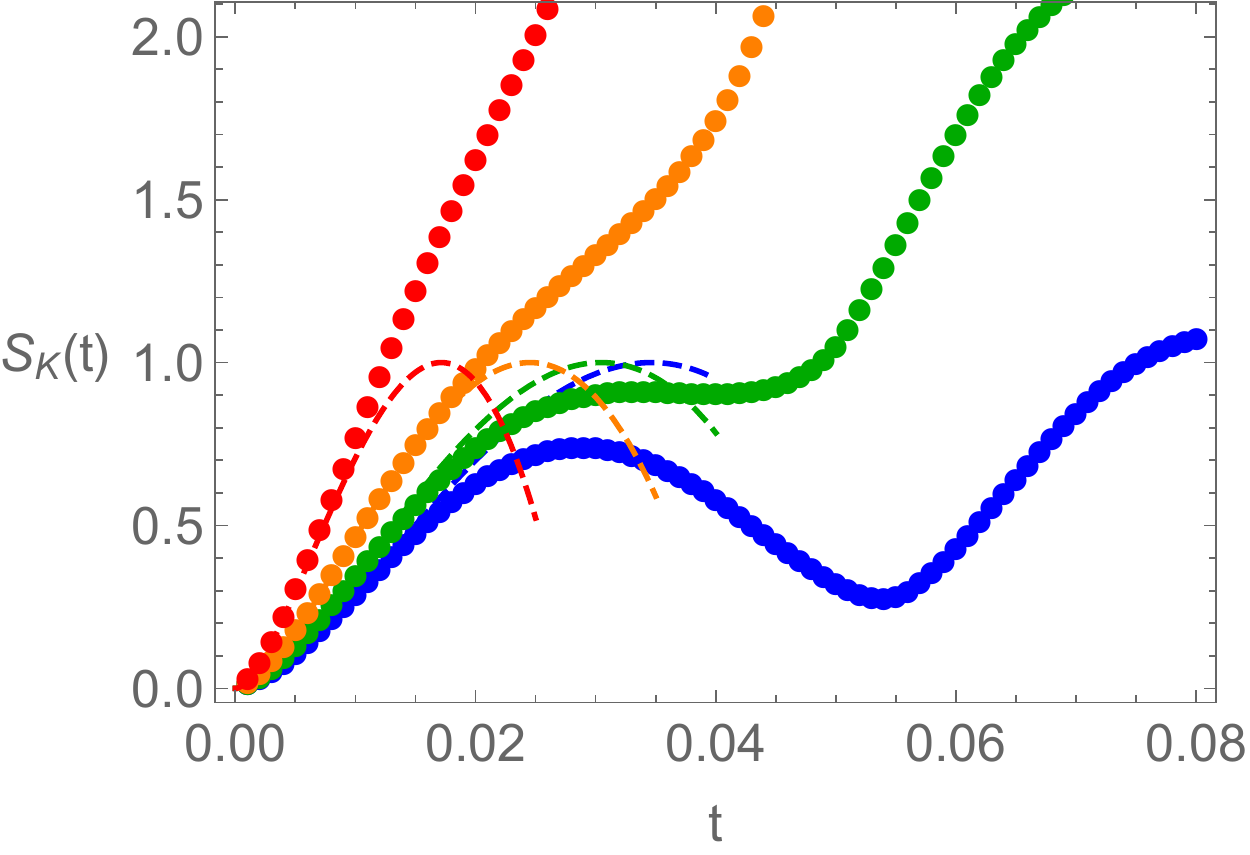} \label{}}
 \caption{Krylov entropy at $T = 10, 20, 40, 100$ (blue, green, orange, red). Dots are numerical data and the dashed lines are the analytic early-time results \eqref{KCF2}.}\label{KRYST2}
\end{figure}

\paragraph{Krylov complexity and entropy at early-times.}
We now study the Krylov complexity and entropy within the time window $0 \leq t \leq 0.08$. In Figs.~\ref{KRYST1} and~\ref{KRYST2} we show the results for Krylov complexity and entropy respectively for different values of temperature and for both types of billiard systems. In particular, the red data sets (corresponding to $T=100$) are qualitatively consistent with the infinite temperature case discussed in~\cite{Hashimoto:2023swv}.
%Dialing the value of temperature for both stadium and circle billiard, we display the Krylov complexity in Fig. \ref{KRYST1} (where the red data is qualitatively consistent with the infinite temperature case in \cite{Hashimoto:2023swv}) and Krylov entropy in Fig. \ref{KRYST2}.
%

Our main findings can be summarized in three aspects. 
Firstly, the operator growth, characterized by $C_K$ and $S_K$, exhibits a temperature-dependent behavior, gradually (or slowly) increasing as the temperature is lowered. Secondly, we have discovered that the stadium billiard exhibits quantitatively different behavior compared to the circle billiard, particularly at sufficiently low temperatures (as indicated by the blue data in the figures). This implies that one may observe distinguishable features in the operator complexity even during the early-time regime by reducing the temperature. Lastly, we have validated our numerical results by showing their agreement with the analytical early-time results presented in Section~\ref{sec.LanczosAlgorithm}: namely equation~\eqref{KCF2}. 

\begin{figure}[]
\centering
     \subfigure[Stadium billiard ($a/R=1$)]
     {\includegraphics[width=7.3cm]{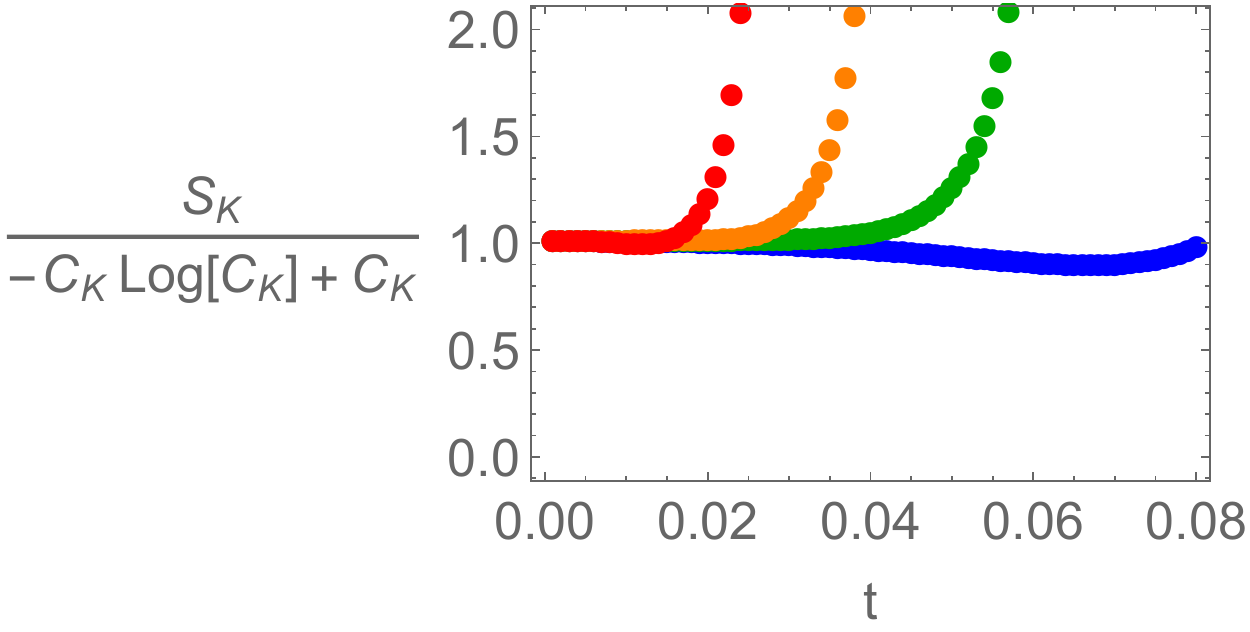} \label{}}
     \subfigure[Circle billiard ($a/R=0$)]
     {\includegraphics[width=7.3cm]{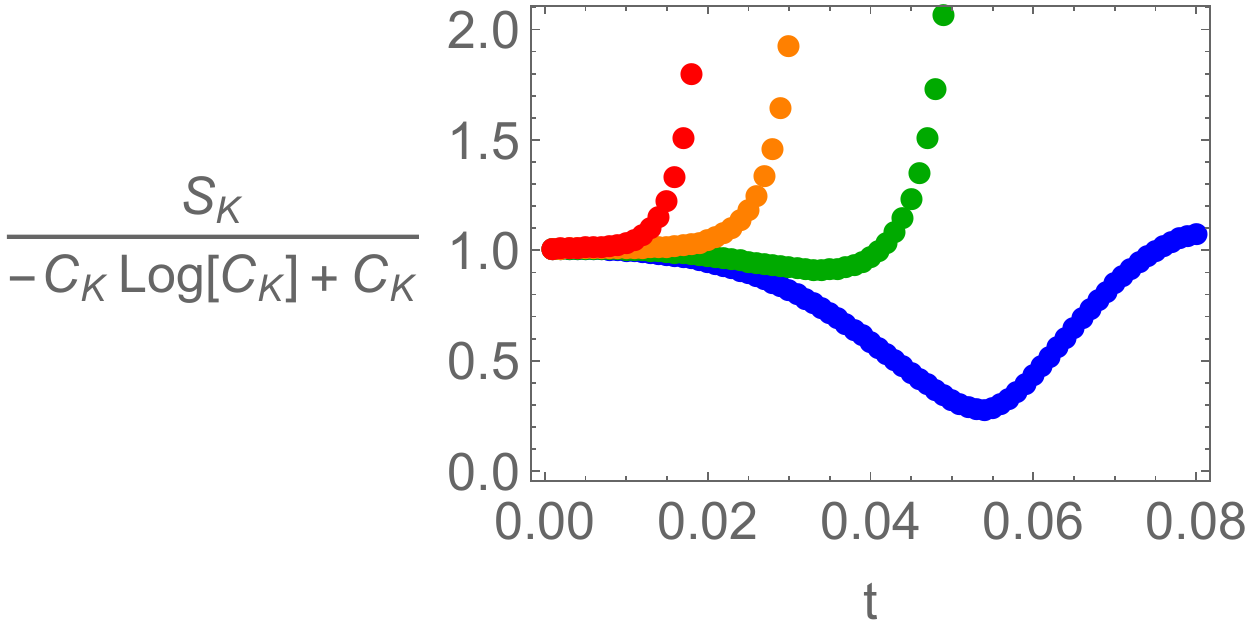} \label{}}
 \caption{The logarithmic relation between Krylov complexity and entropy at $T = 10, 20, 40, 100$ (blue, green, orange, red). We find our data is consistent with \eqref{KCF3} at early-time.}\label{KRYST3}
\end{figure}

Furthermore, we have also confirmed that our numerical results capture the logarithmic relation between Krylov complexity and entropy, equation \eqref{KCF3}, during the early-time regime. This can be seen in Fig. \ref{KRYST3}.

To conclude this section, we remark a connection between spectral complexity and Krylov complexity of states. In~\cite{Erdmenger:2023shk} it was shown that the Krylov state complexity of the time-evolved TFD state was related to the spectral complexity~\eqref{eq:SpectralComplexityDef} via an Ehrenfest theorem in Krylov space. To be precise, the spectral complexity $C_{S}(t)$ can be seen as an approximation to the Krylov complexity of the TFD state at early times for Hamiltonians belonging to the $\tilde{\beta}$-Hermite (Gaussian) ensemble with Dyson index $\tilde{\beta}$ in the limit where the dimension of the Krylov space is large.
 
\begin{figure}[]
\centering
     \subfigure[Left hand side of the Ehrenfest theorem for stadium billiard ($a/R=1$) at $T=10$]
     {\includegraphics[width=7.3cm]{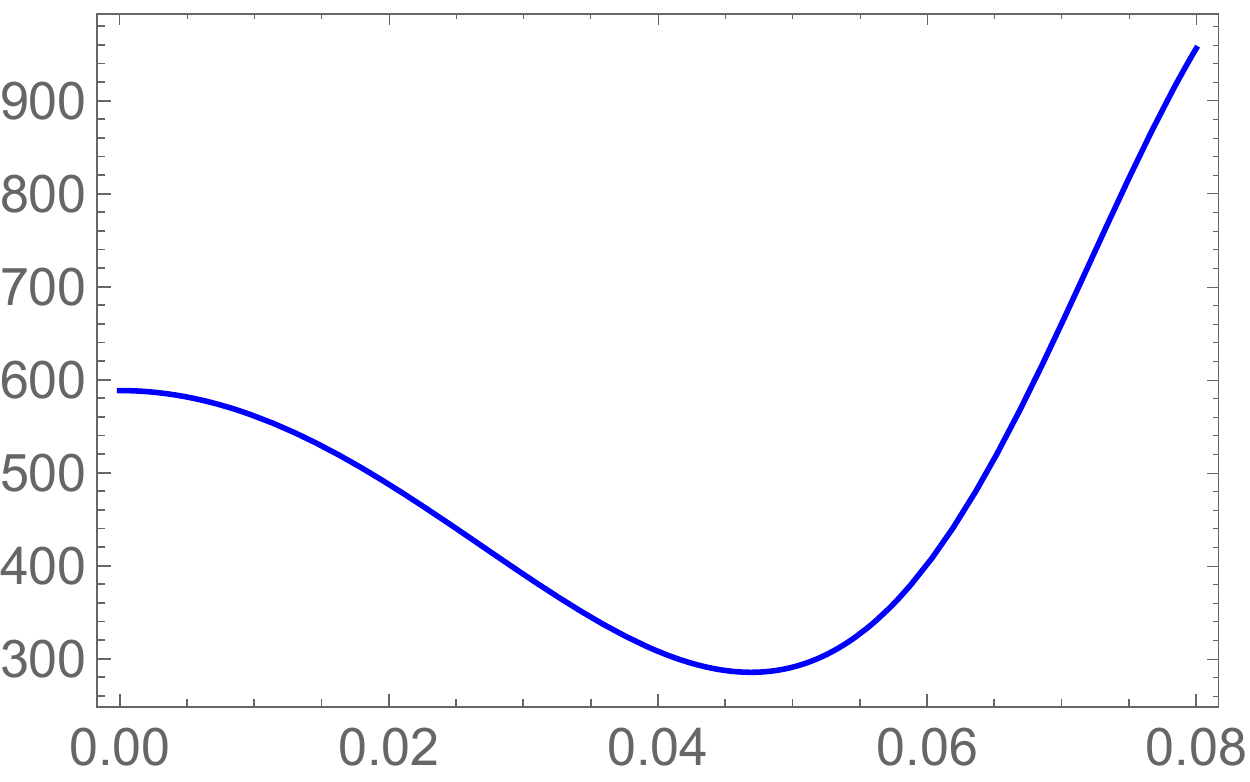}        \put(-5,-10){$t$}
    \put(-205,135){$\partial_t^2C_K(t)$}
    \label{}}
     \subfigure[Right hand side of the Ehrenfest theorem for stadium billiard ($a/R=1$) at $T=10$]
     {\includegraphics[width=7.3cm]{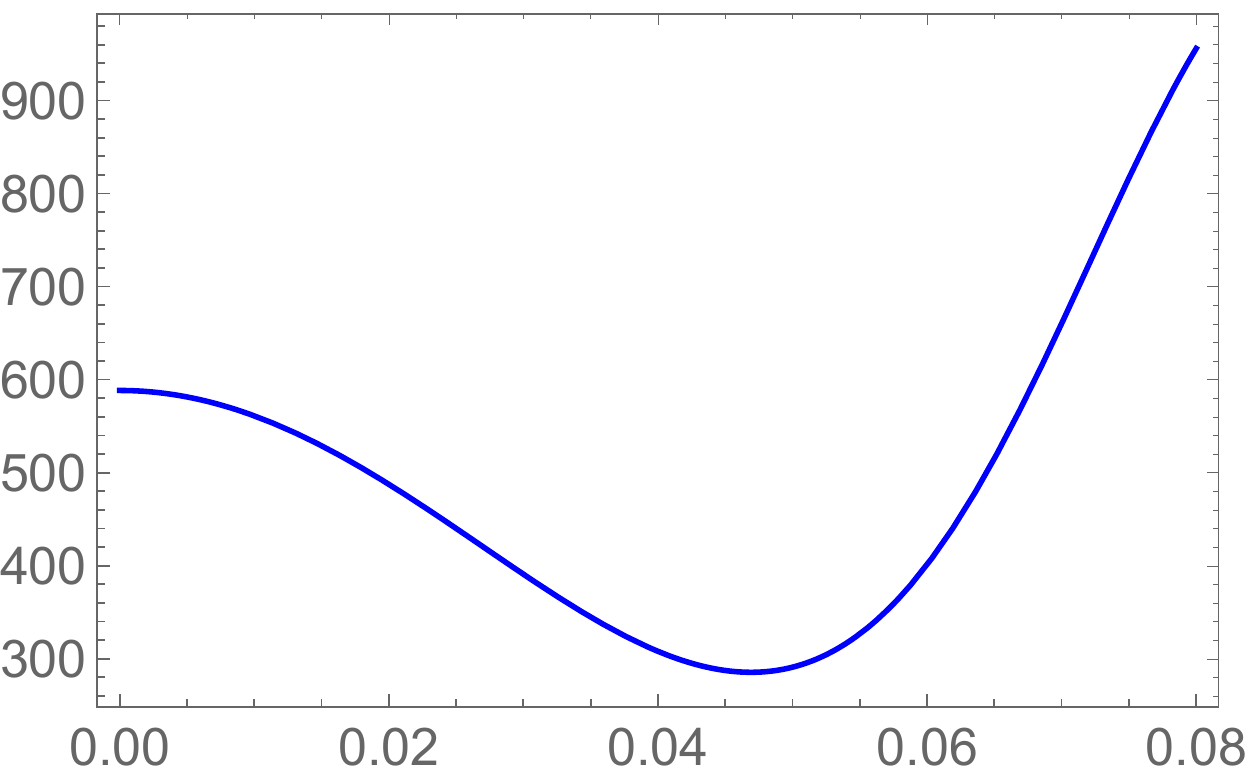} 
     \put(-5,-10){$t$}
    \put(-205,135){$2\sum_n(b_{n+1}^2-b_n^2)\vert\varphi_n(t)\vert^2$}
    \label{}}
     \subfigure[Left hand side of the Ehrenfest theorem for circle billiard ($a/R=0$) at $T=10$]
     {\includegraphics[width=7.3cm]{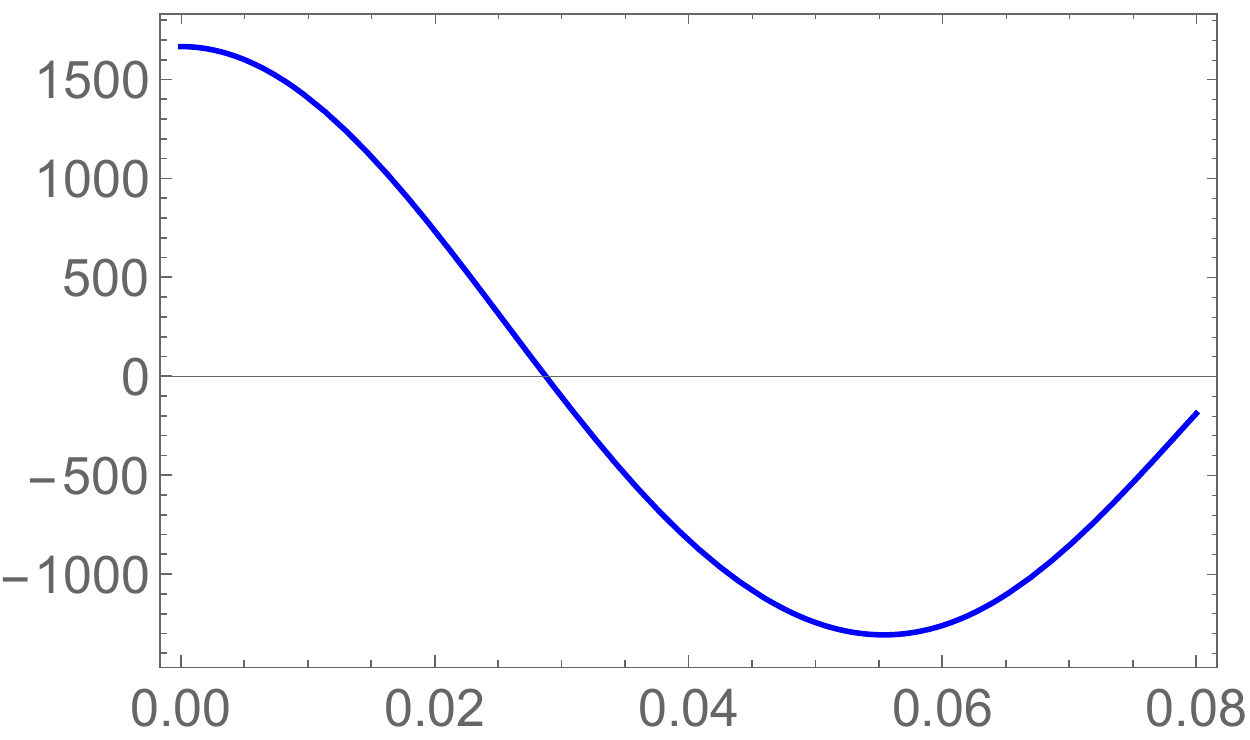} 
     \put(-5,-10){$t$}
    \put(-195,130){$\partial_t^2C_K(t)$}
    \label{}}
     \subfigure[Right hand side of the Ehrenfest theorem for circle billiard ($a/R=0$) at $T=10$]
     {\includegraphics[width=7.3cm]{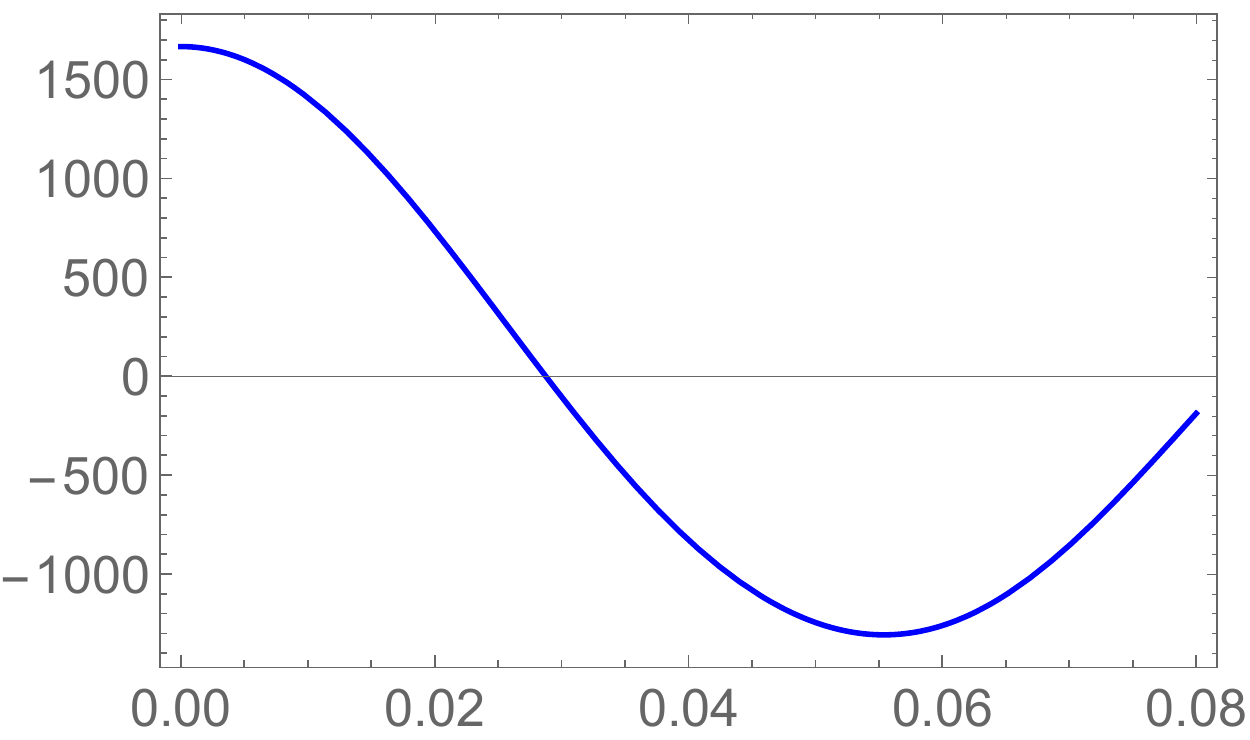} 
      \put(-5,-10){$t$}
    \put(-195,130){$2\sum_n(b_{n+1}^2-b_n^2)\vert\varphi_n(t)\vert^2$}
    \label{}}
 \caption{Both sides of the Ehrenfest theorem (\ref{EhrenfestV2}) for stadium and circle billiards at $T=10$.}\label{fig:Ehrenfest}
\end{figure}
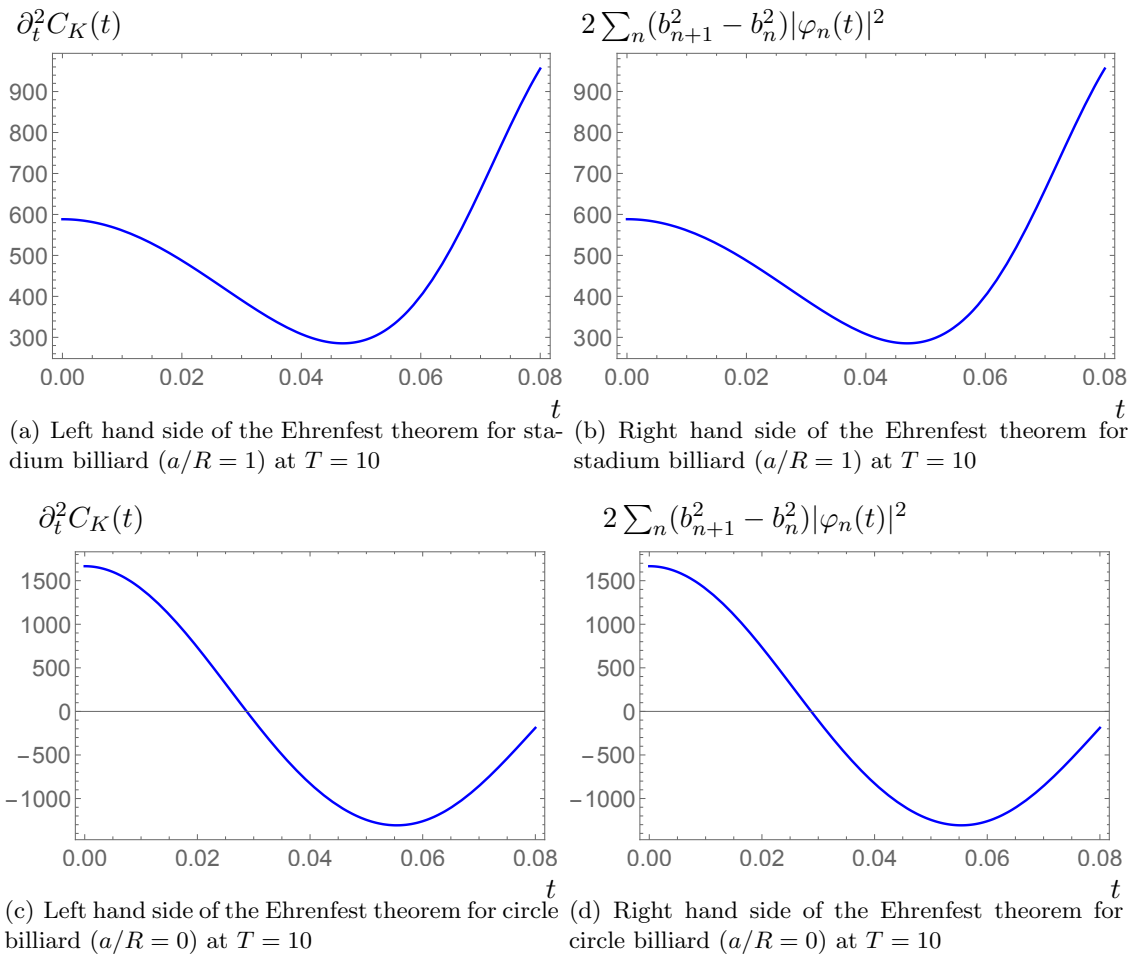

The Ehrenfest theorem of Krylov complexity states the following
\begin{align}\label{Ehrenfest}
\partial_t^2C_K(t)=-[[C_K(t),\mathcal{L}],\mathcal{L}]~,
\end{align}
which in terms of the Lanczos coefficients and transition amplitudes is given by \cite{Muck:2022xfc}
\begin{align}\label{EhrenfestV2}
\partial_t^2C_K(t)=2\sum_n(b_{n+1}^2-b_n^2)\vert\varphi_n(t)\vert^2~.
\end{align}
This expression can be derived from (\ref{DCEQ}) and (\ref{KCF}). To confirm this relation in our numerical computations, we plot both sides of (\ref{EhrenfestV2}) for stadium and circle billiards at $T=10$ in Fig. \ref{fig:Ehrenfest}, which show complete agreement. Therefore, within the precision of our numerical computations, the Ehrenfest theorem of Krylov complexity is valid at early times.
%

%%%%%%%%%%%%%%%%%%%%%%%%%%%%%%
%%%%%%%%%%%%%%%%%%%%%%%%%%%%%
\section{Conclusion}\label{sec:conclusion}

In this paper, we studied the behavior of the Lanczos coefficients, Krylov operator complexity, and spectral complexity for circle and stadium billiards at finite temperature.
The quantum mechanics of dynamical billiard is described in terms of bosonic operators, and the associated Hilbert space is infinite-dimensional. To extract features of chaos with finite degrees of freedom, we truncate the spectrum. Such truncation introduces a saturation of the Lanczos coefficients that implies a (featureless) linear growth of the Krylov complexity as a function of time. However, before saturation, the Lanczos coefficients grow linearly ($b_n = \alpha n+ \gamma$), and one can study the behavior of the slope $\alpha$ as a function of the temperature $T$. For finite and sufficiently low temperatures, we find that the growth rate of the Lanczos coefficients satisfies the bound (\ref{INEQLAT2}) for both circle and stadium billiards. However, as we further decrease the temperature, the behavior of the Lanczos coefficients for the circle billiard (integrable case) becomes erratic and non-linear, while the linear behavior persists for the stadium billiard (chaotic chase). See Figs. \ref{LCF1} and \ref{LCF22}. This is consistent with the results recently reported in \cite{Hashimoto:2023swv}.

We also studied the early-time behavior of the corresponding Krylov operator complexity. The reason to only consider the early-time behavior is as follows. Since we truncate the spectrum of the billiards at some energy $E_\text{max}$, the Lanczos coefficients saturate around $E_\text{max}/2$, and the associated Krylov complexity grows linearly with time after the saturation.
Such a linear behavior is just a reflection of the truncation of the spectrum, and we do not expect it to contain information about the dynamics of the system. Having this in mind, we only studied the early-time behavior of the Krylov operator complexity, because it is only in this region that we expect to extract physical information about the dynamics of the system.

We checked that our numerical results satisfy basic consistency conditions, like the normalization of the wave functions in the specified time window, the Ehrenfest theorem, and the universal early-time relation with Krylov entropy. We also verified that the qualitative features of our results do not depend on the choice of operators (we compare $x$, $x^2$, $p$, and $p^2$). Finally, we observe that the operator growth, Krylov complexity and entropy, tend to increase more slowly as we decrease the system's temperature. The results for circle and stadium billiards are in general qualitatively similar, but they display some differences at very low temperatures.  See Figs. \ref{KRYST1} and \ref{KRYST2}.

We studied the time behavior of the spectral complexity for both circle and stadium billiards, finding a sharp difference between them. For both cases, the spectral complexity initially grows as $C_S(t) \propto c_1 \log \cosh (c_2 t/\beta)$, where $c_1$ and $c_2$ are constants.\footnote{At finite temperature the spectral complexity initially grows as $C_S(t) \propto c_1 \log \cosh (c_2 t/\beta) $, where $c_1$ and $c_2$ are constants, and at infinite temperature ($\beta=0$) as $C_S(t) \propto \tilde{c}_1 \log \cosh (\tilde{c}_2 t) $, where $\tilde{c}_1$ and $\tilde{c}_2$ are different constants.} 
Then it stops growing and starts to oscillate wildly around a constant value. We refer to this behavior as `saturation'. We observe that the spectral complexity saturates much earlier in the chaotic case (stadium billiard), as compared to the integrable case (circle billiard). See Figs. \ref{SCFIG11} and \ref{SCFIG22} . This different behavior may be attributed to the different spectral statistics of integrable (or chaotic) systems. 

It is interesting to compare the time behavior of spectral complexity with the observed behavior for Krylov operator complexity in systems displaying a chaotic-integrable transition. The authors of \cite{Rabinovici:2022beu} observed that, when the dimension of the Krylov space is $K$ the late-time saturation value of Krylov operator complexity for chaotic systems is around $K/2$, while the saturation value is smaller than $K/2$ in integrable systems. The basic idea is that in chaotic systems the Krylov space is fully explored, while only a fraction of the Krylov space is explored in integrable systems. This contrasts with our result for spectral complexity, in which case the saturation occurs a comparatively much larger times for integrable systems. 
{At first, this may not appear to be a problem since the spectral complexity is not directly related to the growth of operators. Therefore, it is not necessary for it to replicate the qualitative behavior of Krylov operator complexity. However, considering that it serves as a measure of complexity and exhibits similar behavior to Krylov state complexity at early times, it might be reasonable to expect that the spectral complexity would also reflect qualitative aspects of other complexity measures.} 

The difference in behavior between spectral and Krylov operator complexity is probably due to the fact that the spectral complexity, in its original definition, has as input the full spectrum of the system, while the analysis performed in \cite{Rabinovici:2022beu} for integrable systems focus on the sector of fixed parity and total magnetization. It might be interesting to investigate the behavior of spectral complexity for some sectors of the circle billiards, for instance, states with a fixed angular momentum, and check if it matches the behavior observed in \cite{Rabinovici:2022beu}. We are currently investigating this. 

This study can be extended in several directions, including, for instance: (i) studying Krylov state complexity for scar states in billiards; (ii) studying the behavior of spectral complexity in systems that display chaotic-integrable transition, like the mixed field Ising model~\cite{Bauls_2011} or the mass-deformed SYK model~\cite{Nosaka:2018iat}.

%%%%%%%%%%%%%%%%%%%%%%%%%%%%%%
%%%%%%%%%%%%%%%%%%%%%%%%%%%%%
\acknowledgments
We would like to thank {Kyoung-Bum Huh, Karl Landsteiner, Ryota Watanabe} for valuable discussions and correspondence.  
This work was supported by the Basic Science Research Program through the National Research Foundation of Korea (NRF) funded by the Ministry of Science, ICT $\&$ Future Planning (NRF-2021R1A2C1006791) and GIST Research Institute(GRI) grant funded by the GIST in 2023.
This work was also supported by Creation of the Quantum Information Science R$\&$D Ecosystem (Grant No. 2022M3H3A106307411) through the National Research Foundation of Korea (NRF) funded by the Korean government (Ministry of Science and ICT).
H.-S Jeong acknowledges the support of the Spanish MINECO ``Centro de Excelencia Severo Ochoa'' Programme under grant SEV-2012-0249. This work is supported through the grants CEX2020-001007-S and PID2021-123017NB-I00, funded by MCIN/AEI/10.13039/501100011033 and by ERDF A way of making Europe.
H.~A. Camargo, V.~Jahnke and M.~Nishida were supported by the Basic Science Research Program through the National Research Foundation of Korea (NRF) funded by the Ministry of Education (NRF-2022R1I1A1A01070589, NRF-2020R1I1A1A01073135, NRF-2020R1I1A1A01072726).
H.~A. Camargo, V.~Jahnke, H.-S. Jeong and M.~Nishida contributed equally to this paper and should be considered co-first authors.

%%%%%%%%%%%%%%%%%%%%%%%%%%%%%
%%%%%%%%%%%%%%%%%%%%%%%%%%%%
\appendix

%%%%%%%%%%%%%%%%%%%%%%%%%%%%%
%    
%%%%%%%%%%%%%%%%%%%%%%%%%%%%%
\section{More on operator growth}\label{}
\subsection{Simple harmonic oscillator: analytical vs. numerical results}\label{shore}
In this section, we examine the Krylov complexity and entropy of a simple harmonic oscillator. Our primary objective is to demonstrate the consistency between our numerical method and the analytic results. By comparing the numerical and analytic outcomes, we aim to establish the reliability and accuracy of our computational approach in the main text.

Consider the Hamiltonian for a simple harmonic oscillator
\begin{align}\label{SHOHAM}
\begin{split}
  H = \frac{p^2}{2} + \frac{\omega^2}{2} x^2 \,,
\end{split}
\end{align}
which gives the energy eigenvalues as 
\begin{align}\label{}
\begin{split}
  E_n = \left( n + \frac{1}{2} \right)\omega \,.
\end{split}
\end{align}
The partition function can be evaluated in a straightforward way
\begin{align}\label{PATC}
\begin{split}
  Z = \text{Tr} \left( e^{-\beta H} \right) = \sum_{n} e^{-\beta E_n} = \frac{\csch\left(\frac{\beta\omega}{2}\right)}{2} \,.
\end{split}
\end{align}

\paragraph{Vanishing Lanczos coefficient.}
We perform the Lanczos algorithm to obtain the Lanczos coefficients $b_n$ of the simple harmonic oscillators. We obtain $b_n$ for $n=0, 1, 2$ analytically (as we will show shortly, the Lanczos algorithm should stop at $n=2$).

As in the main text, we are interested the initial reference operator to be position operator as
\begin{align}\label{A0F}
\begin{split}
  A_0 := x \,.
\end{split}
\end{align}
In order to obtain the first Lanczos coefficient $b_0$ we need to compute the matrix elements $x_{m\ell} = \left<m|x|\ell\right>$.

Writing the position operator in terms of the creation ($a^{\dagger}$) and annihilation operators ($a$) as
\begin{align}\label{}
\begin{split}
  x = \sqrt{\frac{1}{2\omega}} \left(a + a^{\dagger}\right) \,,
\end{split}
\end{align}
one finds
\begin{align}\label{FLC1}
\begin{split}
  x_{m\ell} = \left<m|x|\ell\right> =  \sqrt{\frac{1}{2\omega}} \left( \sqrt{\ell} \, \delta_{m, \ell-1} + \sqrt{\ell+1} \, \delta_{m, \ell+1} \right)\,.
\end{split}
\end{align}

Therefore, using \eqref{WMIP} together with \eqref{FLC1}, the first Lanczos coefficients can be obtained
\begin{align}\label{b0F}
\begin{split}
  b_0=\sqrt{(A_0|A_0)} = \sqrt{\frac{1}{Z} \sum_{m, \ell} e^{-\frac{\beta}{2} (E_m + E_\ell)}  (x_{m\ell})^{2}} = \sqrt{\frac{1}{Z} \frac{\csch^2\left(\frac{\beta\omega}{2}\right)}{4 \omega} } = \sqrt{ \frac{\csch\left(\frac{\beta\omega}{2}\right)}{2\omega} } \,,
\end{split}
\end{align}
where \eqref{PATC} is used in the last equality.
Consequently, the first element of the Krylov basis becomes
\begin{align}\label{RE0CC}
\begin{split}
  \mO_0 := b_{0}^{-1} A_0  =  \sqrt{2 \omega \sinh \left(\frac{\beta \omega}{2}\right)} \, x\,,
\end{split}
\end{align}
where we used \eqref{A0F} and \eqref{b0F}.

Continuing with the Lanczos algorithm, the second Lanczos coefficient, $b_1$, can also be obtained by evaluating
\begin{align}\label{FLC2}
\begin{split}
  \left<m|A_1|\ell\right> = \left<m|[H, \mO_0]|\ell\right> = b_0^{-1} \left(E_m - E_\ell \right)  \left<m|A_0|\ell\right> \,,
\end{split}
\end{align}
where we used $\mO_0 := b_{0}^{-1} A_0$ in the second equality.
Note that we can compute \eqref{FLC2} using \eqref{FLC1} and \eqref{b0F}.
Thus, $b_1$ becomes
\begin{align}\label{b1F}
\begin{split}
  b_1=\sqrt{(A_1|A_1)} = \sqrt{\frac{1}{Z} \sum_{m, \ell} e^{-\frac{\beta}{2} (E_m + E_\ell)}  \left<m|A_1|\ell\right>^{2}} = \sqrt{\frac{\omega^2}{Z} \frac{\csch\left(\frac{\beta\omega}{2}\right)}{2} } = \omega \,.
\end{split}
\end{align}
The corresponding element of the Krylov basis is
\begin{align}\label{}
\begin{split}
   \mO_1 := b_{1}^{-1} A_1  = \omega^{-1} [H, x] \,.
\end{split}
\end{align}

Next, we consider the third (and last) Lanczos coefficient, $b_2$, by
\begin{align}\label{FLC3}
\begin{split}
  \left<m|A_2|\ell\right> &= \left<m| [H, \mO_{1}] - b_{1} \mO_{0} |\ell\right> \\
  &= \left(E_m - E_\ell \right)  \left<m|\mO_1|\ell\right> - b_{1} \left<m|\mO_0|\ell\right> \\ 
  &= b_{1}^{-1}\left(E_m - E_\ell \right)  \left<m|A_1|\ell\right> - b_{1} \, b_{0}^{-1} \left<m|A_0|\ell\right> \,,
\end{split}
\end{align}
which can be evaluated with \eqref{FLC1}-\eqref{b0F} and \eqref{FLC2}-\eqref{b1F}.
Thus, $b_2$ becomes
\begin{align}\label{b2F}
\begin{split}
  b_2=\sqrt{(A_2|A_2)} = \sqrt{\frac{1}{Z} \sum_{m, \ell} e^{-\frac{\beta}{2} (E_m + E_\ell)}  \left<m|A_2|\ell\right>^{2}} = 0 \,.
\end{split}
\end{align}
Since the Lanczos coefficients hit zero, we stop the algorithm.\\

In summary, we obtained the following Lanczos coefficients for a simple harmonic oscillator
\begin{align}\label{LZC}
\begin{split}
  b_0 = \sqrt{ \frac{\csch\left(\frac{\beta\omega}{2}\right)}{2\omega} } \,, \qquad
  b_1 = \omega \,, \qquad
  b_2 = 0 \,,
\end{split}
\end{align}
together with the Krylov basis elements 
\begin{align}\label{OB}
\begin{split}
  \mO_0  =  \sqrt{2 \omega \sinh \left(\frac{\beta \omega}{2}\right)} \, x\,, \qquad 
  \mO_1   = \omega^{-1} [H, x] \,.
\end{split}
\end{align}
Note that by setting $\omega=1$ in \eqref{LZC}-\eqref{OB}, we reproduce the result in \cite{Du:2022ocp}.

\begin{figure}[]
\centering
     \subfigure[$\beta=1$]
     {\includegraphics[width=7.0cm]{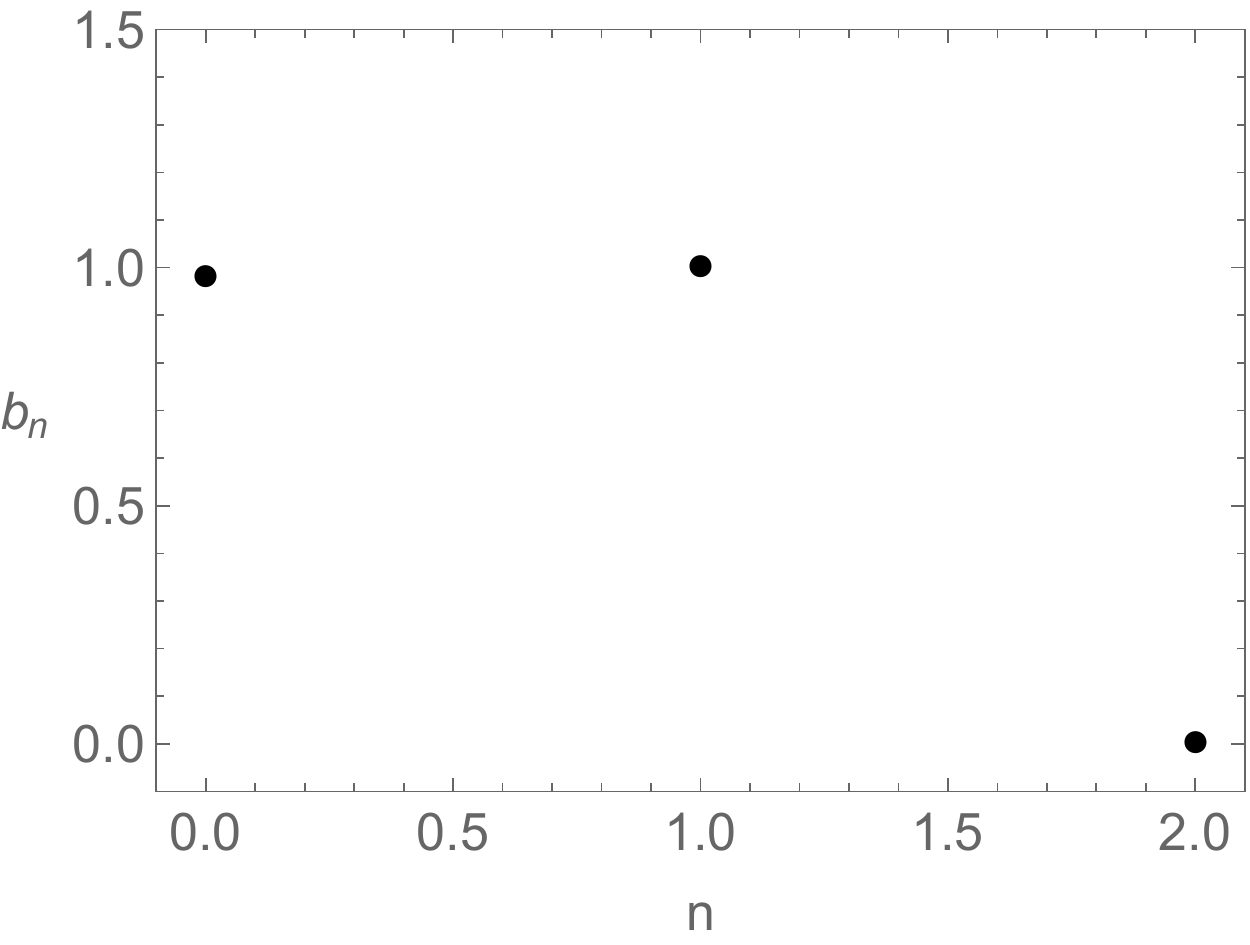} \label{}}
     \subfigure[$\beta=0.5$]
     {\includegraphics[width=7.0cm]{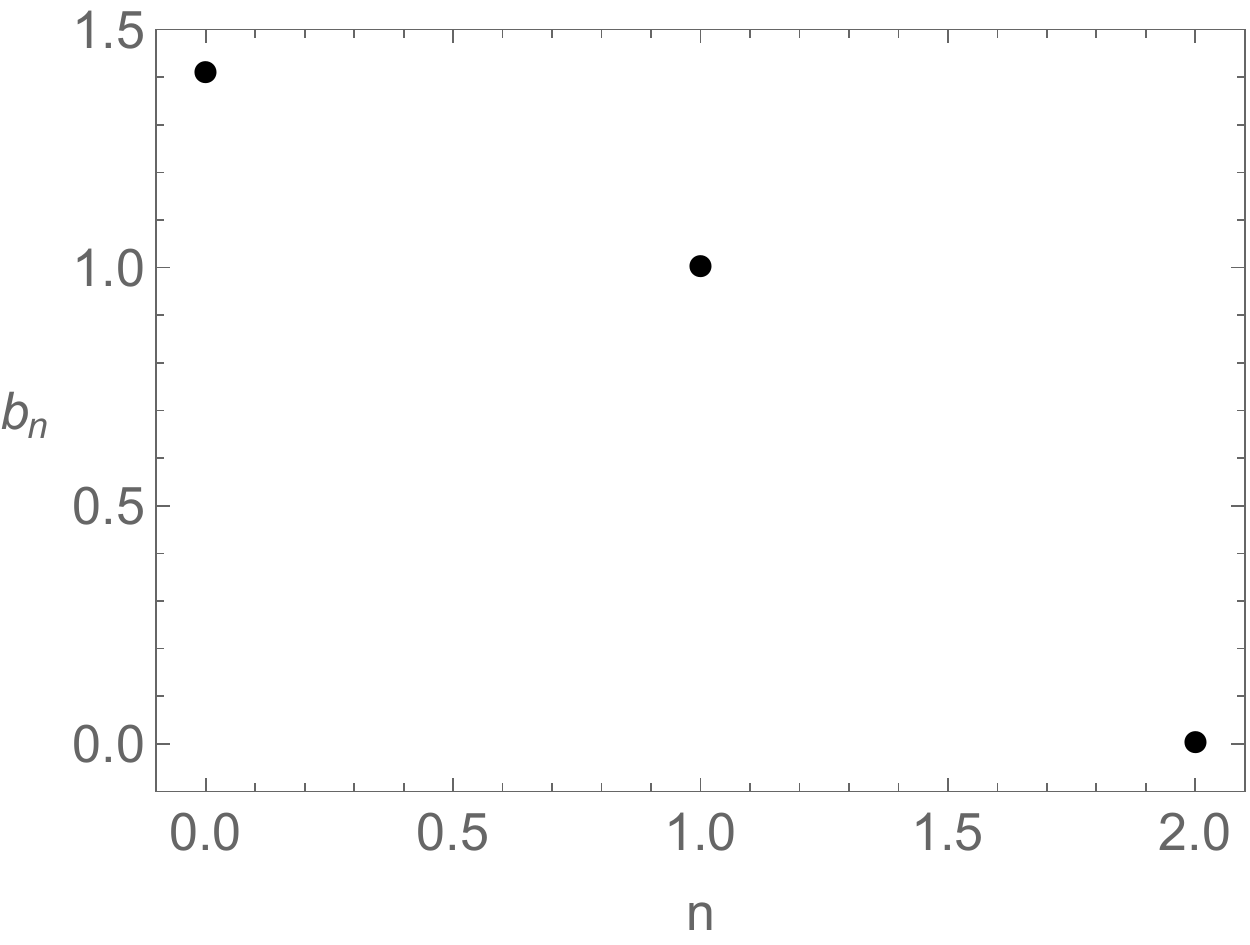} \label{}}
 \caption{Numerically obtained Lanczos coefficient at $\omega = 1$ and $\beta=(1, 0.5)$ (left, right).}\label{LCSHO1}
\end{figure}

Solving the Lanczos algorithm numerically, one can also study $b_n$. In Fig. \ref{LCSHO1},  we display the numerically obtained $b_n$ at $\omega = 1$, which is consistent with our analytic results \eqref{LZC}. For instance
\begin{align}\label{}
\begin{split}
 \text{Analytic results:} \qquad  b_0 |_{\beta=1} \approx 0.97955 \,, \qquad b_0 |_{\beta=0.5} \approx 1.40688 \,.
\end{split}
\end{align}

\paragraph{Periodic Krylov complexity and entropy.}
Next, we evaluate $(\mO_n|\mO(t))$ in \eqref{OnOt} for each $n$ in order to obtain the amplitudes $\varphi_n$ \eqref{EQVA} and consequently Krylov complexity and entropy \eqref{KCF}.

For this purpose, we first compute
\begin{align} \label{WE1}
\begin{split}
\left<m|\mO_0|\ell\right> &=  \sqrt{2 \omega \sinh \left(\frac{\beta \omega}{2}\right)} \, \left<m|x|\ell\right> \\ 
&=  \sqrt{\sinh \left(\frac{\beta \omega}{2}\right)} \, \left( \sqrt{\ell} \, \delta_{m, \ell-1} + \sqrt{\ell+1} \, \delta_{m, \ell+1} \right)  \,,
\end{split}
\end{align}
where we used \eqref{RE0CC} in the first equality and \eqref{FLC1} in the second equality. At the same time,
\begin{align} \label{WE2}
\begin{split}
\left<m|\mO_1|\ell\right> &=  b_1^{-1} \left<m|A_1|\ell\right>\\ 
&= \omega^{-1} b_0^{-1} \left(E_m - E_\ell \right)  \left<m|A_0|\ell\right>  \\
&=  \sqrt{\sinh \left(\frac{\beta \omega}{2}\right)}  \left(E_m - E_\ell \right)   \left( \sqrt{\ell} \, \delta_{m, \ell-1} + \sqrt{\ell+1} \, \delta_{m, \ell+1} \right)  \,,
\end{split}
\end{align}
where we used \eqref{FLC2} and \eqref{FLC1}.

Then, using \eqref{WE1} for $n=0$, we find
\begin{align} \label{K111}
\begin{split}
\varphi_0 &= (\mO_0|\mO(t)) =\frac{1}{Z} \sum_{m, \ell} e^{-\frac{\beta}{2} (E_m + E_\ell)}  e^{it(E_\ell-E_m)}  \left<m|\mO_0|\ell\right>^2  \\
& = \frac{1}{Z} \frac{\csch\left(\frac{\beta\omega}{2}\right)}{2} \cos(\omega t) = \cos(\omega t)\,,\\
\end{split}
\end{align}
where we used \eqref{PATC} in the last equality.
Similarly, using \eqref{WE2} we find for $n=1$
\begin{align} \label{K222}
\begin{split}
\varphi_1 &= i^{-1} (\mO_1|\mO(t)) \\
& =\frac{i^{-1}}{Z} \sum_{m, \ell} e^{-\frac{\beta}{2} (E_m + E_\ell)}  e^{it(E_\ell-E_m)}  \left<m|\mO_1|\ell\right>  \left<\ell|\mO_0|m\right>  \\
&= \frac{-1}{Z} \frac{\csch\left(\frac{\beta\omega}{2}\right)}{2} \sin(\omega t) = -\sin(\omega t)\,.
\end{split}
\end{align}
Therefore, together with \eqref{K111}-\eqref{K222}, the Krylov complexity and entropy \eqref{KCF} of a simple harmonic oscillator can be expressed as 
\begin{align}\label{KYAN}
\begin{split}
  C_K(t) = \sin^2(\omega t) \,,\quad S_K(t) = -\cos^2(\omega t)\, \log (\cos^2(\omega t)) -\sin^2(\omega t)\, \log (\sin^2(\omega t)) \,.
\end{split}
\end{align}
\begin{figure}[]
\centering
     \subfigure[Krylov complexity]
     {\includegraphics[width=7.0cm]{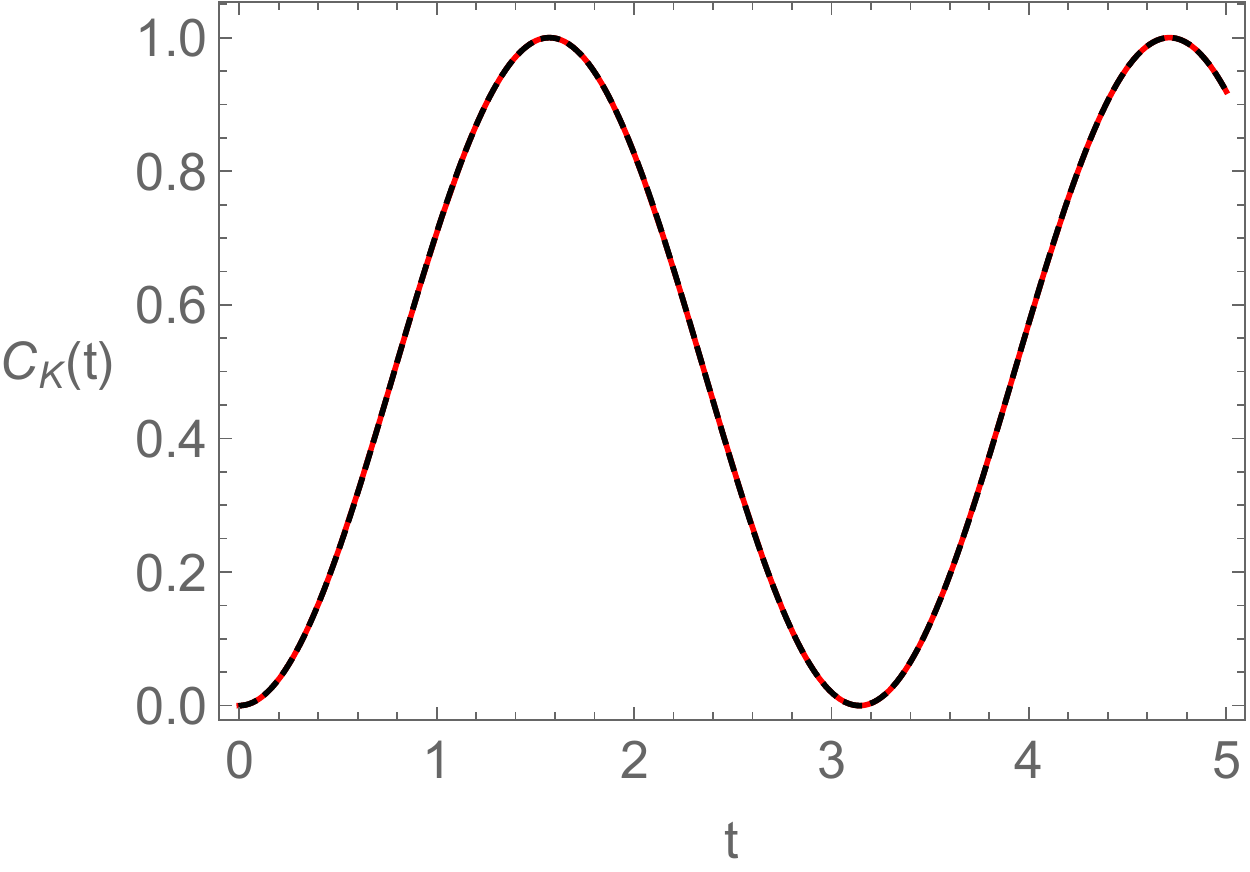} \label{}}
     \subfigure[Krylov entropy]
     {\includegraphics[width=7.0cm]{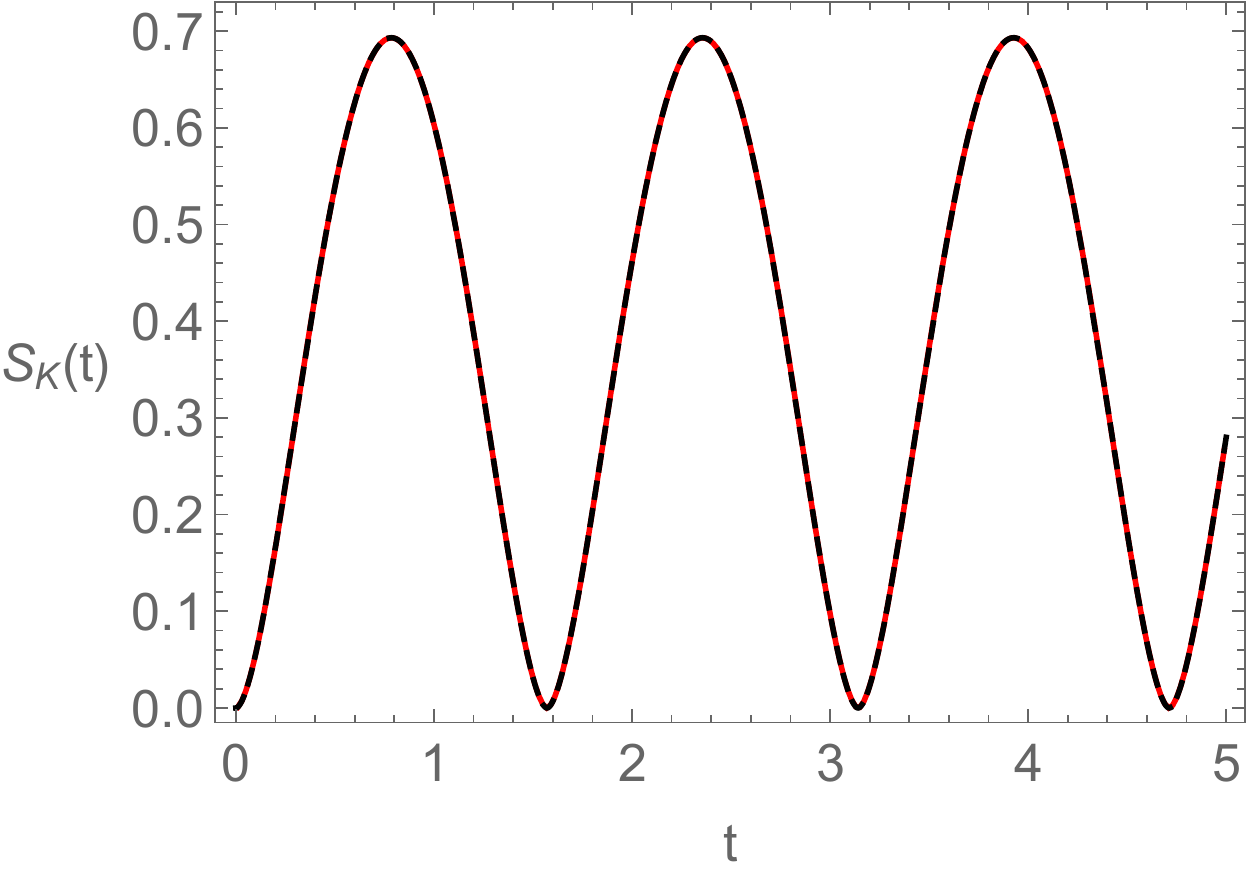} \label{}}
 \caption{Numerically obtained Krylov complexity and entropy at $\omega=\beta = 1$ (red dashed line). The black solid line is analytic formula \eqref{KYAN}.}\label{LCSHO2}
\end{figure}
In Fig. \ref{LCSHO2}, we display the numerically obtained Krylov complexity and entropy consistent with our analytic result \eqref{KYAN}.

%%%%%%%%%%%%%%%%%%%%%%%%%%%%%
%%%%%%%%%%%%%%%%%%%%%%%%%%%%
\paragraph{More on operator dependence.}
Following the procedure described above, one can also find the initial operator dependence for a simple harmonic oscillator.
In particular, we focus on the following three cases:
(I) single operator: ${x}$, ${p}$;
(II) quadratic operator: ${x}^2$, ${p}^2$;
(III) mixed quadratic operator: ${x} {p}$, ${p} {x}$. We summarize our results as follows:
\paragraph{Lanczos coefficients.}
\begin{align}\label{}
\begin{split}
{x} \text{\,\,\,or\,\,\,} {p}&: \quad b_0 = 
\begin{cases}
\sqrt{ \frac{\csch\left(\frac{\beta\omega}{2}\right)}{2\omega} }  \qquad\quad\, (\text{for\,\,} {x}) \\ 
\sqrt{\frac{- \omega\csch\left(\frac{\beta\omega}{2}\right)}{2} } \qquad (\text{for\,\,} {p})
\end{cases}
\,, \qquad\qquad\quad b_1 = \omega \,, \\
& \qquad b_2 = 0 \,, \\ \\
{x}^2 \text{\,\,\,or\,\,\,} {p}^2&: \quad b_0 = 
\begin{cases}
\frac{1}{2 \omega}\sqrt{ 1+3 \csch^2\left(\frac{\beta\omega}{2}\right)}  \quad\, (\text{for\,\,} {x}^2) \\
\frac{\omega}{2} \sqrt{ 1+3 \csch^2\left(\frac{\beta\omega}{2}\right) } \quad\,\,\,\, (\text{for\,\,} {p}^2)
\end{cases}
\,, \qquad b_1 = \sqrt{\frac{8 \omega^2}{5+\cosh(\beta \omega)}} \,, \\
& \qquad b_2 = 2\omega \sqrt{ \frac{3+\cosh(\beta \omega)}{5+\cosh(\beta \omega)} }  \,, \qquad\qquad\qquad\qquad\quad\,\,\,\, b_3 = 0 \,, \\ \\
{x}{p} \text{\,\,\,or\,\,\,} {p}{x}&: \quad b_0 = \frac{i \coth\left(\frac{\beta\omega}{2}\right)}{2} \,, \qquad\qquad\qquad\qquad\qquad\qquad\,\,\  b_1 = 2\omega \sech\left(\frac{\beta\omega}{2}\right) \,,  \\
& \qquad b_2 = 2\omega \tanh\left(\frac{\beta\omega}{2}\right) \,, \qquad\qquad\qquad\qquad\qquad\quad\, b_3 = 0 \,.
\end{split}
\end{align}
We find that $b_n = 0$ at $n=3$ for the quadratic/mixed operator case, unlike the single operator case.
Furthermore, we also report on the case of a more general mixed quadratic operator, $c_1 {x} {p} + c_2 {p} {x}$
\begin{align}\label{}
\begin{split}
c_1 \,{x} {p} \,+\, c_2 \,{p} {x}: \quad b_0 &= \sqrt{-\frac{1}{4} (c_1-c_2)^2 - \frac{1}{4} (c_1+c_2)^2 \csch^2 \left(\frac{\beta \omega}{2}\right) } \, \\
b_1 &= \frac{2\sqrt{2} (c_1+c_2) \omega}{\sqrt{c_1^2 + 6 c_1 c_2 + c_2^2 +(c_1-c_2)^2 \cosh(\beta \omega) }} \,,  \\
b_2 &= \frac{2\omega}{\sqrt{1+ \frac{(c_1+c_2)^2 \csch^2 \left(\frac{\beta \omega}{2}\right) }{(c_1-c_2)^2} }} \,, \qquad\qquad
b_3 = 0 \,.
\end{split}
\end{align}

\paragraph{Krylov complexity.}

\begin{align}\label{ANIDPE21}
\begin{split}
{x} \text{\,\,\,or\,\,\,} {p}&: \quad C_K(t) = \sin^2(\omega t) \,, \\ \\
{x}^2 \text{\,\,\,or\,\,\,} {p}^2&: \quad C_K(t) = \frac{2 }{5+\cosh(\beta \omega)} \left[ \sin^2(2 \omega t) + 8\frac{3+\cosh (\beta \omega)}{5+\cosh(\beta \omega)}\sin^4(\omega t) \right ]\,, \\ \\
{x}{p} \text{\,\,\,or\,\,\,} {p}{x}&: \quad C_K(t) = \sech^2\left(\frac{\beta \omega}{2}\right) \left[ \sin^2(2\omega t) + 8 \tanh^2\left(\frac{\beta \omega}{2}\right) \sin^4(\omega t) \right] \,.
\end{split}
\end{align}
We also present a more general crossed operator case, $c_1 {x} {p} + c_2 {p} {x}$, as
\begin{align}\label{}
\begin{split}
c_1 \,{x} {p} \,+\, c_2 \,{p} {x}: \quad C_K(t) =\,\, &\frac{2(c_1+c_2)^2}{c_1^2+6c_1c_2+c_2^2+(c_1-c_2)^2\cosh(\beta\omega)} \sin^2(2\omega t) \\
&\qquad +\frac{32(c_1^2-c_2^2)\sinh^2\left(\frac{\beta \omega}{2}\right)}{\left( c_1^2+6c_1c_2+c_2^2+(c_1-c_2)^2\cosh(\beta \omega) \right)^2} \sin^4(\omega t) \,.
\end{split}
\end{align}
We also found the analytic expression for the Krylov entropy. However, its analytic expression is not so illuminating so we do not display it here.

\begin{figure}[]
\centering
     \subfigure[${x}$ or ${p}$]
     {\includegraphics[width=4.8cm]{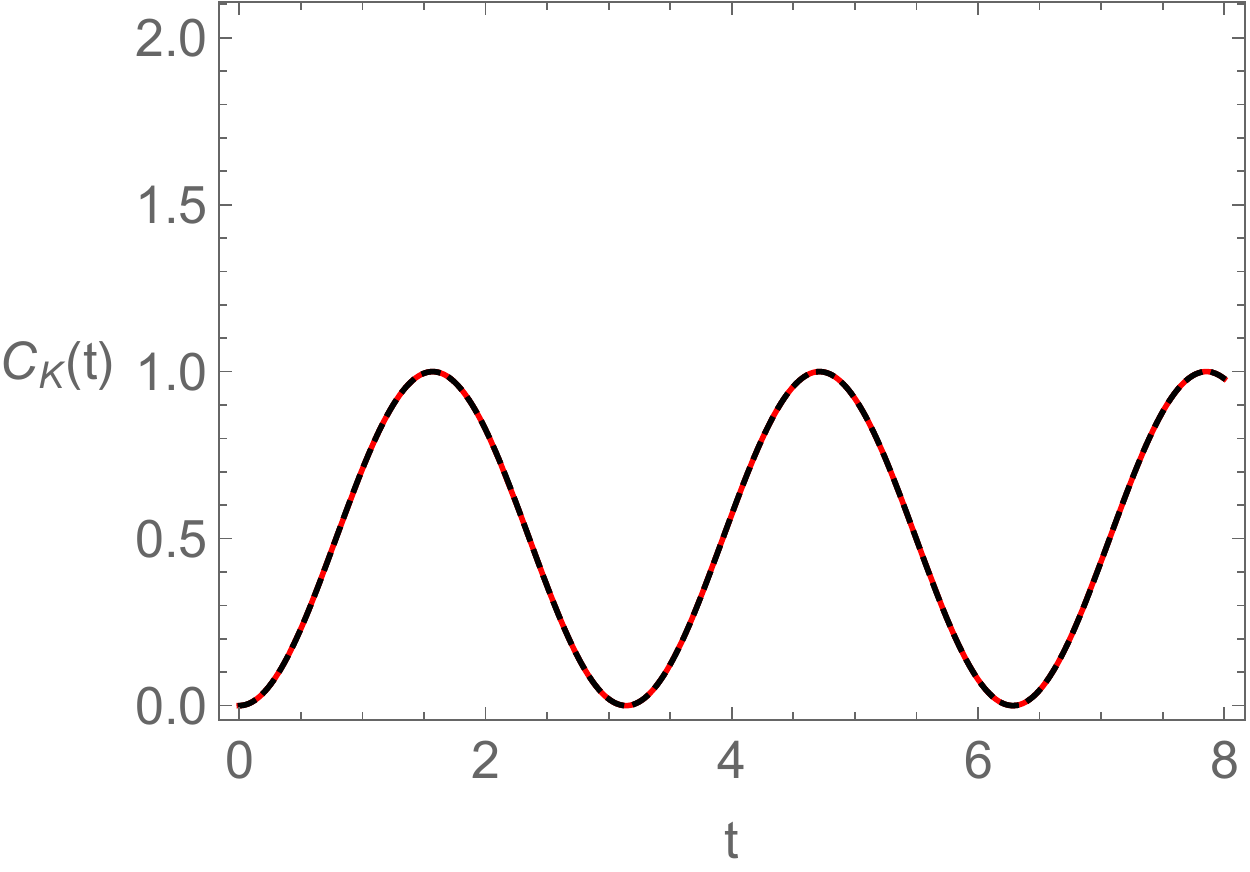} \label{}}
     \subfigure[${x}^2$ or ${p}^2$]
     {\includegraphics[width=4.8cm]{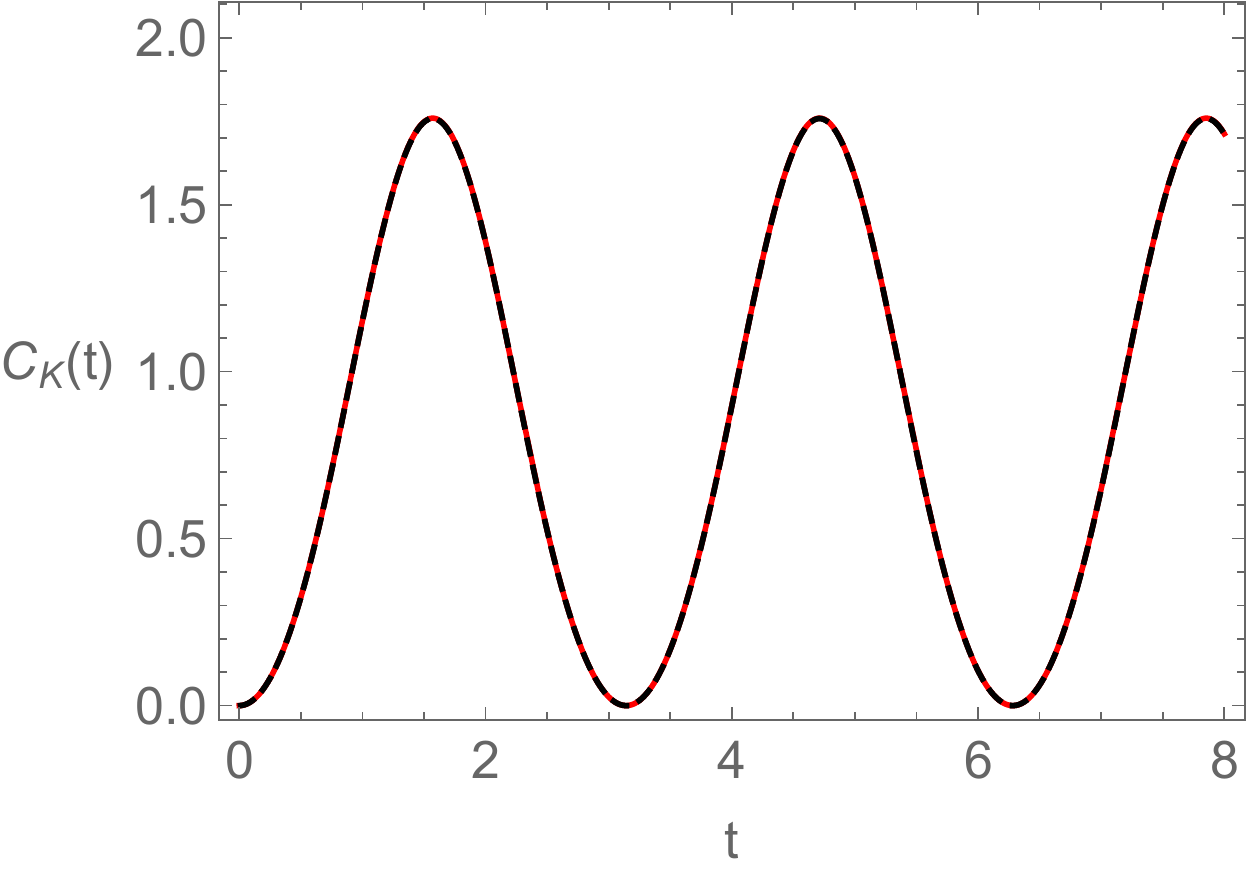} \label{}}
     \subfigure[${x}{p}$ or ${p}{x}$]
     {\includegraphics[width=4.8cm]{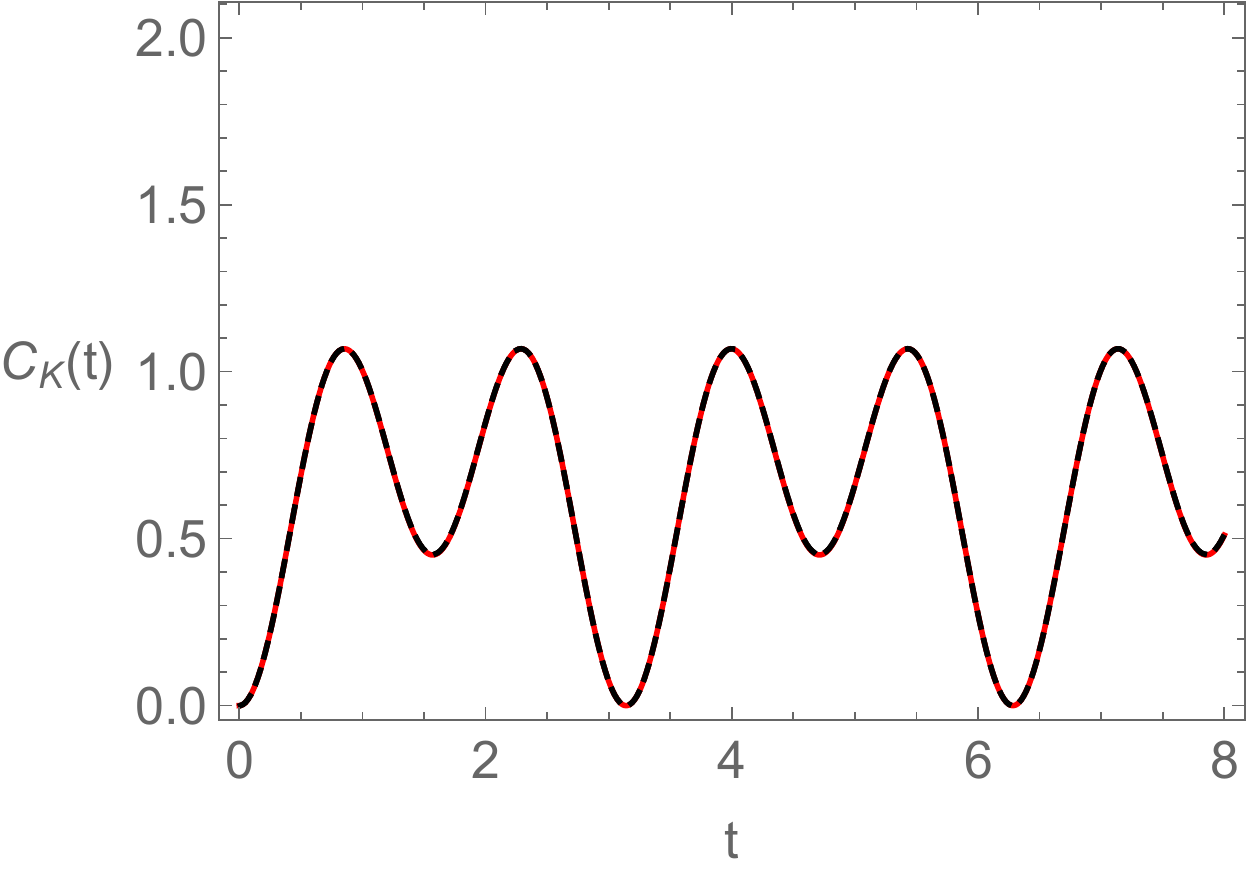} \label{}}
 \caption{Krylov complexity at $\omega = 1, \beta =1/2$. Black dashed lines are analytic result \eqref{ANIDPE21} and red solid lines are numerical result. \textbf{Left:} Single operator case, ${x}$ or ${p}$.  \textbf{Center:} Quadratic operator case, ${x}^2$ or ${p}^2$. \textbf{Right:} Crossed operator case, ${x}{p}$ or ${p}{x}$.}\label{TPK}
\end{figure}

Basically, our observations are twofold. First, Krylov complexity is periodic in time in all cases. Second, a $\beta$ dependence appears for the quadratic/mixed operator case. In Fig.~\ref{TPK} we show the Krylov complexity.
%

%%%%%%%%%%%%%%%%%%%%%%%%%%%%%
%%%%%%%%%%%%%%%%%%%%%%%%%%%%
\subsection{Billiard systems}\label{appmom}

\subsubsection{Lanczos coefficients by moment method}\label{appmoma}

In this section, we briefly review another method to compute the Lanczos coefficients, which we call the moment method~\cite{Parker:2018yvk,Rabinovici:2021qqt,Barbon:2019wsy,RecursionBook,Camargo:2022rnt}.

Using the auto-correlation function from the Lanczos algorithm \eqref{EQVA}, $\varphi_{0}(t)$, one can define the moment $\mu_{2n}$ as the coefficients of the Tayler series of $\varphi_{0}(t)$ as
\begin{align}\label{MOMFOR}
    \varphi_{0}(t):=\sum_{n=0}^{\infty} \mu_{2n}\frac{(it)^{2n}}{(2n)!} \,, \qquad \mu_{2n}:=\frac{1}{i^{2n}}\frac{\textrm{d}^{2n}\varphi_{0}(t)}{\textrm{d}t^{2n}}\Big\vert_{t=0} \,.
\end{align}
Then, using the definition of the inner product \eqref{OnOt} for $\varphi_{0}(t)$, the moments \eqref{MOMFOR} can be written as
\begin{align} \label{MOMENFF1}
\mu_{2n} = \frac{1}{Z} \sum_{m, \ell} (E_\ell-E_m)^{2n} e^{-\frac{\beta}{2} (E_m + E_\ell)}   \left<m|\mO^\dagger_0|\ell\right> \left<\ell|\mO_0|m\right>.
\end{align}

Furthermore, using the determinant of the Hankel matrix of the moments, there exists a non-linear relation between the Lanczos coefficients $b_{n}$ and the moments $\mu_{2n}$ given by
\begin{align}\label{}
    b_{1}^{2n}\cdots b_{n}^{2}=\det\left(\mu_{i+j}\right)_{0\leq i,j\leq n} \,,
\end{align}
where $\mu_{i+j}$ is the Hankel matrix constructed from the moments. Alternatively, this relation can also be expressed via a recursion relation
\begin{align}\label{MOMENFF2}
\begin{split}
M^{(j)}_{2\ell}&=\frac{M^{(j-1)}_{2\ell}}{b_{j-1}^{2}}-\frac{M^{(j-2)}_{2\ell-2}}{b_{j-2}^{2}}~\quad \textrm{with} \quad \ell=j,\ldots,n \,, \\
M^{(0)}_{2\ell}&=\mu_{2\ell} \,, \quad b_{-1} = b_{0}:=1\,, \quad  M^{(-1)}_{2\ell}=0 \,, \\
b_{n}&=\sqrt{M^{(n)}_{2n}} \,,
\end{split}
\end{align}
where the initial reference operator is assumed to be normalized.\\

\begin{figure}[]
\centering
     \subfigure[Stadium billiard ($a/R=1$)]
     {\includegraphics[width=7.0cm]{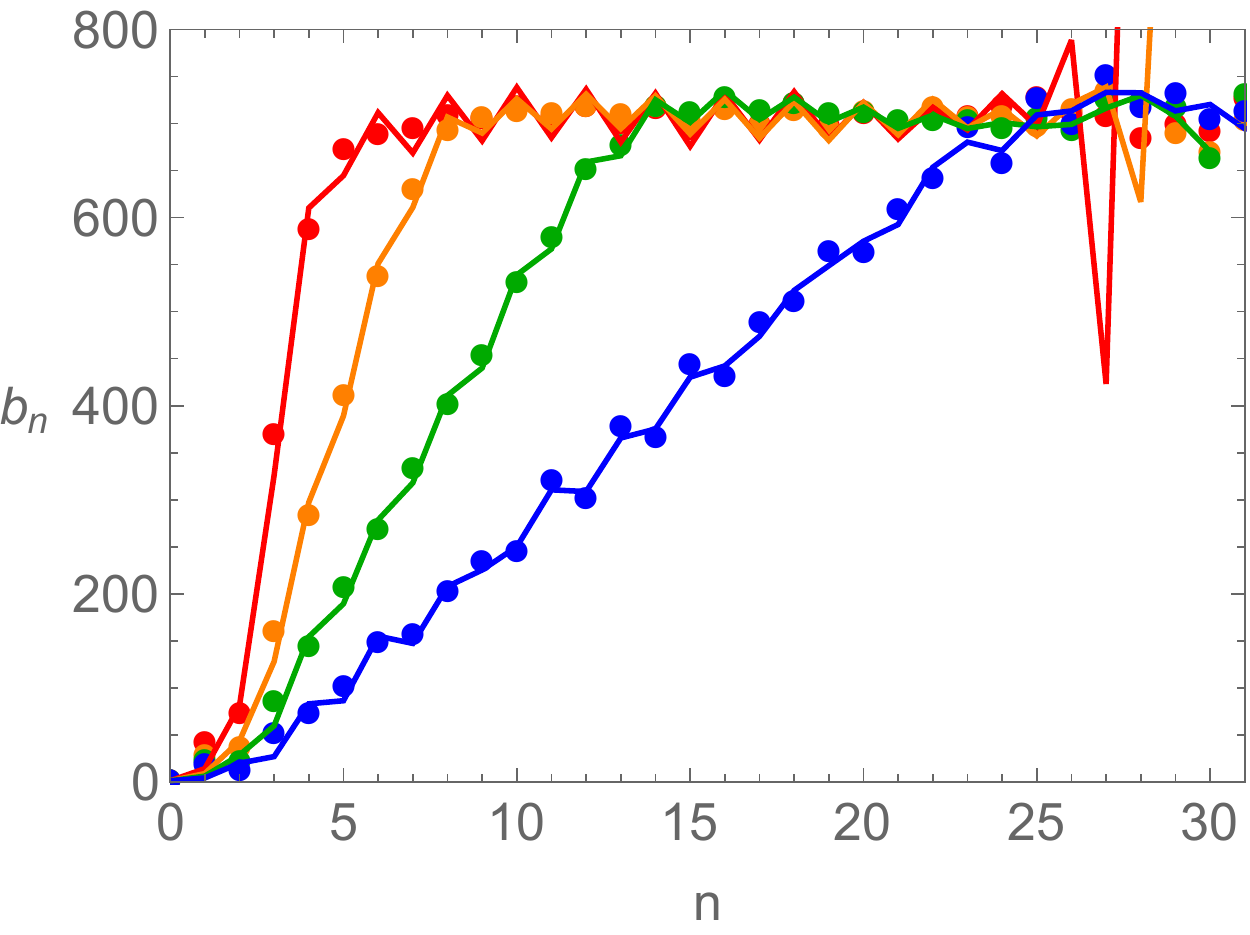} \label{STACIRLT2a}}
     \subfigure[Circle billiard ($a/R=0$)]
     {\includegraphics[width=7.0cm]{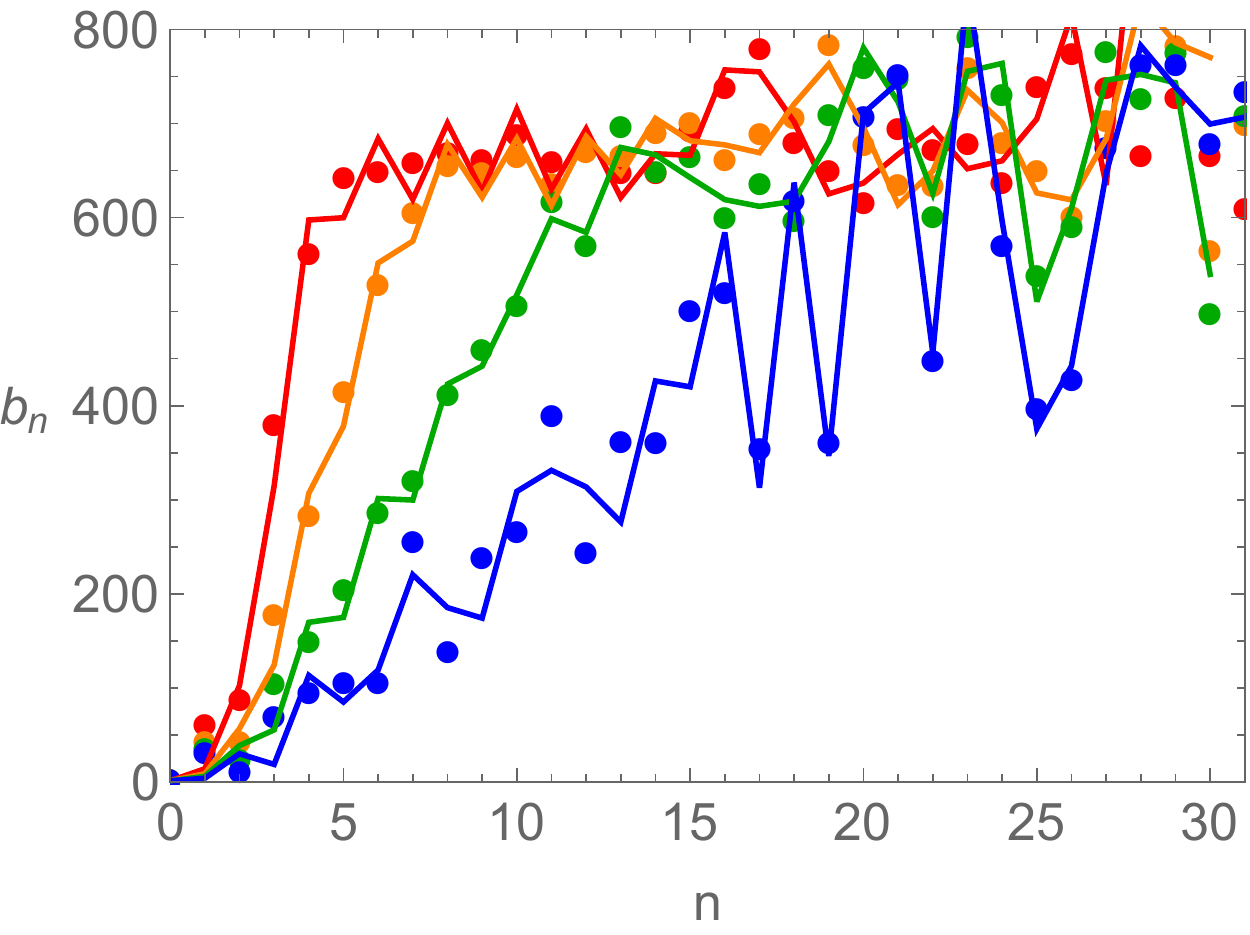} \label{}}
 \caption{Lanczos coefficients at $T = 10, 20, 40, 100$ (blue, green, orange, red). Dots are obtained by Lanczos algorithm, while the solid lines are evaluated by the moment method.}\label{STACIRLT2}
\end{figure}

Utilizing the moment method above, \eqref{MOMENFF1}-\eqref{MOMENFF2}, we display the Lanczos coefficients for the billiards problems \eqref{BILLMO} in Fig. \ref{STACIRLT2}.
One can see that Lanczos coefficients from Lanczos algorithm (dots) are consistent with the one from the moment method (lines). 
%
%We also find that in our numerical code, the Lanczos algorithm may produce the better numerical results than the moment method. For instance, the moment method of billiard problems produces the unstable data in a large $n$ regime at higher temperature: see the red (or orange) line in Fig. \ref{STACIRLT2a}.

%%%%%%%%%%%%%%%%%%%%%%%%%%%%%
%    
%%%%%%%%%%%%%%%%%%%%%%%%%%%%%
\subsubsection{Enlarged time window}\label{appmomb}
\begin{figure}[]
\centering
     \subfigure[$b_n$]
     {\includegraphics[width=4.5cm]{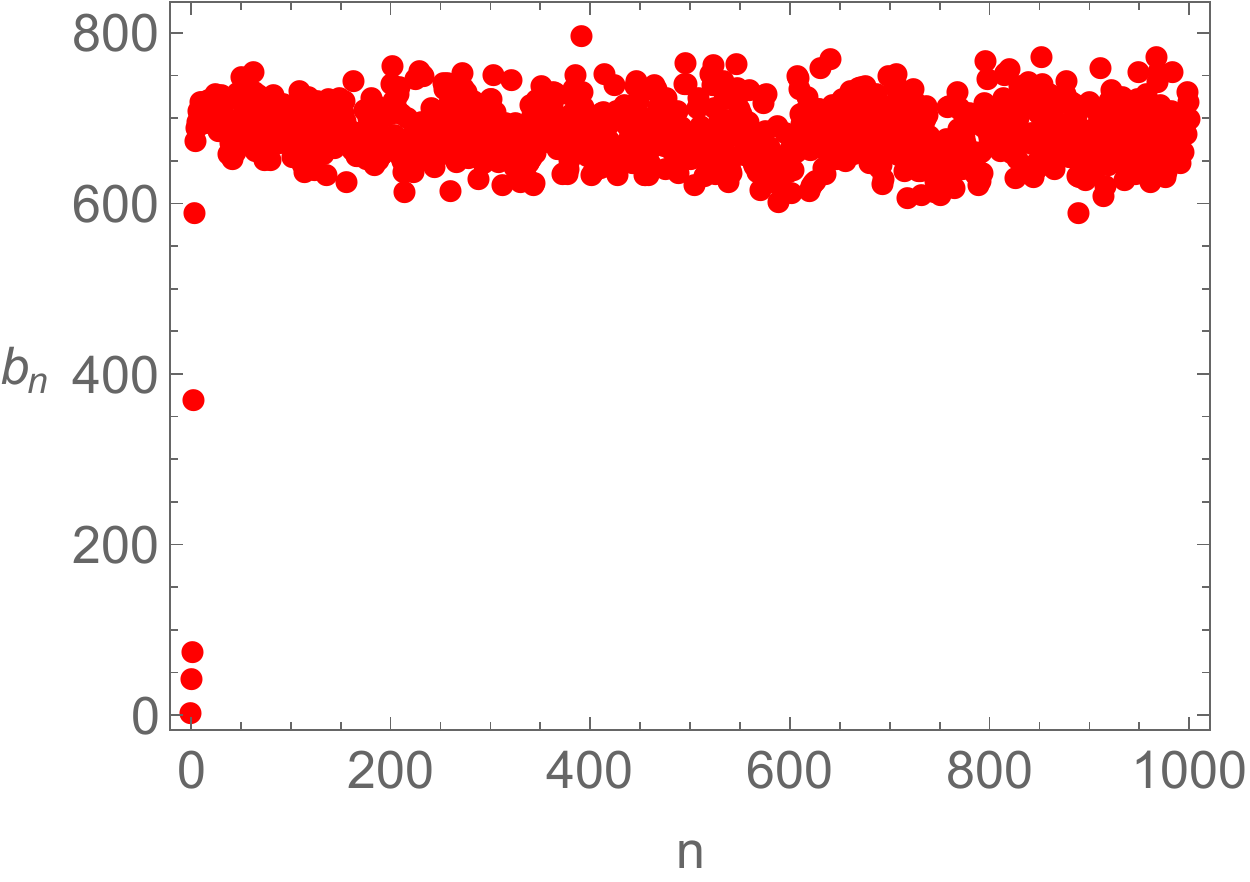} \label{}}
     \subfigure[Normalization condition]
     {\includegraphics[width=5.2cm]{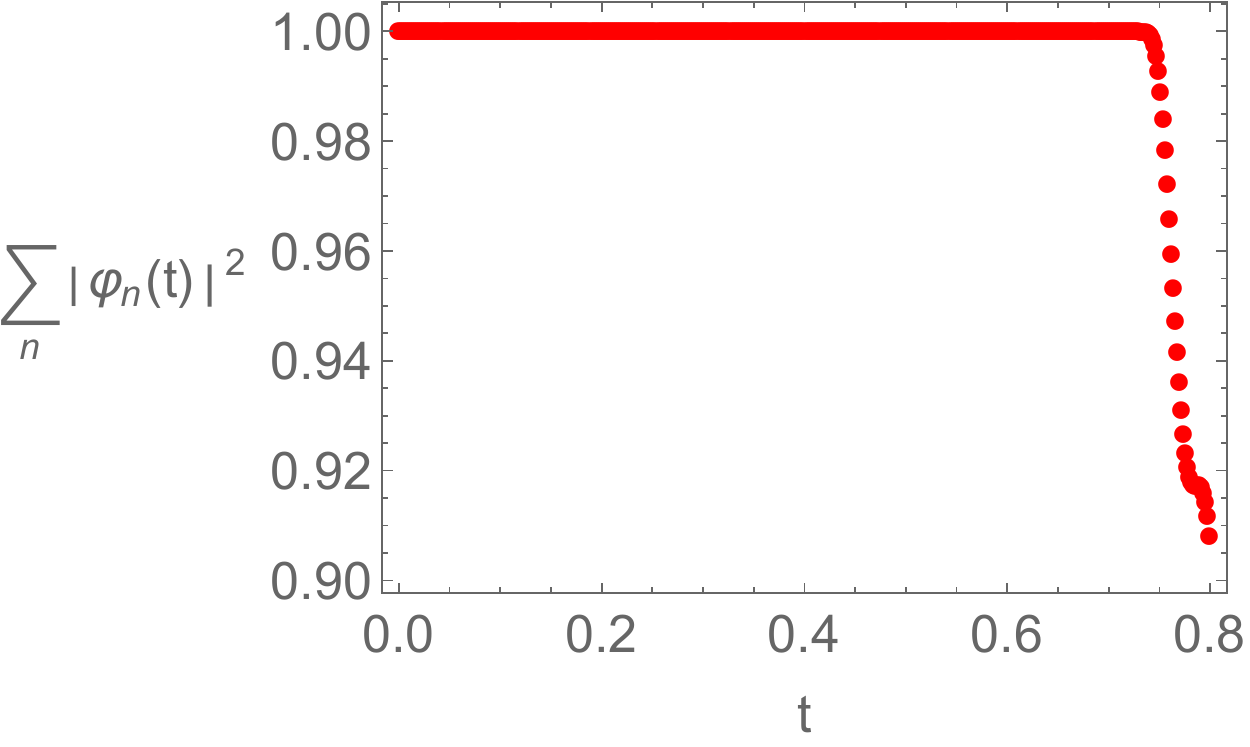} \label{}}
     \subfigure[Krylov complexity]
     {\includegraphics[width=4.5cm]{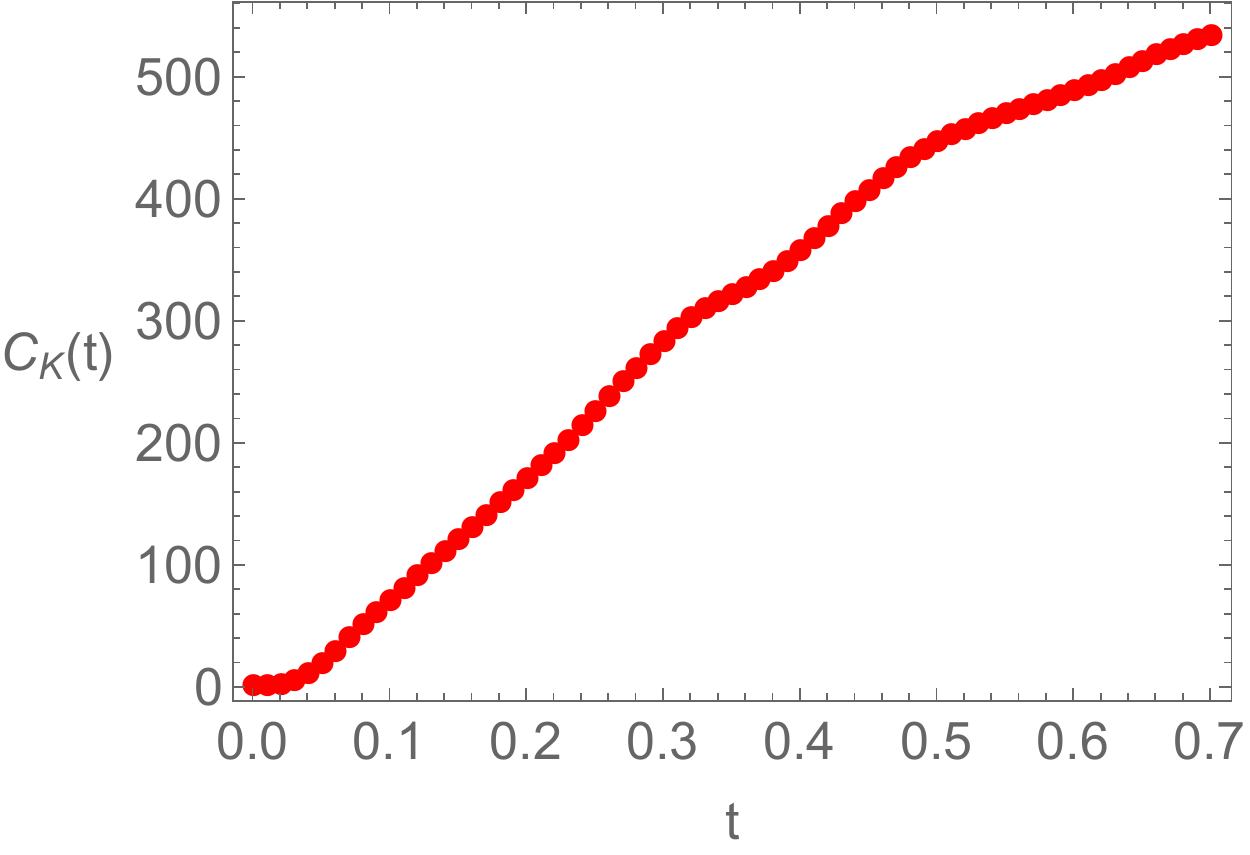} \label{}}
 \caption{Stadium billiard ($a/R=1$) at $T=100$.}\label{APPFIG1112}
\end{figure}

In this section, we examine the dependence of the choice of $n_{\text{max}}$ on the Lanczos coefficient $b_n$. In the main text, we utilized a fixed value of $n_{\text{max}}=100$ and observed that the normalization condition is satisfied up to $t \approx 0.08$. Here, we present results demonstrating that by increasing the value of $n_{\text{max}}$, the time window for which the normalization condition holds can be expanded. For instance, in Fig. \ref{APPFIG1112}, we set $n_{\text{max}}=1000$ (left figure) and observe that the normalization condition can now be satisfied up to $t \approx 0.7$ (center figure), allowing for the computation of the Krylov complexity within an extended time range (right figure).
This suggests that by selecting a larger value of $n_{\text{max}}=1000$, as explored in \cite{Hashimoto:2023swv}, one can investigate the late-time behavior of the Krylov complexity.

%%%%%%%%%%%%%%%%%%%%%%%%%%%%%
%%%%%%%%%%%%%%%%%%%%%%%%%%%%
%\bibliography{HyunSikRefs}

\begin{thebibliography}{10}

\bibitem{einstein1917a}
A.~Einstein, \emph{Zum quantensatz von sommerfeld und epstein}, {\emph{Deutsche
  physikalische Gesellschaft, Verhandlungen} {\bf 19} (1917) 82--92}.

\bibitem{gutzwiller1971}
M.~C. Gutzwiller, \emph{Periodic orbits and classical quantization conditions},
  {\emph{Journal of Mathematical Physics} {\bf 12} (1971) 343}.

\bibitem{gutzwiller1990}
M.~C. Gutzwiller, \emph{Chaos in Classical and Quantum Mechanics}.
\newblock Springer-Verlag, New York, 1990.

\bibitem{berry1977}
M.~V. Berry and M.~Tabor, \emph{Level clustering in the regular spectrum},
  {\emph{Proceedings of the Royal Society of London A} {\bf 356} (1977) 375}.

\bibitem{berry1985}
M.~V. Berry, \emph{Semiclassical theory of spectral rigidity},
  {\emph{Proceedings of the Royal Society of London A} {\bf 400} (1985) 229}.

\bibitem{PhysRevLett.69.1477}
E.~B. Bogomolny, B.~Georgeot, M.-J. Giannoni and C.~Schmit, \emph{Chaotic
  billiards generated by arithmetic groups},
  \href{http://dx.doi.org/10.1103/PhysRevLett.69.1477}{\emph{Phys. Rev. Lett.}
  {\bf 69} (Sep, 1992) 1477--1480}.

\bibitem{PhysRevLett.69.2188}
J.~Bolte, G.~Steil and F.~Steiner, \emph{Arithmetical chaos and violation of
  universality in energy level statistics},
  \href{http://dx.doi.org/10.1103/PhysRevLett.69.2188}{\emph{Phys. Rev. Lett.}
  {\bf 69} (Oct, 1992) 2188--2191}.

\bibitem{PhysRevResearch.3.L012019}
A.~Prakash, J.~H. Pixley and M.~Kulkarni, \emph{Universal spectral form factor
  for many-body localization},
  \href{http://dx.doi.org/10.1103/PhysRevResearch.3.L012019}{\emph{Phys. Rev.
  Res.} {\bf 3} (Feb, 2021) L012019}.

\bibitem{Winer:2020mdc}
M.~Winer, S.-K. Jian and B.~Swingle, \emph{{An exponential ramp in the
  quadratic Sachdev-Ye-Kitaev model}},
  \href{http://dx.doi.org/10.1103/PhysRevLett.125.250602}{\emph{Phys. Rev.
  Lett.} {\bf 125} (2020) 250602}, [\href{http://arxiv.org/abs/2006.15152}{{\tt
  2006.15152}}].

\bibitem{Larkin1969QuasiclassicalMI}
A.~I. Larkin and Y.~N. Ovchinnikov, \emph{Quasiclassical method in the theory
  of superconductivity}, {\emph{Journal of Experimental and Theoretical
  Physics} (1969) }.

\bibitem{Xu:2022vko}
S.~Xu and B.~Swingle, \emph{{Scrambling Dynamics and Out-of-Time Ordered
  Correlators in Quantum Many-Body Systems: a Tutorial}},
  \href{http://arxiv.org/abs/2202.07060}{{\tt 2202.07060}}.

\bibitem{Richter:2022sik}
K.~Richter, J.~D. Urbina and S.~Tomsovic, \emph{{Semiclassical roots of
  universality in many-body quantum chaos}},
  \href{http://dx.doi.org/10.1088/1751-8121/ac9e4e}{\emph{J. Phys. A} {\bf 55}
  (2022) 453001}, [\href{http://arxiv.org/abs/2205.02867}{{\tt 2205.02867}}].

\bibitem{Rozenbaum:2016mmv}
E.~B. Rozenbaum, S.~Ganeshan and V.~Galitski, \emph{{Lyapunov Exponent and
  Out-of-Time-Ordered Correlator\textquoteright{}s Growth Rate in a Chaotic
  System}}, \href{http://dx.doi.org/10.1103/PhysRevLett.118.086801}{\emph{Phys.
  Rev. Lett.} {\bf 118} (2017) 086801},
  [\href{http://arxiv.org/abs/1609.01707}{{\tt 1609.01707}}].

\bibitem{Xu:2019lhc}
T.~Xu, T.~Scaffidi and X.~Cao, \emph{{Does scrambling equal chaos?}},
  \href{http://dx.doi.org/10.1103/PhysRevLett.124.140602}{\emph{Phys. Rev.
  Lett.} {\bf 124} (2020) 140602}, [\href{http://arxiv.org/abs/1912.11063}{{\tt
  1912.11063}}].

\bibitem{Hashimoto:2020xfr}
K.~Hashimoto, K.-B. Huh, K.-Y. Kim and R.~Watanabe, \emph{{Exponential growth
  of out-of-time-order correlator without chaos: inverted harmonic
  oscillator}}, \href{http://dx.doi.org/10.1007/JHEP11(2020)068}{\emph{JHEP}
  {\bf 11} (2020) 068}, [\href{http://arxiv.org/abs/2007.04746}{{\tt
  2007.04746}}].

\bibitem{Dowling:2023hqc}
N.~Dowling, P.~Kos and K.~Modi, \emph{{Scrambling is Necessary but Not
  Sufficient for Chaos}},  \href{http://arxiv.org/abs/2304.07319}{{\tt
  2304.07319}}.

\bibitem{Huang:2017fng}
Y.~Huang, F.~G. S.~L. Brand\~ao and Y.-L. Zhang, \emph{{Finite-size scaling of
  out-of-time-ordered correlators at late times}},
  \href{http://dx.doi.org/10.1103/PhysRevLett.123.010601}{\emph{Phys. Rev.
  Lett.} {\bf 123} (2019) 010601}, [\href{http://arxiv.org/abs/1705.07597}{{\tt
  1705.07597}}].

\bibitem{Parker:2018yvk}
D.~E. Parker, X.~Cao, A.~Avdoshkin, T.~Scaffidi and E.~Altman, \emph{{A
  Universal Operator Growth Hypothesis}},
  \href{http://dx.doi.org/10.1103/PhysRevX.9.041017}{\emph{Phys. Rev. X} {\bf
  9} (2019) 041017}, [\href{http://arxiv.org/abs/1812.08657}{{\tt
  1812.08657}}].

\bibitem{Avdoshkin:2022xuw}
A.~Avdoshkin, A.~Dymarsky and M.~Smolkin, \emph{{Krylov complexity in quantum
  field theory, and beyond}},  \href{http://arxiv.org/abs/2212.14429}{{\tt
  2212.14429}}.

\bibitem{Maldacena:2015waa}
J.~Maldacena, S.~H. Shenker and D.~Stanford, \emph{{A bound on chaos}},
  \href{http://dx.doi.org/10.1007/JHEP08(2016)106}{\emph{JHEP} {\bf 08} (2016)
  106}, [\href{http://arxiv.org/abs/1503.01409}{{\tt 1503.01409}}].

\bibitem{Dymarsky:2021bjq}
A.~Dymarsky and M.~Smolkin, \emph{{Krylov complexity in conformal field
  theory}}, \href{http://dx.doi.org/10.1103/PhysRevD.104.L081702}{\emph{Phys.
  Rev. D} {\bf 104} (2021) L081702},
  [\href{http://arxiv.org/abs/2104.09514}{{\tt 2104.09514}}].

\bibitem{Camargo:2022rnt}
H.~A. Camargo, V.~Jahnke, K.-Y. Kim and M.~Nishida, \emph{{Krylov complexity in
  free and interacting scalar field theories with bounded power spectrum}},
  \href{http://dx.doi.org/10.1007/JHEP05(2023)226}{\emph{JHEP} {\bf 05} (2023)
  226}, [\href{http://arxiv.org/abs/2212.14702}{{\tt 2212.14702}}].

\bibitem{Kundu:2023hbk}
A.~Kundu, V.~Malvimat and R.~Sinha, \emph{{State Dependence of Krylov
  Complexity in $2d$ CFTs}},  \href{http://arxiv.org/abs/2303.03426}{{\tt
  2303.03426}}.

\bibitem{Balasubramanian:2022tpr}
V.~Balasubramanian, P.~Caputa, J.~M. Magan and Q.~Wu, \emph{{Quantum chaos and
  the complexity of spread of states}},
  \href{http://dx.doi.org/10.1103/PhysRevD.106.046007}{\emph{Phys. Rev. D} {\bf
  106} (2022) 046007}, [\href{http://arxiv.org/abs/2202.06957}{{\tt
  2202.06957}}].

\bibitem{Erdmenger:2023shk}
J.~Erdmenger, S.-K. Jian and Z.-Y. Xian, \emph{{Universal chaotic dynamics from
  Krylov space}},  \href{http://arxiv.org/abs/2303.12151}{{\tt 2303.12151}}.

\bibitem{Iliesiu:2021ari}
L.~V. Iliesiu, M.~Mezei and G.~S\'arosi, \emph{{The volume of the black hole
  interior at late times}},
  \href{http://dx.doi.org/10.1007/JHEP07(2022)073}{\emph{JHEP} {\bf 07} (2022)
  073}, [\href{http://arxiv.org/abs/2107.06286}{{\tt 2107.06286}}].

\bibitem{Barbon:2019wsy}
J.~L.~F. Barb\'on, E.~Rabinovici, R.~Shir and R.~Sinha, \emph{{On The Evolution
  Of Operator Complexity Beyond Scrambling}},
  \href{http://dx.doi.org/10.1007/JHEP10(2019)264}{\emph{JHEP} {\bf 10} (2019)
  264}, [\href{http://arxiv.org/abs/1907.05393}{{\tt 1907.05393}}].

\bibitem{Avdoshkin:2019trj}
A.~Avdoshkin and A.~Dymarsky, \emph{{Euclidean operator growth and quantum
  chaos}},
  \href{http://dx.doi.org/10.1103/PhysRevResearch.2.043234}{\emph{Phys. Rev.
  Res.} {\bf 2} (2020) 043234}, [\href{http://arxiv.org/abs/1911.09672}{{\tt
  1911.09672}}].

\bibitem{Dymarsky:2019elm}
A.~Dymarsky and A.~Gorsky, \emph{{Quantum chaos as delocalization in Krylov
  space}}, \href{http://dx.doi.org/10.1103/PhysRevB.102.085137}{\emph{Phys.
  Rev. B} {\bf 102} (2020) 085137},
  [\href{http://arxiv.org/abs/1912.12227}{{\tt 1912.12227}}].

\bibitem{Rabinovici:2020ryf}
E.~Rabinovici, A.~S\'anchez-Garrido, R.~Shir and J.~Sonner, \emph{{Operator
  complexity: a journey to the edge of Krylov space}},
  \href{http://dx.doi.org/10.1007/JHEP06(2021)062}{\emph{JHEP} {\bf 06} (2021)
  062}, [\href{http://arxiv.org/abs/2009.01862}{{\tt 2009.01862}}].

\bibitem{Cao:2020zls}
X.~Cao, \emph{{A statistical mechanism for operator growth}},
  \href{http://dx.doi.org/10.1088/1751-8121/abe77c}{\emph{J. Phys. A} {\bf 54}
  (2021) 144001}, [\href{http://arxiv.org/abs/2012.06544}{{\tt 2012.06544}}].

\bibitem{Kim:2021okd}
J.~Kim, J.~Murugan, J.~Olle and D.~Rosa, \emph{{Operator delocalization in
  quantum networks}},
  \href{http://dx.doi.org/10.1103/PhysRevA.105.L010201}{\emph{Phys. Rev. A}
  {\bf 105} (2022) L010201}, [\href{http://arxiv.org/abs/2109.05301}{{\tt
  2109.05301}}].

\bibitem{Rabinovici:2021qqt}
E.~Rabinovici, A.~S\'anchez-Garrido, R.~Shir and J.~Sonner, \emph{{Krylov
  localization and suppression of complexity}},
  \href{http://dx.doi.org/10.1007/JHEP03(2022)211}{\emph{JHEP} {\bf 03} (2022)
  211}, [\href{http://arxiv.org/abs/2112.12128}{{\tt 2112.12128}}].

\bibitem{Trigueros:2021rwj}
F.~B. Trigueros and C.-J. Lin, \emph{{Krylov complexity of many-body
  localization: Operator localization in Krylov basis}},
  \href{http://arxiv.org/abs/2112.04722}{{\tt 2112.04722}}.

\bibitem{Fan:2022xaa}
Z.-Y. Fan, \emph{{Universal relation for operator complexity}},
  \href{http://dx.doi.org/10.1103/PhysRevA.105.062210}{\emph{Phys. Rev. A} {\bf
  105} (2022) 062210}, [\href{http://arxiv.org/abs/2202.07220}{{\tt
  2202.07220}}].

\bibitem{Heveling:2022hth}
R.~Heveling, J.~Wang and J.~Gemmer, \emph{{Numerically Probing the Universal
  Operator Growth Hypothesis}},  \href{http://arxiv.org/abs/2203.00533}{{\tt
  2203.00533}}.

\bibitem{Bhattacharjee:2022vlt}
B.~Bhattacharjee, X.~Cao, P.~Nandy and T.~Pathak, \emph{{Krylov complexity in
  saddle-dominated scrambling}},
  \href{http://dx.doi.org/10.1007/JHEP05(2022)174}{\emph{JHEP} {\bf 05} (2022)
  174}, [\href{http://arxiv.org/abs/2203.03534}{{\tt 2203.03534}}].

\bibitem{Caputa:2022eye}
P.~Caputa and S.~Liu, \emph{{Quantum complexity and topological phases of
  matter}},  \href{http://arxiv.org/abs/2205.05688}{{\tt 2205.05688}}.

\bibitem{Muck:2022xfc}
W.~M\"uck and Y.~Yang, \emph{{Krylov complexity and orthogonal polynomials}},
  \href{http://dx.doi.org/10.1016/j.nuclphysb.2022.115948}{\emph{Nucl. Phys. B}
  {\bf 984} (2022) 115948}, [\href{http://arxiv.org/abs/2205.12815}{{\tt
  2205.12815}}].

\bibitem{Rabinovici:2022beu}
E.~Rabinovici, A.~S\'anchez-Garrido, R.~Shir and J.~Sonner, \emph{{Krylov
  complexity from integrability to chaos}},
  \href{http://dx.doi.org/10.1007/JHEP07(2022)151}{\emph{JHEP} {\bf 07} (2022)
  151}, [\href{http://arxiv.org/abs/2207.07701}{{\tt 2207.07701}}].

\bibitem{He:2022ryk}
S.~He, P.~H.~C. Lau, Z.-Y. Xian and L.~Zhao, \emph{{Quantum chaos, scrambling
  and operator growth in $ T\overline{T} $ deformed SYK models}},
  \href{http://dx.doi.org/10.1007/JHEP12(2022)070}{\emph{JHEP} {\bf 12} (2022)
  070}, [\href{http://arxiv.org/abs/2209.14936}{{\tt 2209.14936}}].

\bibitem{Hornedal:2022pkc}
N.~H\"ornedal, N.~Carabba, A.~S. Matsoukas-Roubeas and A.~del Campo,
  \emph{{Ultimate Physical Limits to the Growth of Operator Complexity}},
  \href{http://arxiv.org/abs/2202.05006}{{\tt 2202.05006}}.

\bibitem{Alishahiha:2022nhe}
M.~Alishahiha, \emph{{On Quantum Complexity}},
  \href{http://arxiv.org/abs/2209.14689}{{\tt 2209.14689}}.

\bibitem{Alishahiha:2022anw}
M.~Alishahiha and S.~Banerjee, \emph{{A universal approach to Krylov State and
  Operator complexities}},  \href{http://arxiv.org/abs/2212.10583}{{\tt
  2212.10583}}.

\bibitem{Bhattacharyya:2023dhp}
A.~Bhattacharyya, D.~Ghosh and P.~Nandi, \emph{{Operator growth and Krylov
  Complexity in Bose-Hubbard Model}},
  \href{http://arxiv.org/abs/2306.05542}{{\tt 2306.05542}}.

\bibitem{Magan:2020iac}
J.~M. Mag\'an and J.~Sim\'on, \emph{{On operator growth and emergent Poincar\'e
  symmetries}}, \href{http://dx.doi.org/10.1007/JHEP05(2020)071}{\emph{JHEP}
  {\bf 05} (2020) 071}, [\href{http://arxiv.org/abs/2002.03865}{{\tt
  2002.03865}}].

\bibitem{Jian:2020qpp}
S.-K. Jian, B.~Swingle and Z.-Y. Xian, \emph{{Complexity growth of operators in
  the SYK model and in JT gravity}},
  \href{http://dx.doi.org/10.1007/JHEP03(2021)014}{\emph{JHEP} {\bf 03} (2021)
  014}, [\href{http://arxiv.org/abs/2008.12274}{{\tt 2008.12274}}].

\bibitem{Rabinovici:2023yex}
E.~Rabinovici, A.~S\'anchez-Garrido, R.~Shir and J.~Sonner, \emph{{A bulk
  manifestation of Krylov complexity}},
  \href{http://arxiv.org/abs/2305.04355}{{\tt 2305.04355}}.

\bibitem{Caputa:2021ori}
P.~Caputa and S.~Datta, \emph{{Operator growth in 2d CFT}},
  \href{http://dx.doi.org/10.1007/JHEP12(2021)188}{\emph{JHEP} {\bf 12} (2021)
  188}, [\href{http://arxiv.org/abs/2110.10519}{{\tt 2110.10519}}].

\bibitem{Caputa:2021sib}
P.~Caputa, J.~M. Magan and D.~Patramanis, \emph{{Geometry of Krylov
  complexity}},
  \href{http://dx.doi.org/10.1103/PhysRevResearch.4.013041}{\emph{Phys. Rev.
  Res.} {\bf 4} (2022) 013041}, [\href{http://arxiv.org/abs/2109.03824}{{\tt
  2109.03824}}].

\bibitem{Patramanis:2021lkx}
D.~Patramanis, \emph{{Probing the entanglement of operator growth}},
  \href{http://dx.doi.org/10.1093/ptep/ptac081}{\emph{PTEP} {\bf 2022} (2022)
  063A01}, [\href{http://arxiv.org/abs/2111.03424}{{\tt 2111.03424}}].

\bibitem{Patramanis:2023cwz}
D.~Patramanis and W.~Sybesma, \emph{{Krylov complexity in a natural basis for
  the Schr\"odinger algebra}},  \href{http://arxiv.org/abs/2306.03133}{{\tt
  2306.03133}}.

\bibitem{Chattopadhyay:2023fob}
A.~Chattopadhyay, A.~Mitra and H.~J.~R. van Zyl, \emph{{Spread complexity as
  classical dilaton solutions}},  \href{http://arxiv.org/abs/2302.10489}{{\tt
  2302.10489}}.

\bibitem{Iizuka:2023pov}
N.~Iizuka and M.~Nishida, \emph{{Krylov complexity in the IP matrix model}},
  \href{http://arxiv.org/abs/2306.04805}{{\tt 2306.04805}}.

\bibitem{Pal:2023yik}
K.~Pal, K.~Pal, A.~Gill and T.~Sarkar, \emph{{Time evolution of spread
  complexity and statistics of work done in quantum quenches}},
  \href{http://arxiv.org/abs/2304.09636}{{\tt 2304.09636}}.

\bibitem{Bhattacharya:2022gbz}
A.~Bhattacharya, P.~Nandy, P.~P. Nath and H.~Sahu, \emph{{Operator growth and
  Krylov construction in dissipative open quantum systems}},
  \href{http://arxiv.org/abs/2207.05347}{{\tt 2207.05347}}.

\bibitem{Liu:2022god}
C.~Liu, H.~Tang and H.~Zhai, \emph{{Krylov Complexity in Open Quantum
  Systems}},  \href{http://arxiv.org/abs/2207.13603}{{\tt 2207.13603}}.

\bibitem{Bhattacharjee:2022lzy}
B.~Bhattacharjee, X.~Cao, P.~Nandy and T.~Pathak, \emph{{An operator growth
  hypothesis for open quantum systems}},
  \href{http://arxiv.org/abs/2212.06180}{{\tt 2212.06180}}.

\bibitem{Bhattacharya:2023zqt}
A.~Bhattacharya, P.~Nandy, P.~P. Nath and H.~Sahu, \emph{{On Krylov complexity
  in open systems: an approach via bi-Lanczos algorithm}},
  \href{http://arxiv.org/abs/2303.04175}{{\tt 2303.04175}}.

\bibitem{Hashimoto:2023swv}
K.~Hashimoto, K.~Murata, N.~Tanahashi and R.~Watanabe, \emph{{Krylov complexity
  and chaos in quantum mechanics}},
  \href{http://arxiv.org/abs/2305.16669}{{\tt 2305.16669}}.

\bibitem{Susskind:2014moa}
L.~Susskind, \emph{{Entanglement is not enough}},
  \href{http://dx.doi.org/10.1002/prop.201500095}{\emph{Fortsch. Phys.} {\bf
  64} (2016) 49--71}, [\href{http://arxiv.org/abs/1411.0690}{{\tt 1411.0690}}].

\bibitem{Susskind:2014rva}
L.~Susskind, \emph{{Computational Complexity and Black Hole Horizons}},
  \href{http://dx.doi.org/10.1002/prop.201500093,
  10.1002/prop.201500092}{\emph{Fortsch. Phys.} {\bf 64} (2016) 44--48},
  [\href{http://arxiv.org/abs/1403.5695}{{\tt 1403.5695}}].

\bibitem{Balasubramanian:2019wgd}
V.~Balasubramanian, M.~Decross, A.~Kar and O.~Parrikar, \emph{{Quantum
  Complexity of Time Evolution with Chaotic Hamiltonians}},
  \href{http://dx.doi.org/10.1007/JHEP01(2020)134}{\emph{JHEP} {\bf 01} (2020)
  134}, [\href{http://arxiv.org/abs/1905.05765}{{\tt 1905.05765}}].

\bibitem{Balasubramanian:2021mxo}
V.~Balasubramanian, M.~DeCross, A.~Kar, Y.~C. Li and O.~Parrikar,
  \emph{{Complexity growth in integrable and chaotic models}},
  \href{http://dx.doi.org/10.1007/JHEP07(2021)011}{\emph{JHEP} {\bf 07} (2021)
  011}, [\href{http://arxiv.org/abs/2101.02209}{{\tt 2101.02209}}].

\bibitem{Haferkamp:2021uxo}
J.~Haferkamp, P.~Faist, N.~B.~T. Kothakonda, J.~Eisert and N.~Y. Halpern,
  \emph{{Linear growth of quantum circuit complexity}},
  \href{http://dx.doi.org/10.1038/s41567-022-01539-6}{\emph{Nature Phys.} {\bf
  18} (2022) 528--532}, [\href{http://arxiv.org/abs/2106.05305}{{\tt
  2106.05305}}].

\bibitem{Sinai_1970}
Y.~G. Sinai, \emph{Dynamical systems with elastic reflections},
  \href{http://dx.doi.org/10.1070/rm1970v025n02abeh003794}{\emph{Russian
  Mathematical Surveys} {\bf 25} (apr, 1970) 137--189}.

\bibitem{Bunimovich_1975}
L.~A. Bunimovich, \emph{On ergodic properties of certain billiards},
  \href{http://dx.doi.org/10.1007/bf01075700}{\emph{Functional Analysis and Its
  Applications} {\bf 8} (1975) 254--255}.

\bibitem{Bunimovich_1979}
L.~A. Bunimovich, \emph{On the ergodic properties of nowhere dispersing
  billiards}, \href{http://dx.doi.org/10.1007/bf01197884}{\emph{Communications
  in Mathematical Physics} {\bf 65} (oct, 1979) 295--312}.

\bibitem{Bunimovich_1991}
L.~A. Bunimovich, Y.~G. Sinai and N.~I. Chernov, \emph{Statistical properties
  of two-dimensional hyperbolic billiards},
  \href{http://dx.doi.org/10.1070/rm1991v046n04abeh002827}{\emph{Russian
  Mathematical Surveys} {\bf 46} (aug, 1991) 47--106}.

\bibitem{Benettin:1978aa}
G.~Benettin, \emph{Numerical experiments on the free motion of a point mass
  moving in a plane convex region: Stochastic transition and entropy},
  \href{http://dx.doi.org/10.1103/PhysRevA.17.773}{\emph{Physical Review A}
  {\bf 17} (1978) 773--785}.

\bibitem{Hashimoto:2017oit}
K.~Hashimoto, K.~Murata and R.~Yoshii, \emph{{Out-of-time-order correlators in
  quantum mechanics}},
  \href{http://dx.doi.org/10.1007/JHEP10(2017)138}{\emph{JHEP} {\bf 10} (2017)
  138}, [\href{http://arxiv.org/abs/1703.09435}{{\tt 1703.09435}}].

\bibitem{BenettinStochastic}
G.~Benettin and J.~M. Strelcyn, \emph{Numerical experiments on the free motion
  of a point mass moving in a plane convex region: Stochastic transition and
  entropy}, \href{http://dx.doi.org/10.1103/PhysRevA.17.773}{\emph{Phys. Rev.
  A} {\bf 17} (Feb, 1978) 773--785}.

\bibitem{McDonaldSpectrum}
S.~W. McDonald and A.~N. Kaufman, \emph{Spectrum and eigenfunctions for a
  hamiltonian with stochastic trajectories},
  \href{http://dx.doi.org/10.1103/PhysRevLett.42.1189}{\emph{Phys. Rev. Lett.}
  {\bf 42} (Apr, 1979) 1189--1191}.

\bibitem{CasatiSpectra}
G.~Cassati, F.~Valz-Gris and I.~Guarnieri, \emph{On the connection between
  quantization of nonintegrable systems and statisctical theory of spectra},
  \href{http://dx.doi.org/10.1007/BF02798790}{\emph{Lett. Nuovo Cim.} {\bf 28}
  (1980) 279--282}.

\bibitem{Bauls_2011}
M.~C. Ba{\~{n}}uls, J.~I. Cirac and M.~B. Hastings, \emph{Strong and weak
  thermalization of infinite nonintegrable quantum systems},
  \href{http://dx.doi.org/10.1103/physrevlett.106.050405}{\emph{Physical Review
  Letters} {\bf 106} (feb, 2011) }.

\bibitem{Nosaka:2018iat}
T.~Nosaka, D.~Rosa and J.~Yoon, \emph{{The Thouless time for mass-deformed
  SYK}}, \href{http://dx.doi.org/10.1007/JHEP09(2018)041}{\emph{JHEP} {\bf 09}
  (2018) 041}, [\href{http://arxiv.org/abs/1804.09934}{{\tt 1804.09934}}].

\bibitem{Du:2022ocp}
B.-n. Du and M.-x. Huang, \emph{{Krylov Complexity in Calabi-Yau Quantum
  Mechanics}},  \href{http://arxiv.org/abs/2212.02926}{{\tt 2212.02926}}.

\bibitem{RecursionBook}
V.~S. Viswanath and G.~M{\"u}ller, \emph{The Recursion Method: Application to
  Many-Body Dynamics}.
\newblock Springer Berlin, Heidelberg, Germany, 1994.

\end{thebibliography}
\bibliographystyle{JHEP}
\providecommand{\href}[2]{#2}\begingroup\raggedright\endgroup

\end{document}